\documentclass[
 superscriptaddress,
 reprint,
 amsmath,amssymb,
 aps,
]{revtex4-1}

\usepackage{graphicx}
\usepackage{dcolumn}
\usepackage{bm}

\usepackage{amsmath}
\usepackage{amsfonts}
\usepackage{amssymb}
\usepackage{dsfont}
\usepackage{tikz}
\usepackage{wrapfig}
\usepackage{braket}
\usepackage[caption=false]{subfig}
\usepackage{float}
\usepackage{placeins}
\usepackage{bm}
\usepackage{epstopdf}
\usepackage{hyperref}
\usepackage[percent]{overpic}
\usepackage{xcolor}

\begin{document}

\title{Extended Bose-Hubbard models with Rydberg macrodimer dressing}

\author{Mathieu Barbier}
\email{barbier@itp.uni-frankfurt.de}
\affiliation{Institut f\"ur Theoretische Physik, Goethe-Universit\"at, 60438 Frankfurt/Main, Germany}
\author{Simon Hollerith}
\affiliation{Max-Planck-Institut f\"ur Quantenoptik, 85748 Garching, Germany}
\author{Walter Hofstetter}
\affiliation{Institut f\"ur Theoretische Physik, Goethe-Universit\"at, 60438 Frankfurt/Main, Germany}

\date{\today}

\begin{abstract}
Extended Hubbard models have proven to bear novel phases of matter, but their experimental realization remains challenging. In this work we propose to use bosonic quantum gases dressed with molecular bound states in Rydberg interaction potentials for the observation of these quantum states. We study the molecular Rabi coupling with respect to the effective principal quantum number and trapping frequency of the ground state atoms for various molecular potentials of Rubidium and Potassium, and the hereby resulting dressed interaction strength. Additionally, we propose a two-color excitation scheme which significantly increases the dressed interaction and cancels otherwise limiting AC Stark shifts. We study the various equilibrium phases of the corresponding extended Bose-Hubbard model by means of the Cluster Gutzwiller approach and perform time evolution simulations via the Lindblad master equation. We find a supersolid phase by slowly ramping the molecular Rabi coupling of an initially prepared superfluid and discuss the role of dissipation.
\end{abstract}

\maketitle

\section{Introduction}
In recent years, the strong interactions of Rydberg atoms were discovered as a fruitful platform to engineer extend-range interactions in optical lattices and tweezers ~\cite{SpinLattice1,SpinLattice2,DressingTweezer}. Rydberg dressing - the admixing of the interactions to the ground state by off-resonantly coupling to a Rydberg state - provides a way to increase the experimental timescales beyond the typical lifetime of the Rydberg states ~\cite{Cryogenic}. Theoretical studies of Rydberg-dressed quantum gases have shown rich equilibrium phase diagrams comprising quantum phases such as Mott insulating, superfluid and density wave phases and even supersolids - quantum phases simultaneously exhibiting frictionless flow of superfluids and broken lattice translational symmetry of crystalline structures ~\cite{RydbergPhasediagram1,RydbergSupersolidsIII,RydbergSupersolidsV,RydbergPhasediagram3,RydbergPhasediagram4,RydbergPhasediagram5,RydbergPhasediagram6}.\\
Unfortunately, admixed scattering rates limit the acces\-sible timescales ~\cite{SpinLattice1,AvalancheII}, especially since measured lifetimes of Rydberg-dressed atomic clouds were observed to be significantly below the single particle lifetime~~\cite{AtomLoss2}.
The collective character of the observed loss rates was found to depend on the Rabi frequency and the detuning. It is suspected to be induced by black-body transitions to neighboring Rydberg states, possibly followed by an on-resonant excitation avalanche, and can be reduced by working with lower densities \cite{AtomLoss1,SpinLattice1,Cryogenic,AvalancheIV}.
Furthermore the AC Stark shift induced by the coupling laser requires unreasonably high tunneling rates \cite{ACStark1,ACStark2}, especially perpendicular to the propagation direction of the excitation laser.  Due to these difficulties, there exists so far only one recent publication realizing Rydberg-dressed interactions in the itinerant regime \cite{ItinerantRydberg}.\\
A recent theoretical study has proposed to dress to the minima and maxima of Rydberg interaction potentials ~\cite{MacrodimerDressing1}. Coupling to such potential curves induces a distance-specific interaction with tunable strength and an increased dressing quality - the ratio of dressed interaction and decoherence (assuming the absence of collective losses). In this study we complete the description of the coupling to these potentials by including macrodimer states - molecular bound states within the wells of these interaction potentials \cite{MacrodimerScaling,MacrodimerPrediction2,MacrodimerPrediction3,MacrodimerCs1,MacrodimerCs2,MacrodimerCs3} and propose a new coupling scheme, which enhances the admixed interaction.\\
This work is divided in two main sections: First, we study the scaling properties of the dressed interaction, the dressing quality and the AC Stark shift with the help of the packages \texttt{Alkali Rydberg Calculator} \cite{ARC} and \texttt{Pair Interaction} \cite{PI}, and compare these scaling laws to the ones obtained from conventional Rydberg dressing schemes. We propose a two-color dressing scheme with which the dressing quality can be optimized while the AC Stark shift vanishes. Furthermore, it allows us to increase the otherwise weak absolute interaction strength. We find a substantial ratio between the dressed interaction and scattering rate, which is promising for optical lattice experiments. We then investigate the equilibrium phases of the corresponding extended Bose-Hubbard Hamiltonian. We obtain spatially modulated equilibrium ground states. We show that slowly ramping up the coupling of an initial superfluid state to the macrodimer state leads to a broken translational symmetry. If the tunability of the presented scheme provides a regime, where collective losses are absent such that experimental loss rates approach the single particle and macrodimer loss limit, spatial ordering is shown to be achievable on experimental timescales.\\

\begin{figure*}
\begin{minipage}{0.23\linewidth}
\begin{overpic}[width = 1\textwidth, trim = {0 0 0 0}, clip]{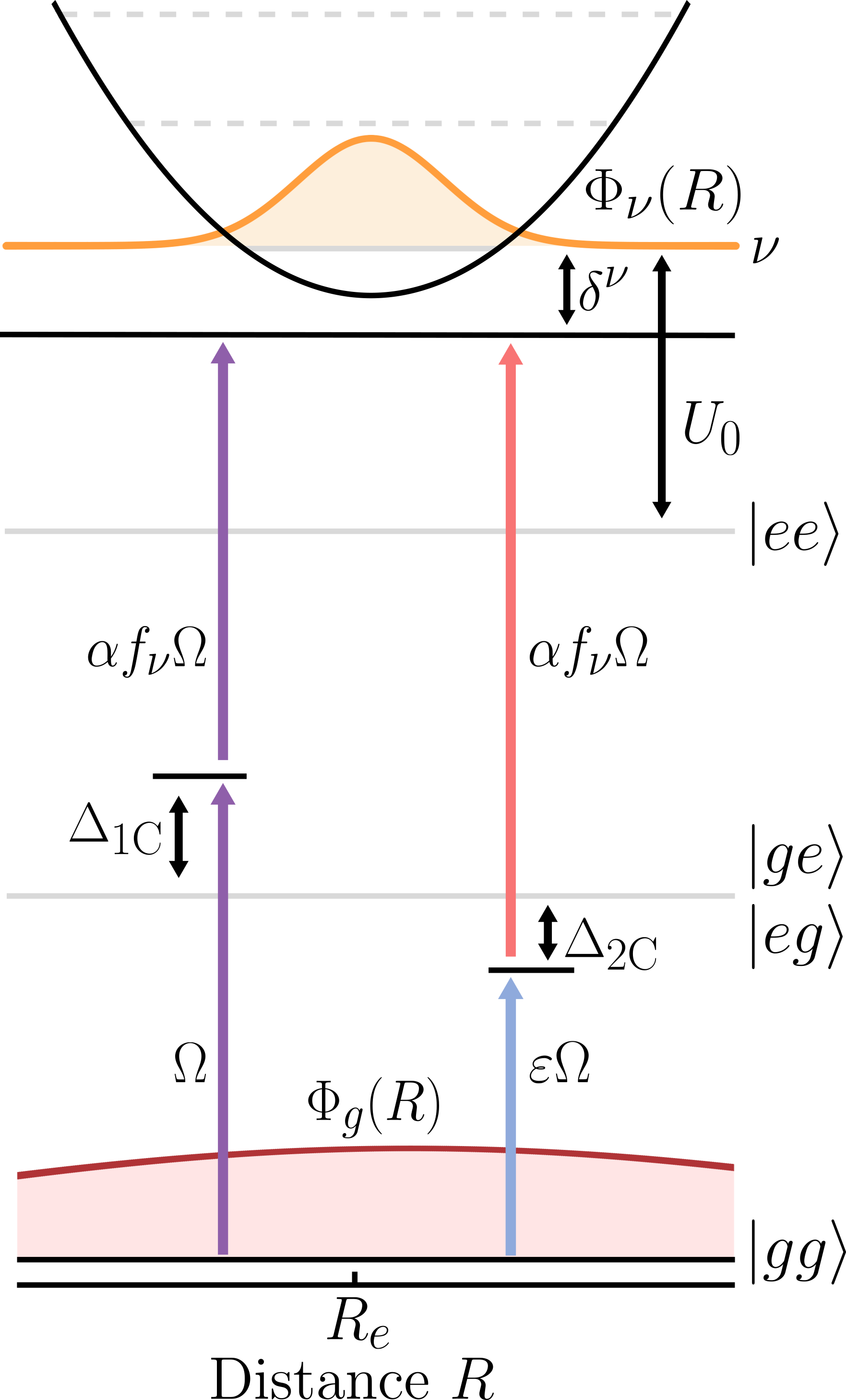}
\put (-5,103) {\small(a)}
\end{overpic}
\end{minipage}
\begin{minipage}{0.25\linewidth}
\begin{overpic}[width = 0.98\textwidth, trim = {0 0 0 5}, clip]{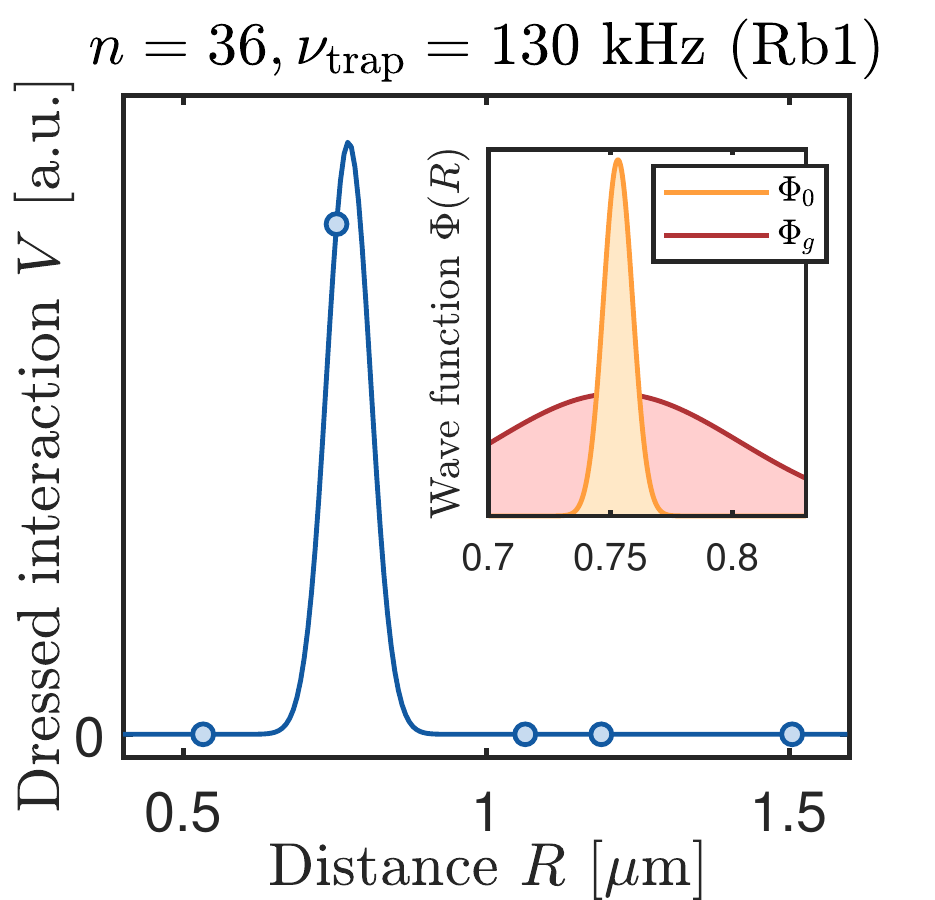}
\put (-2,93) {\small(b)}
\end{overpic}
\begin{overpic}[width = 0.9\textwidth, trim = {180 20 180 -30}, clip]{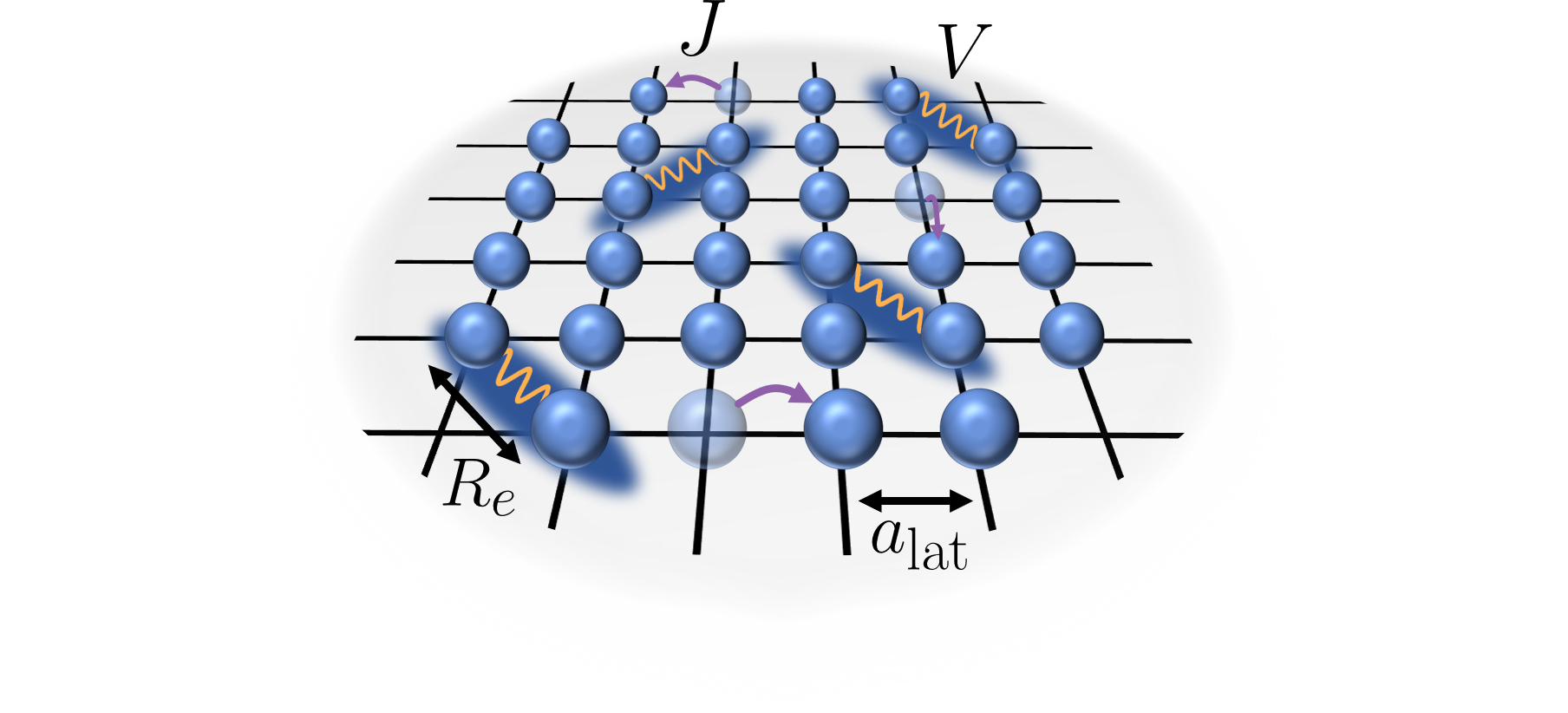}
\put (-2,74) {\small(c)}
\end{overpic}
\end{minipage}
\begin{minipage}{0.38\linewidth}
\begin{overpic}[width = 0.99\textwidth, trim = {0 0 40 5}, clip]{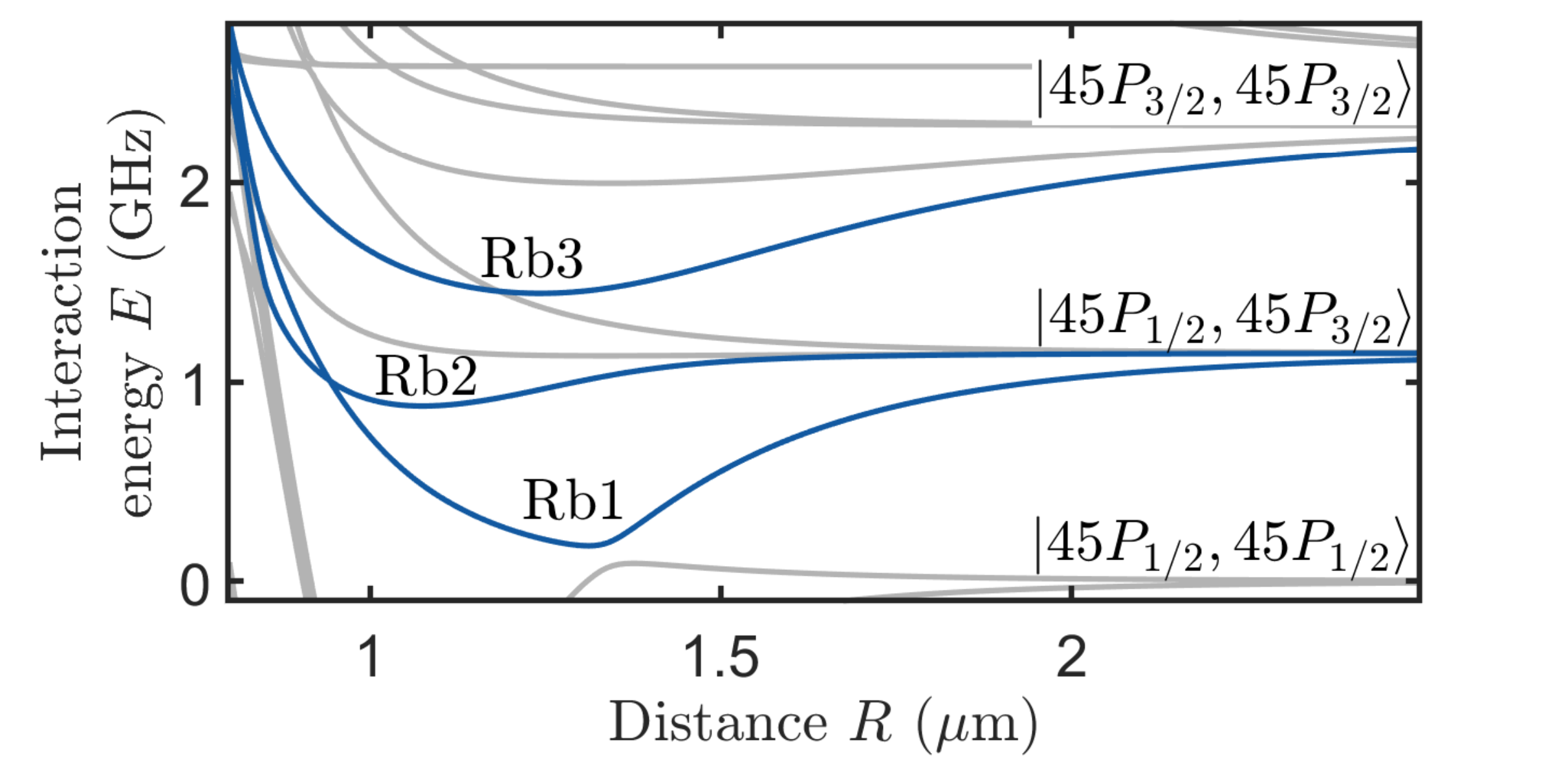}
\put (3,53) {\small(d)}
\end{overpic}
\includegraphics[width = 0.99\textwidth, trim = {0 0 40 5}, clip]{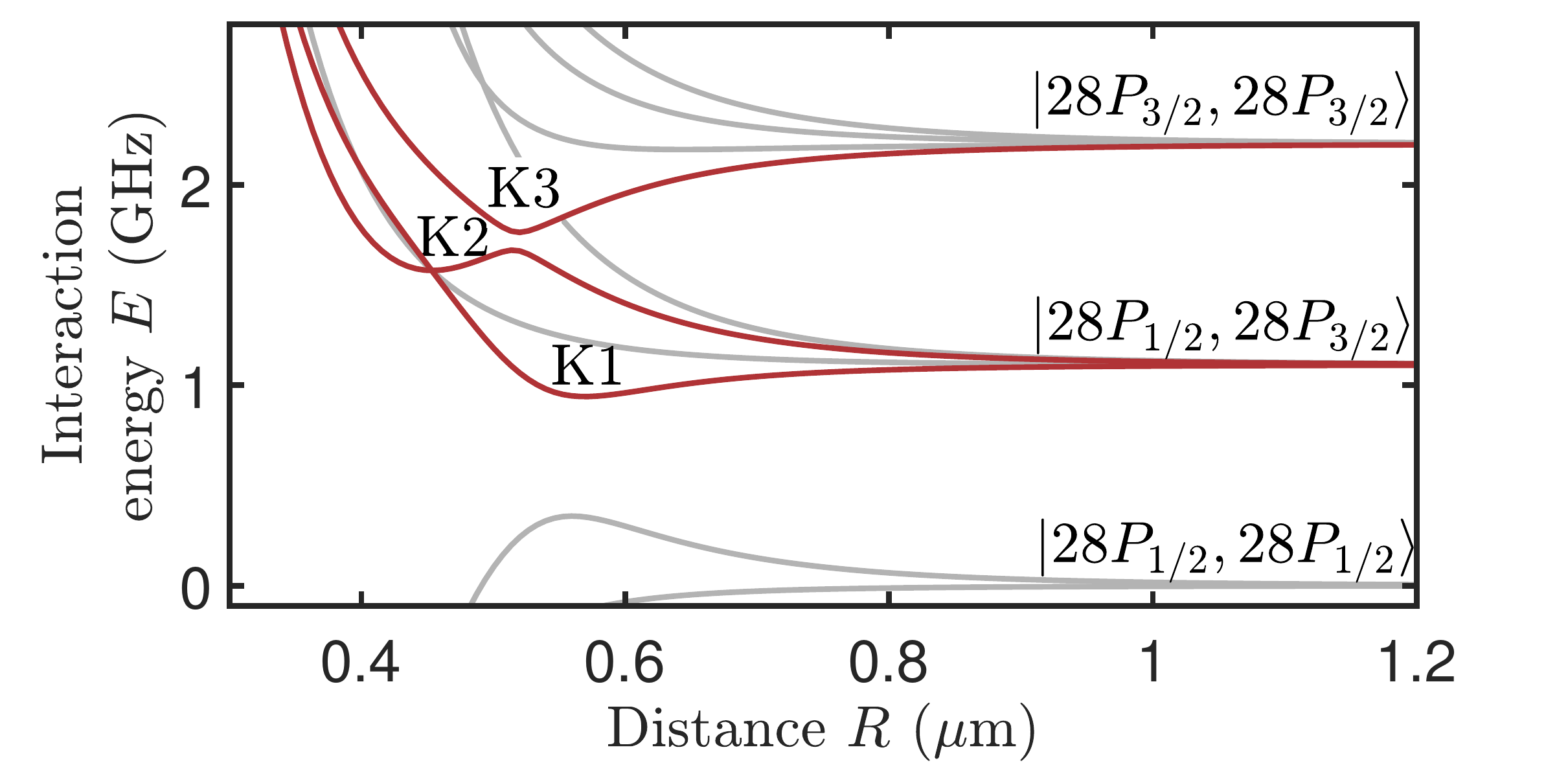}
\end{minipage}
%
%
\caption{(a) Schematic of single-color and two-color coupling scheme for the realization of the macrodimer dressing. While the intermediate state detuning $\Delta_\text{1C}$ is approximately given by half of the energy shift of the potential $U_0$ in the single-color scheme, the two-color scheme benefits from the additional tunability. The potential well is further described by the position of the minimum $R_e$ and its eigenstates are approximately given by the vibrational states with vibrational quantum number $\nu$ and their corresponding vibrational wave function $\Phi_\nu(R)$. The atoms in the electronic ground state are described by the initial relative wave function $\Phi_g(R)$, which depends on the trapping frequency $\nu_\text{trap}$. (b) Dressed interaction curve for the potential well Rb1 at $n = 36$ and a trapping frequency $\nu_\text{trap} = 130$ kHz. The blue dots represent typical distances in an optical lattice with lattice constant $a_\text{lat} = 532$ nm. (c) Bosonic atoms trapped in a two-dimensional optical lattice with lattice constant $a_\text{lat}$ and tunneling with rate $J$. The dressing results in a distance-specific interaction with tunable spatial profile given by the distance $R_e$ and strength $V$. (d) In this work, we focus on three potential wells for $^\text{87}$Rb (upper diagram) and $^\text{39}$K (lower diagram), which arise from avoided crossing energetically located between the asymptotic pair states $|nP_{1/2} nP_{1/2}\rangle$, $|nP_{1/2} nP_{3/2}\rangle$ and $|nP_{3/2} nP_{3/2}\rangle$ for $n \in [25,75]$.}
\label{fig:scheme}
\end{figure*}

\section{Macrodimer-dressing and model}
Rydberg dressing close to the Rydberg resonance has been shown to induce van-der-Waals interactions to pairs of ground state atoms, which saturate to a soft-core potential \cite{DressingDephasing2,RydbergDressing,Softcore,Softcore2,Softcore3,SpinLattice1}. Given a single-photon transition with Rabi coupling $\Omega = \langle e | \hat{H} | g \rangle$ and frequency detuning $\Delta$, the soft-core potential depth becomes $V = \Omega^4/(2\Delta)^3$, with admixture $P_\text{Ryd} = \Omega^2/(2\Delta)^2$ of the Rydberg state to the ground state.
Alternatively, dressing with far-off resonant avoided crossings between different pair potentials, which occur at closer distance (see FIG \ref{fig:scheme}. (b)), also yields effective interatomic interactions with modified scaling behaviors.
The upper potential well of these avoided crossings harbors a multitude of vibrational bound states, which are coupled through a two-photon transition via an intermediate state containing one bare Rydberg state and one ground state atom (see FIG. \ref{fig:scheme}. (a)). In the case of a single-color excitation scheme - one coupling laser for both transitions with Rabi frequency $\Omega$ and the frequency detuning $\Delta_\text{1C}$ to the intermediate state -  the effective two-photon Rabi coupling to a molecular state with vibrational quantum number $\nu$ after adiabatic elimination becomes $\tilde{\Omega}^\text{1C}_\nu = \alpha f_\nu \Omega^2/\Delta_\text{1C}$. The prefactor $\alpha$ describes the difference between the single particle Rabi frequency coupling $|gg\rangle$ to the intermediate states $|ge\rangle$ and $|eg\rangle$, and the coupling between the intermediate states and the molecular state, and depends on the electronic structure of the molecular state. It can be optimized through the polarization of the excitation light and the quantization axis of the ground state atoms relative to the molecular orientation ~\cite{Alpha}.
The Franck-Condon factor $f_\nu$ is defined through the overlap integral between the initial relative wave function $\Phi_g(R)$ of the ground state \cite{Wannier} and the vibrational wave function $\Phi_\nu(R)$ as $f_\nu = \int \Phi^*_\nu(R) \Phi_g(R) \text{d}R$. Here, we assume the asymptotic pair state decomposition to vary little over distance $R$, which allows us to separate the overlap of interatomic motion from the electronic coupling.
Within this dressing scheme, the intermediate state detuning $\Delta_\text{1C}$ is set by the energy shift $U_0$ of the potential well and gets slightly modified by the chosen two-photon detuning $\delta$ as $2\Delta_\text{1C} = U_0 + \delta$ (see FIG \ref{fig:scheme}. (a)).
Going to a two-color coupling scheme - one coupling laser for the transition between the intermediate and the molecular state with frequency $\Omega$ and one weaker laser coupling the ground to the intermediate state with frequency $\varepsilon \Omega$ ($\varepsilon < 1$) - allows for further tunability, as the detuning $\Delta_\text{2C}$ of the intermediate state becomes independent of the shift $U_0$. Within adiabatic elimination, the effective coupling strength is then given by $\tilde{\Omega}^\text{2C}_\nu = \alpha \varepsilon f_\nu \Omega^2/\Delta_\text{2C}$.

\begin{figure*}
\begin{minipage}{0.325\linewidth}
\flushleft
\begin{overpic}[width = 0.9156\textwidth, trim = {0 -22 30 -5}, clip]{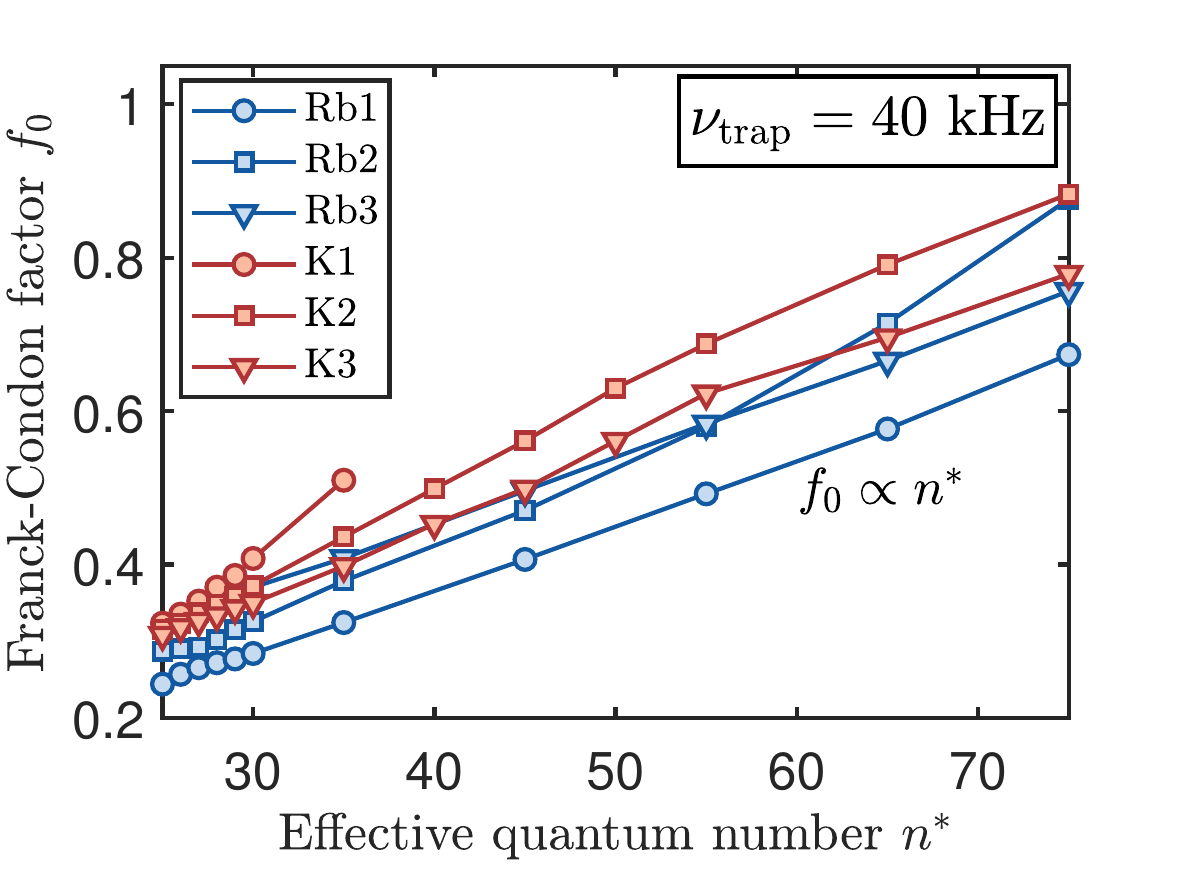}
\put (0,91) {\small(a)}
\end{overpic}
\begin{overpic}[width = 0.995\textwidth, trim = {0 0 5 7}, clip]{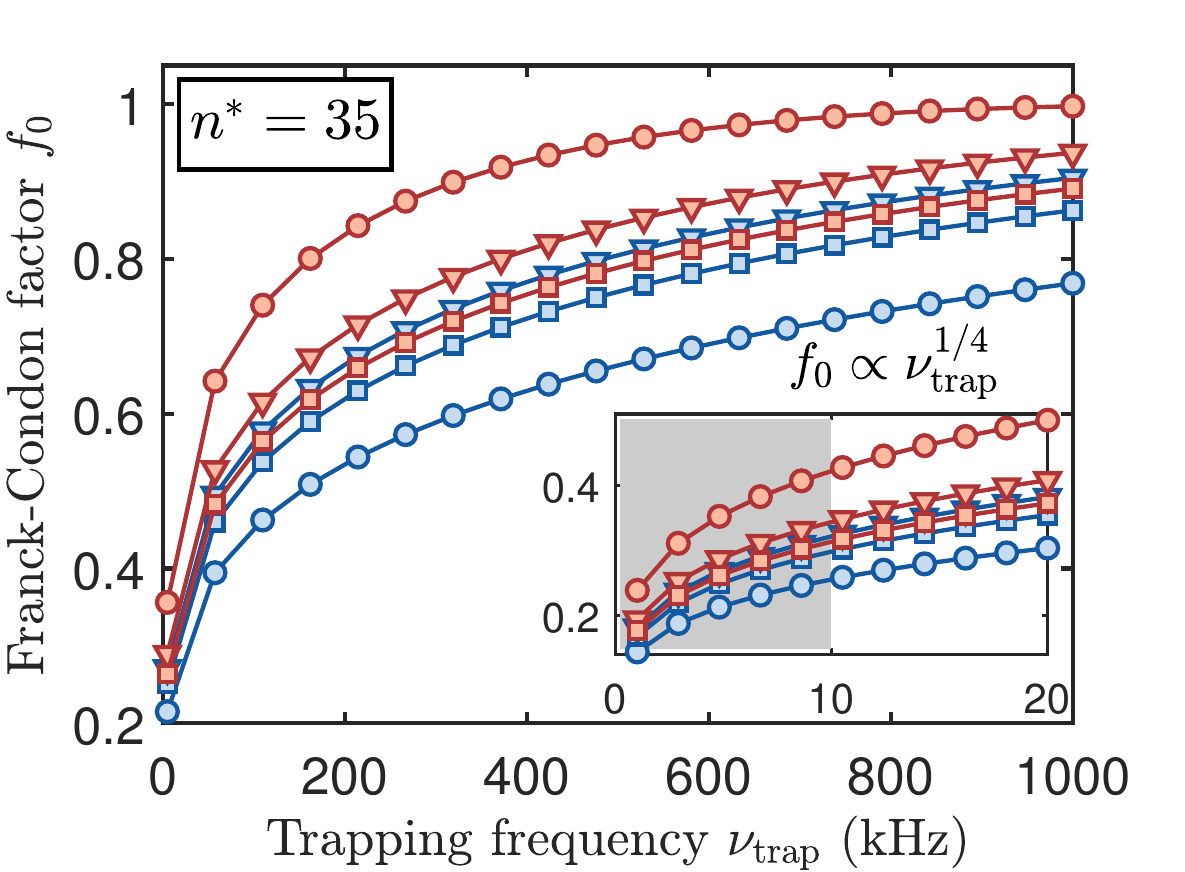}
\put (0,77) {\small(b)}
\end{overpic}
\end{minipage}
\begin{minipage}{0.325\linewidth}
\centering
\begin{overpic}[width = 0.99\linewidth, trim = {0 -14 0 0}, clip]{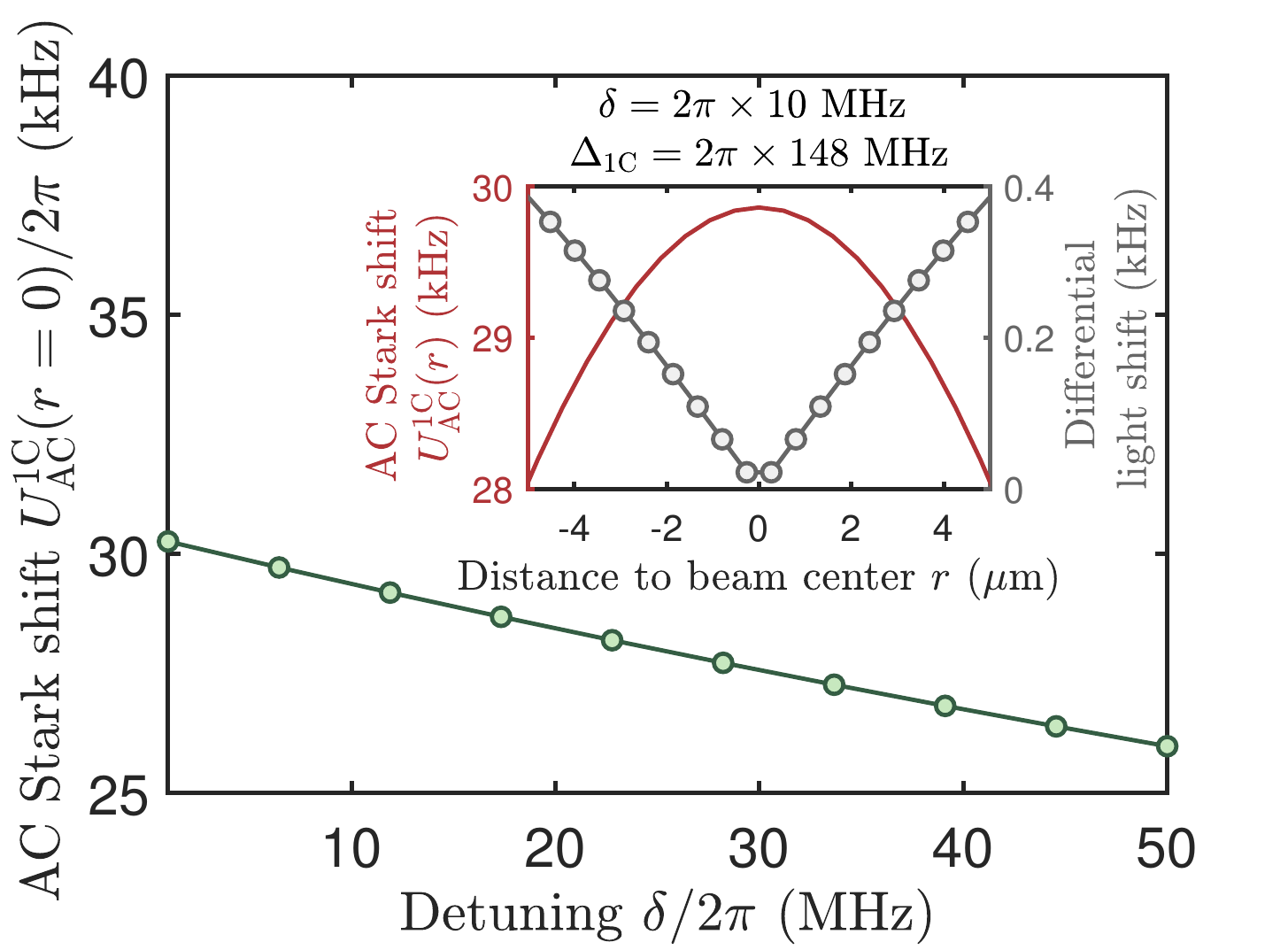}
\put (0,80.5) {\small(c)}
\end{overpic}
\begin{overpic}[width = 1\linewidth, trim = {0 0 0 -10}, clip]{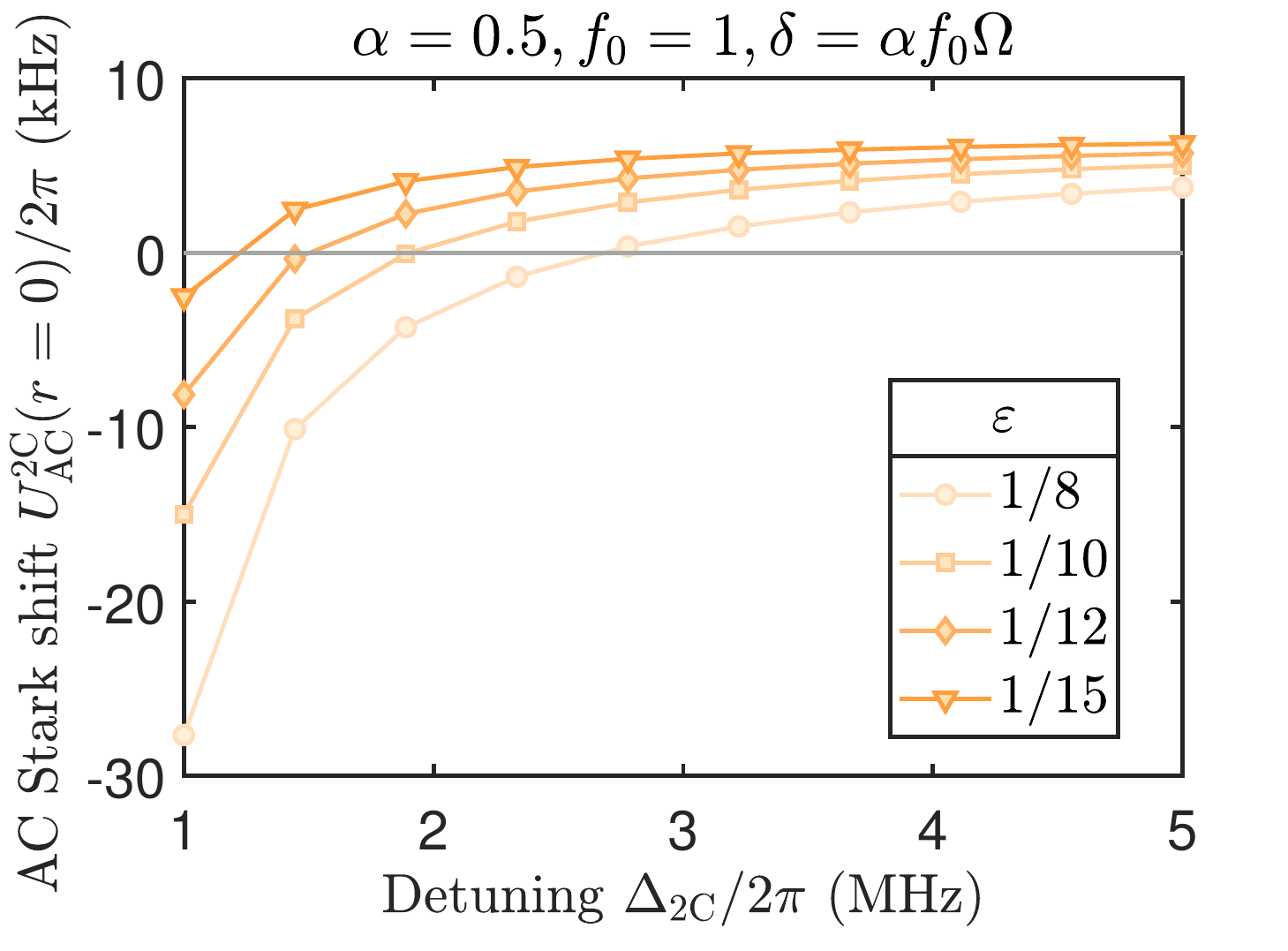}
\put (0,76) {\small(d)}
\end{overpic}
\end{minipage}
\begin{minipage}{0.325\linewidth}
\begin{overpic}[width = 1\linewidth, trim = {0 -14 0 0}, clip]{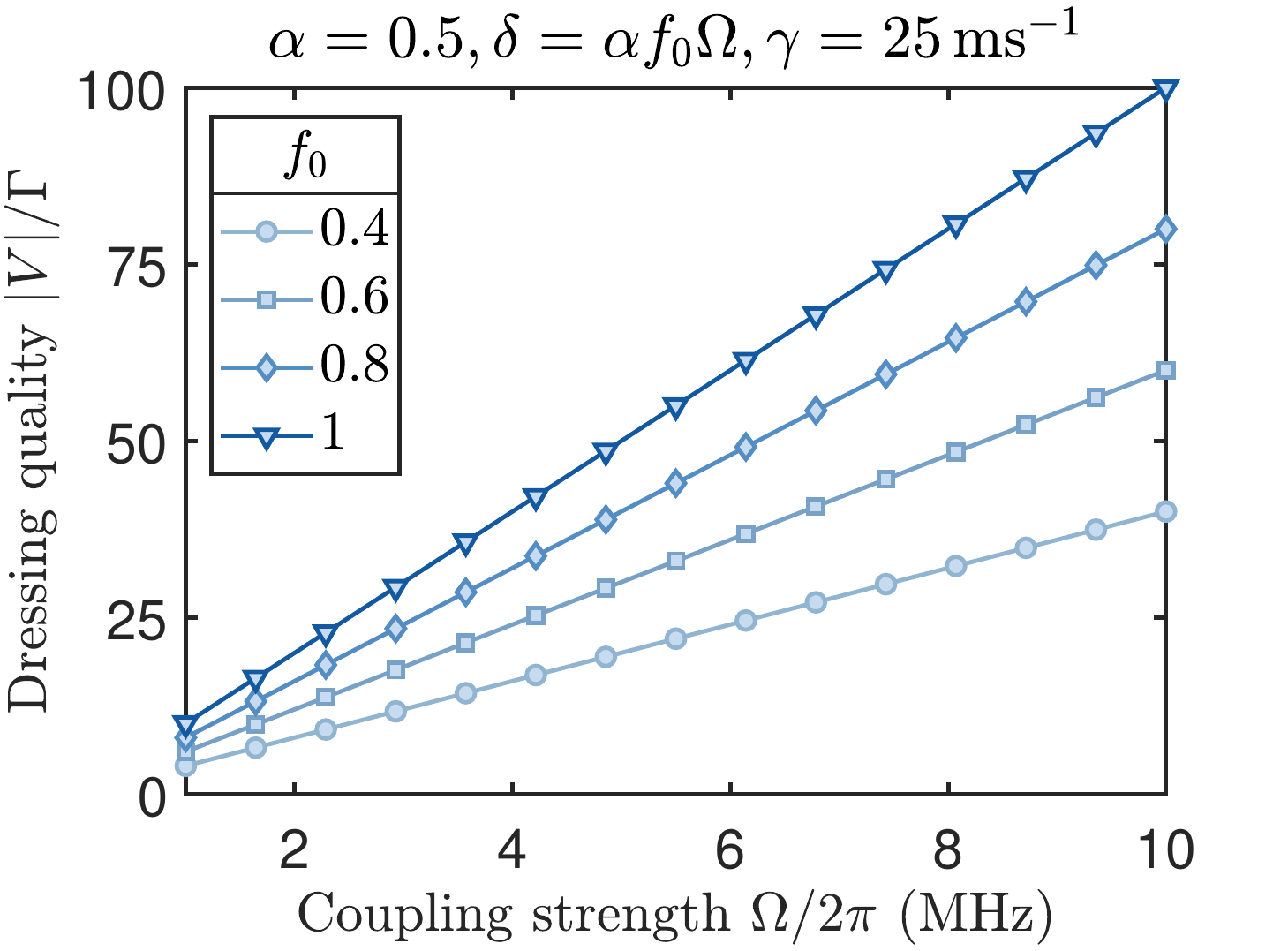}
\put (0,80) {\small(e)}
\end{overpic}
\begin{overpic}[width = 1\linewidth, trim = {0 0 0 -2}, clip]{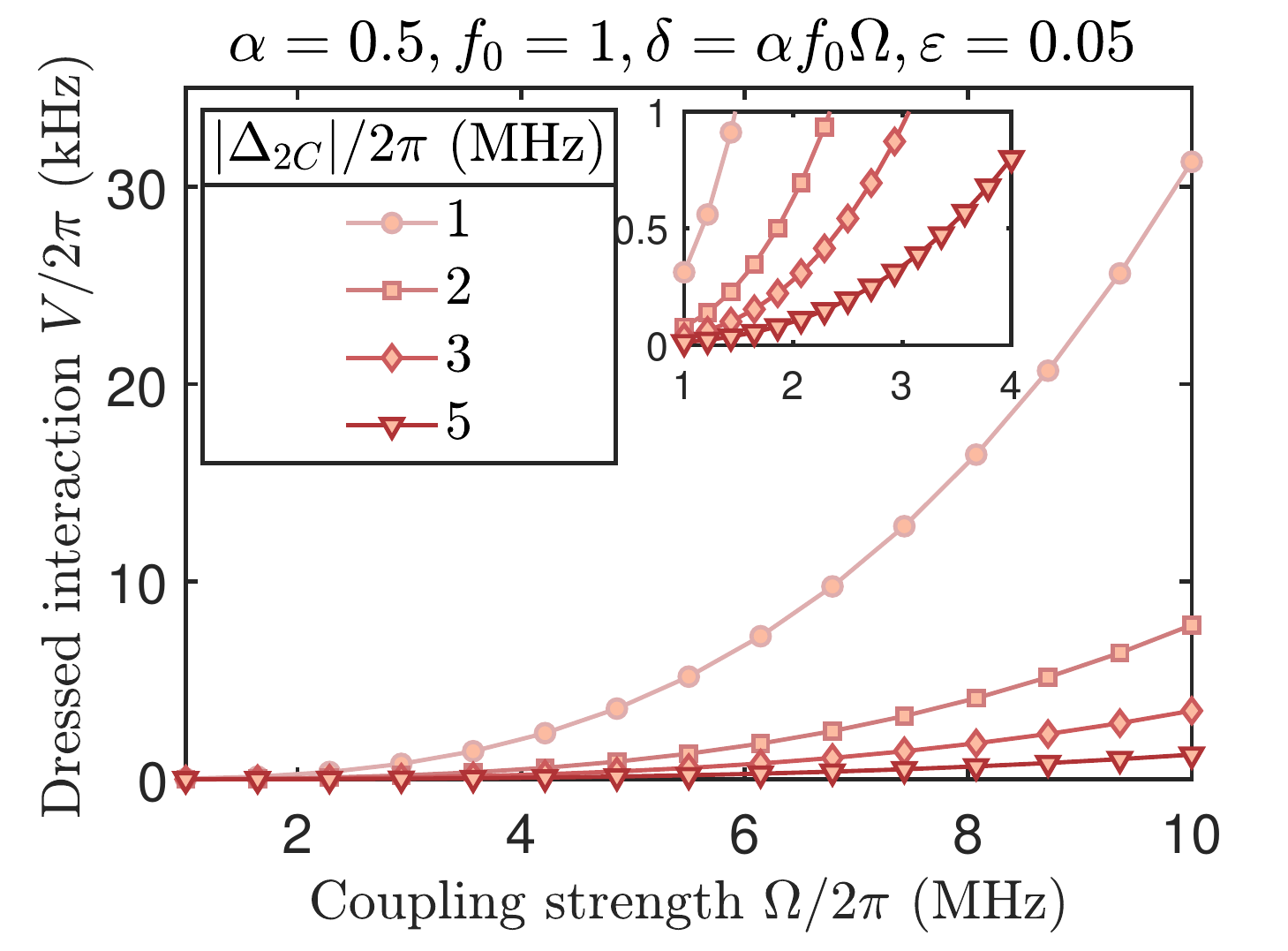}
\put (0,75) {\small(f)}
\end{overpic}
\end{minipage}
\caption{(a,b) Franck-Condon factor $f_0$ versus effective principal quantum number $n^*$ for fixed trapping frequency $\nu_{\text{trap}}$ (a) and versus trapping frequency $\nu_{\text{trap}}$ for fixed effective principal quantum number $n = 35$ (b). For all macrodimer potentials, we find a linear growth of the Franck-Condon factor for increasing effective principal quantum numbers $f_0 \propto n^*$ which scales with the trapping frequency as $f_0 \propto \nu_\mathrm{trap}^{1/4}$. The grey area denotes the typical range of trapping frequencies in the itinerant regime. (c-f) For the following plots, we use the potential well Rb1 at $n = 35$, a Rabi frequency $\Omega = 2\pi \times 3$ MHz and the optimum detuning $\delta = \alpha f_0 \Omega$ unless mentioned otherwise. For the studied interaction potential of Rubidium, we choose typical values for the electronic coefficient $\alpha = 0.5$ \cite{Alpha} and a Rydberg decay rate $\gamma = 25$ ms$^\text{-1}$ \cite{DissipationI}. (c,d) Single-color AC Stark shift $U^\text{1C}_\text{AC}(r = 0)$ versus $\delta$ (c) and two-color AC Stark shift $U^\text{2C}_\text{AC}(r = 0)$ versus $\Delta_\text{1C}$ (d) at the beam center $r = 0$. Away from the beam center, the differential light shift induced by the AC Stark shift between neighboring lattice sites (circle markers in inset of (c)) is significant in the single-color scheme. In the two-color scheme, the additional tunability gained through the independent intermediate state detuning $\Delta_\text{2C}$ and coupling strength $\varepsilon \Omega$ allows to suppress the total AC Stark effect. (e,f) The dressing quality (e) for various Franck-Condon factors $f_0$ and the dressed interaction (f) versus coupling strength assuming single particle and macrodimer losses. The dressing quality is independent of the coupling between ground and intermediate state, while the interaction profits from smaller detunings.}
\label{fig:potentials}
\end{figure*}
\noindent
The dressed interaction strength of a single vibrational mode can be obtained from the Hamiltonian of the three level system with the two body ground state, the intermediate state and the molecular state and is given by $V_\nu = \hbar \tilde{\Omega}^2_\nu/4\delta_\nu$ for $\delta_\nu \gg \Omega_\nu$.
For both coupling schemes, through a combination of the larger detuning $\delta_\nu$ to higher vibrational modes and the decreasing Franck-Condon factors $f_\nu$, the contribution $V_0$ of the lowest vibrational state dominates the full interaction $V = \sum_\nu V_\nu$ (see Appendix A).
We obtain a tunable interaction between particles at the specific distance $R_e$ matching the avoided crossing (see FIG \ref{fig:scheme}. (b)).

\subsection{Scaling properties and tunability}
In this section, we study the dependence of the Franck-Condon factor $f_0$ of the lowest vibrational bound state on the choice of potential curves, the effective principal quantum number $n^*$ \cite{QuantumDefect1,QuantumDefect2} and the on-site trapping frequency $\nu_{\text{trap}}$ for a given atom species in order to determine whether strong dressed interactions are attainable. Furthermore, we discuss the AC Stark shift $U_\text{AC}$ and the dressing quality $|V|/\Gamma$ with respect to the chosen dressing scheme. We focus on three potential wells energetically located between the fine structure split states composed of $|e\rangle \equiv |nP_{1/2}\rangle$ and $|e'\rangle \equiv |nP_{3/2}\rangle$ (see FIG. \ref{fig:scheme}. (c)) for Potassium ($^{\text{39}}$K) and Rubidium ($^{\text{87}}$Rb). Among alkali atoms, these species show the clearest evidence of binding potentials within the fine structure. Further macrodimer potentials can be found in the energy regime between different principal and orbital quantum numbers. Additionally, macrodimer potentials can be induced through an external electric field, as shown for Cesium ~\cite{MacrodimerCs1,MacrodimerCs2,MacrodimerCs3,MacrodimerPrediction3}. Promising candidates for macrodimer dressing schemes may also be Strontium and Ytterbium, since these possess metastable states with larger spatial overlap with Rydberg states, which provides enhanced coupling strengths significantly exceeding the ones obtained with Alkali atoms ~\cite{RydbergSr1,RydbergSr2,RydbergSr3,RydbergYt,ACStark2}.\\
We first vary the effective principal quantum number $n^* \in [25,75]$ and study the scaling of the potential well position $R_e$, the potential well depth $D_e$, the shift $U_0$, the vibrational spacing $\Delta_\nu$ and the Franck-Condon factor $f_0$. By fitting a power law, we confirm the relations $R_e \propto (n^*)^{8/3}$ and $D_e \propto (n^*)^{-3} + \epsilon (n^*)^{-4}$, found in another study ~\cite{MacrodimerScaling}. Additionally, we obtain the power law of the energy shift $U_0 \propto (n^*)^{-3}$ and the spacing $\Delta_\nu \propto (n^*)^{-3}$(see Appendix B).\\
The Franck-Condon factor $f_0$ depends on the trapping frequency $\nu_\text{trap}$ of the potential well and the effective principal quantum number $n^*$, and is bounded by $1$. Variation of the effective principal quantum numbers leads to a linear behavior, i.e. $f_0 \propto n^*$ (see FIG. \ref{fig:potentials}. (a)). For larger $n^*$ the potential well becomes more shallow, leading to higher overlap with the typically broader ground state wave function.
Varying the trapping frequencies $\nu_\text{trap} \in [10,1000]$ kHz for fixed effective principal quantum number $n^* = 35$, we find a scaling law of the Franck-Condon factor $f_0 \propto \nu_{\text{trap}}^{1/4}$ with increasing trapping frequency $\nu_\text{trap}$ (see FIG. \ref{fig:potentials}. (b)). For trapping frequencies up to $\nu_\text{trap} = 10$ kHz - for which superfluid phases of Rubidium atoms trapped in an optical lattice can be found \cite{Wannier,SuperfluidMott} - we obtain Franck-Condon factors of $f_0 \in [0.15, 0.4]$ (see FIG. \ref{fig:potentials}. (b) inset).
For a higher $n^* = 65$ more suitable for tweezer experiments \cite{OpticalTweezers1,OpticalTweezers2,OpticalTweezers3,OpticalTweezers4}, we find that the Franck-Condon factor can reach the optimum value of $f_0 = 1$ within the range of accessible trapping frequencies and decrease as $f_0 \propto \nu_{\text{trap}}^{-1/4}$ after the maximum (see Appendix B). Both characteristic power laws result from the scaling of the width of the ground state wave function with the trapping frequency $\nu_\text{trap}$.\\
In order to access the itinerant regime, inhomogeneous AC Stark shifts have to be lower than the hopping rate. The Gaussian profile of the laser beam induces a site-dependent shift rendering neighboring sites off-resonant. In the single-color dressing scheme, the AC Stark shift is given by $U^\text{1C}_\text{AC}(r) = \frac{1}{4}\Omega^2(r)/\Delta_\text{1C} = \frac{1}{2}\Omega^2(r)/(U_0 + \delta)$ with the distance $r$ to the beam center. Here, the intensity profile of the laser beam with beam waist $w$ leads to $\Omega(r) = \Omega_0 \text{exp}(-r^2/w^2)$. For a typical value of $w = 20$ $\mu$m and a lattice spacing of $a_\text{lat} = 532$ nm, the differential light shift between neighboring sites becomes substantial, especially further away from the center of the beam (see FIG. \ref{fig:potentials}. (c)). This inhibits coherent tunneling if the hopping rate $J$ is below the energy difference. Because this problem mainly appears perpendicularly to the laser propagation direction, coherent hopping is still available along one dimension of the optical lattice \cite{ItinerantRydberg}.\\
In contrast to the single-color dressing scheme, the two-color dressing scheme allows for additional tunability of the total AC Stark shift. The shift is then given by $U^\text{2C}_\text{AC}(r) = \frac{1}{4}\Omega^2(r)(\varepsilon^2/\Delta_\text{2C} + 1/(U_0+\delta-\Delta_\text{2C}))$ and is tunable through the intermediate state detuning $\Delta_\text{2C}$ and both coupling strengths (see FIG. \ref{fig:potentials}. (d)). By appropriate choice of the coupling to the intermediate state, the total AC Stark shift cancels, such that tunneling along all dimensions of the optical lattice becomes possible.\\
An additional motivation for the macrodimer dressing instead of a conventional dressing comes from the dressing quality $|V|/\Gamma$, the ratio between the dressed interaction $V$ and the decoherence rate $\Gamma$, which indicates timescales on which coherent dynamics take place. Assuming only single atom loss, the ratio for the conventional Rydberg dressing is $|V|/\Gamma_\text{Ryd} = \Omega^2/(2|\Delta| \gamma)$ with the Rydberg decay rate $\gamma$ \cite{MacrodimerDressing1}. Its value can be optimized by increasing the Rabi coupling $\Omega$, though the dressing regime requires $\Omega/\Delta \ll 1$ in order for the Rydberg admixture to be small.\\
The situation is different for macrodimer-dressing schemes. In the following, we also include the vibrational states into the description. The previously defined two-photon coupling yields the optimal ratio $|V|/\Gamma = \alpha f_0 \Omega/(2\gamma)$ with $\Gamma = \Gamma_\text{Ryd} + \Gamma_\text{mol}$ obtained at $\delta_\text{opt} = \alpha f_0 \Omega$ for both the single and two-color dressing schemes (see Appendix C). This result is consistent with \cite{MacrodimerDressing1} and is independent of $\Delta_\text{1C}/\Delta_\text{2C}$. At this optimal detuning, the scattering rates at the intermediate state and the molecular state are equal. For reasonable values of the Rabi coupling and for typical scattering rates ~\cite{DissipationI,DissipationIII}, the dressing quality calculated without accounting for collective losses is sufficiently large to observe coherent dynamics (see FIG. \ref{fig:potentials}. (e)). The strength of the dressed interaction however strongly depends on the choice of the dressing scheme. For the single-color scheme we obtain an interaction strength $V^\text{1C} = \alpha^2 f_0^2 \Omega^4/(4 \Delta_\mathrm{1C}^2 \delta) \overset{\delta_\text{opt}}{=} \alpha f_0 \Omega^3/(4 \Delta_\mathrm{1C}^2)$, which is suppressed since the detuning $\Delta_\text{1C}$ given by $U_0/2$ takes on large values. This forces one to decrease $\delta$ away from the optimum to boost $V$ towards experimental time scales into a regime with smaller dressing quality. In contrast, the independent intermediate state detuning $\Delta_\text{2C}$ in the two-color scheme allows one to tune and increase the interaction strength $V^\text{2C} = \alpha^2 \varepsilon^2 f_0^2 \Omega^4/(4 \Delta_\text{2C}^2 \delta) \overset{\delta_\text{opt}}{=} \alpha \varepsilon^2 f_0 \Omega^3/(4 \Delta_\text{2C}^2)$. Within the dressing condition $\varepsilon\Omega/\Delta_\text{2C} \ll 1$, we are now able to achieve large values (see FIG. \ref{fig:potentials}. (f)) at the optimum detuning $\delta_\text{opt}$.\\
Combining the scalings of the Franck-Condon factor $f_0 \propto n^*$ and the single-photon coupling strength $\Omega \propto (n^*)^{-3/2}$ ~\cite{Polarizability}, we obtain the dressed interaction $V \propto (n^*)^{-4}$ for both dressing schemes, which implies strong dressed interactions for smaller $n^*$. On the other hand, decay rates decrease with higher quantum numbers as the radiative and the black-body lifetime given by their scattering rates $\Gamma_\text{rad} \propto 1/(n^*)^{3}$ and $\Gamma_\text{BBR} \propto 1/(n^*)^{2}$ increase ~\cite{DissipationI,DissipationII,Lifetimes}. At low (high) values of $n^*$ for which spontaneous (black-body induced) transitions dominate the effective lifetime, the scaling of the dressing quality becomes $|V|/\Gamma \propto (n^*)^{-1}$ ($|V|/\Gamma \propto (n^*)^{-2}$).\\
We again want to emphasize that Rydberg induced losses were found to be significantly above the expected single particle losses and density-dependent \cite{AtomLoss1,AtomLoss2,AvalancheIII,AvalancheIV}. Since these loss signatures were found to be weaker further detuned from the Rydberg resonance, we expect that our tunable scheme will also be able to find a regime where collective losses are smaller.\\
In the following, we use the obtained results to study Rydberg macrodimer dressing for a two-dimensional optical lattice in the optimized parameter regime. We discuss the corresponding equilibrium phase diagram and the possible preparation of itinerant states of such a system with respect to the dressed interaction $V$ and scattering at the bare Rydberg state and macrodimer state with scattering rates $\Gamma_\mathrm{Ryd}$ and $\Gamma_\mathrm{mol}$.

\subsection{Hamiltonian and methods}
The single-species extended Bose-Hubbard model realized by our dressing scheme reads
\begin{equation}
\begin{split}
\hat{H} = &-J\sum_{\langle ij \rangle}(\hat{b}^\dag_i \hat{b}_j + \hat{b}^\dag_j \hat{b}_i) + \frac{U}{2} \sum_i \hat{b}^\dag_i \hat{b}^\dag_i \hat{b}_i \hat{b}_i\\
&- \mu \sum_i \hat{b}^\dag_i \hat{b}_i + \sum_{ij} V_{ij} \hat{b}^\dag_i \hat{b}^\dag_j \hat{b}_i \hat{b}_j,\\
\end{split} \label{Hamiltonian}
\end{equation}
with the hopping rate $J$, the on-site interaction $U$, the chemical potential $\mu$ and the dressed interaction strength $V_{ij}$, which is non-vanishing only at distances close to the molecular bond length. 
Due to the Gaussian envelope of the coupling laser, $V_{ij}$ is also spatially dependent, which we neglect in the subsequent calculations. The interaction appears between two particles at sites $i$ and $j$ separated by distance $R_e$ and can be written as $V_{ij} = V\delta_{R_{ij},R_e}$, where $\delta_{i,j}$ is the Kronecker delta and $R_{ij}$ is the distance between sites $i$ and $j$.
In the following, we focus on molecular potentials for principal quantum numbers of roughly $n = 30$ ($n = 36$).
With a typical lattice spacing of $a_\text{lat} = 532$ nm of an optical square lattice, $V_{ij}$ peaks at interatomic distances of $R_e = a_\text{lat}$ ($R_e = \sqrt{2} a_\text{lat}$), corresponding to nearest-neighbour NN (next-nearest neighbour NNN) interactions, for both Rubidium and Potassium (see FIG. \ref{fig:scalings} in Appendix B). The corresponding long-range interaction term of the Hamiltonian \eqref{Hamiltonian} becomes $V \sum_{\langle ij \rangle} \hat{b}^\dag_i \hat{b}^\dag_j \hat{b}_i \hat{b}_j$ ($ V \sum_{\langle \langle ij \rangle \rangle} \hat{b}^\dag_i \hat{b}^\dag_j \hat{b}_i \hat{b}_j$). Both specific models have been investigated in numerous studies and predict spatially ordered ground states for a suitable choice of the models parameters ~\cite{NN1,NN3,NN4,NN5}.\\
The extended Bose-Hubbard model \ref{Hamiltonian} is valid within the single-band approximation. This requires, that the system is sufficiently cold and interactions weak enough such that transitions to higher bands are prohibited. For typical values of the lattice depth $V_0 \in [5,20]$ $E_R$ with the recoil energy $E_R = \hbar^2 \pi^2/(2 a^2_\text{lat} m)$ of atoms with mass $m$, the band gap $E_\text{gap}$ is usually in the order of magnitude of several recoil energies, i.e. $E_\text{gap} = 2\sqrt{V_0 E_R}$  \cite{SingleBand2}. For a lattice spacing of $a_\text{lat} = 532$ nm and Rubidium ($^{\text{87}}$Rb), the corresponding recoil energy $E_R \approx 2\pi \times 3$ kHz leads to a band gap around $E_\text{gap}/2\pi \in [5,30]$ kHz. As the tunable parameters of the Hamiltonian \eqref{Hamiltonian} are typically below a kHz, we expect them to be much smaller than the band gap, i.e. $E_\text{gap} \gg J,U,V$.\\
\begin{figure*}
\begin{minipage}{0.365\linewidth}
\begin{overpic}[width = 1\textwidth, trim = {0 0 5 5}, clip]{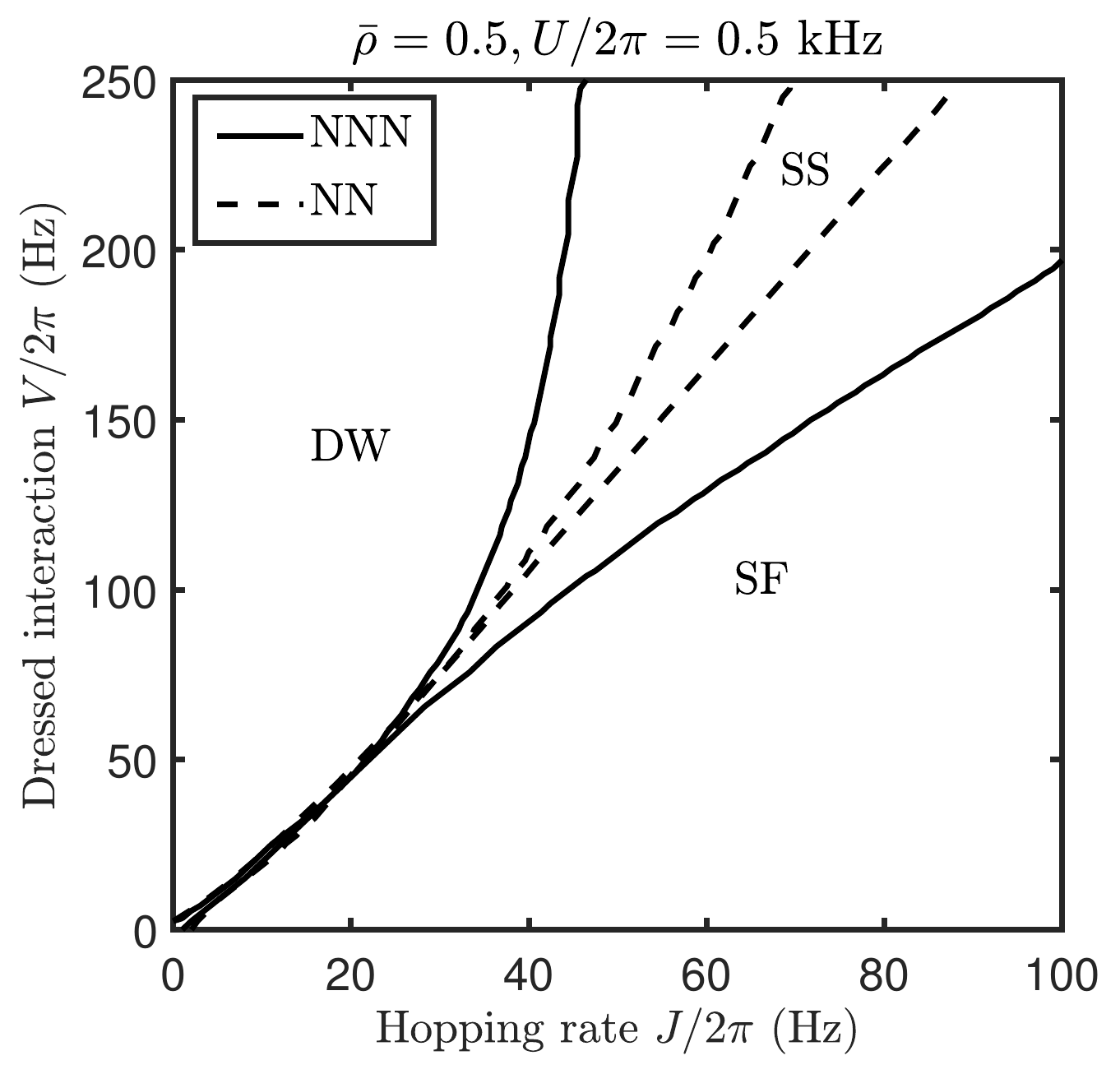}
\put (0,96) {\small(a)}
\end{overpic}
\end{minipage}
\begin{minipage}{0.24\linewidth}
\begin{overpic}[width = 0.75\textwidth, trim = {20 0 0 5}, clip]{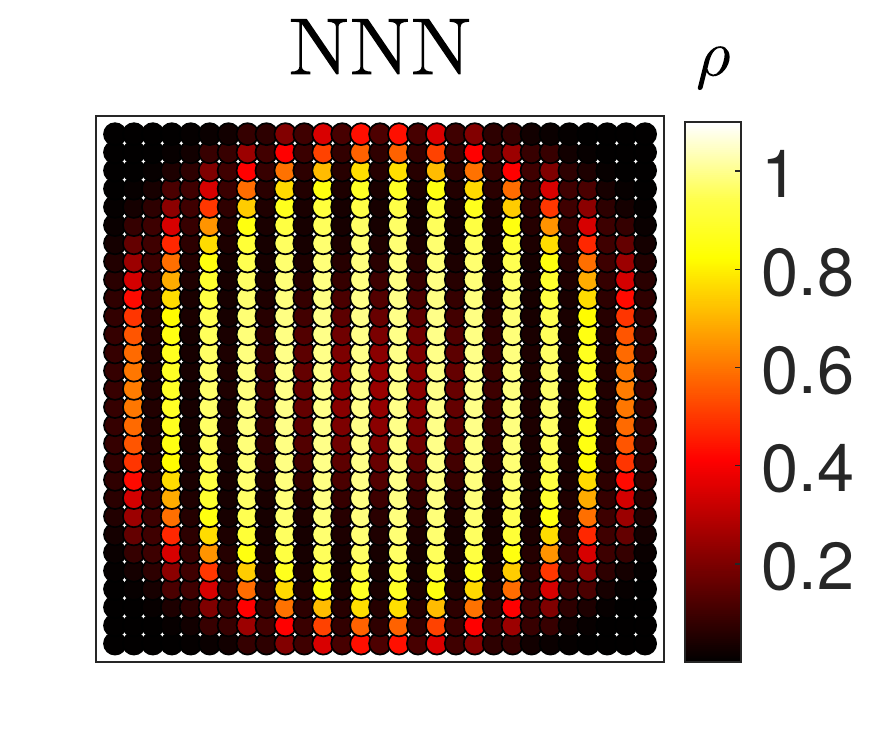}
\put (0,95) {\small(b)}
\end{overpic}
\begin{overpic}[width = 0.75\textwidth, trim = {20 0 0 5}, clip]{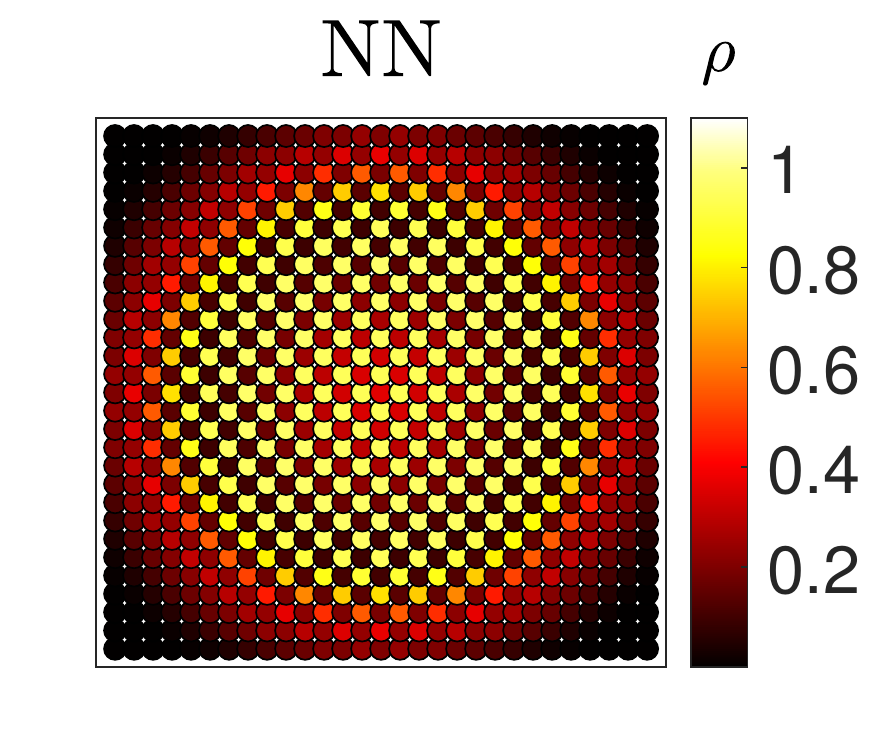}
\put(-8,40){\rotatebox{90}{\scriptsize Position $y$ (units of $a_\mathrm{lat}$)}}
\put (0,0) {\scriptsize Position $x$ (units of $a_\mathrm{lat}$)}
\end{overpic}
\vspace*{0.26cm}
\end{minipage}
\begin{minipage}{0.365\linewidth}
\begin{overpic}[width = 1\textwidth, trim = {0 0 0 5}, clip]{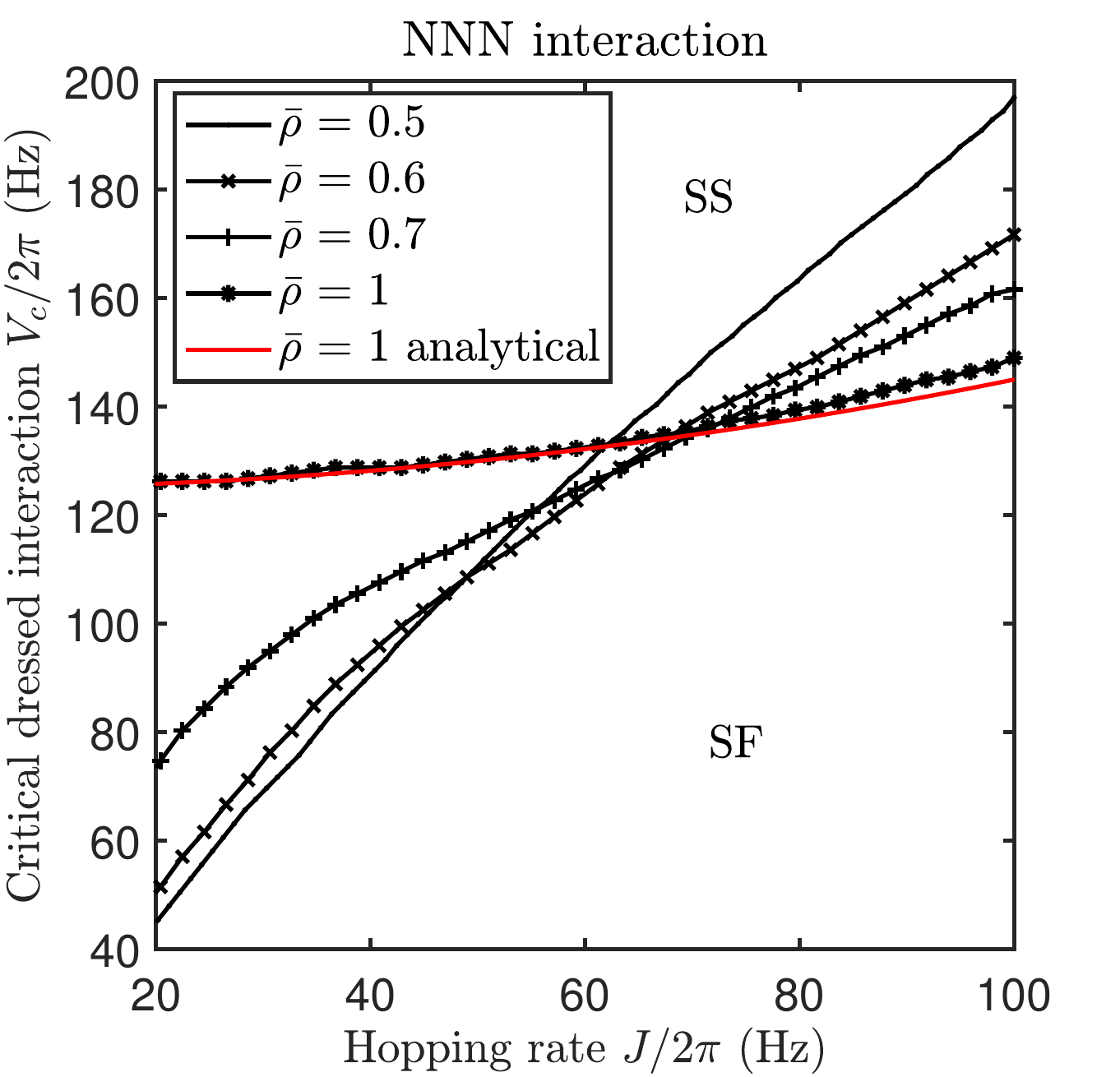}
\put (0,96) {\small(c)}
\end{overpic}
\end{minipage}
\caption{(a) Equilibrium phase diagram of the extended Bose-Hubbard model for $U = 2\pi \times 0.5$ kHz and $\bar{\rho} = 0.5$ for both NN (dashed line) and NNN (solid line) interactions. We obtain a density wave (DW) and a supersolid (SS) as well as a homogeneous superfluid (SF) regime. NN interaction results in a checkerboard ordering while NNN interaction leads to striped ordering. (b) Ground state density pattern of exemplary supersolid states of the phase diagram with additional harmonic potential. The upper figure represents a striped ordered supersolid obtained with NNN interaction, the lower figure a checkerboard ordered supersolid obtained with NN interaction. Here, the average filling $\bar \rho = 0.5$ refers to the density at the center of the harmonic potential. (c) Phase transition for various fillings with NNN interaction. The solid red line corresponds to the phase boundary obtained through an analytic approach for $\bar{\rho} = 1$.}
\label{fig:equilibrium}
\end{figure*}
\noindent
We first investigate the equilibrium phase diagram by performing numerical simulations with the variational cluster Gutzwiller approach (CGA) ~\cite{ClusterPairing,ClusterDisorder,ClusterDipolar,ClusterGauge,ClusterTriangular,ClusterHofstadter,ClusterAnisotropic,ClusterTwoSpecies}. Within this method the system is described by a lattice of clusters with each cluster embedded in a self-consistent mean field, hereby factorizing the wave function of the full system into cluster wave functions as
\begin{equation}
|\Psi\rangle = \prod_{\mathcal{C}} |\Psi\rangle_{\mathcal{C}}
\end{equation}
which satisfy the newly defined cluster Schrödinger equation $\hat{H}_\mathcal{C} |\Psi\rangle_\mathcal{C} = E_\mathcal{C} |\Psi\rangle_\mathcal{C}$ with $\hat{H} = \sum_\mathcal{C} \hat{H}_\mathcal{C}$. The single-cluster Hamiltonian $\hat{H}_{\mathcal{C}}$ itself can be written as
\begin{equation}
\begin{split}
\hat{H}_\mathcal{C} = &-J\sum_{\langle ij \rangle \in \mathcal{C}}(\hat{b}^\dag_i \hat{b}_j + \hat{b}^\dag_j \hat{b}_i) + \frac{U}{2} \sum_{i \in \mathcal{C}} \hat{b}^\dag_i \hat{b}^\dag_i \hat{b}_i \hat{b}_i\\
&- \mu \sum_{i \in \mathcal{C}} \hat{b}^\dag_i \hat{b}_i + V \sum_{ij \in \mathcal{C}} \hat{b}^\dag_i \hat{b}^\dag_j \hat{b}_i \hat{b}_j \delta_{R_{ij},R_e}\\
&-J\sum_{i \in \mathcal{\partial C}}(\hat{b}^\dag_i \varphi_i+ \hat{b}_i \varphi_i^*) + V \sum_{i \in \mathcal{\partial C}} \hat{b}^\dag_i \hat{b}_i \eta_i + E(\varphi,\eta)
\end{split} \label{Cluster}
\end{equation}
with $\mathcal{C}$ denoting the set of sites within a cluster, $\mathcal{\partial C}$ the sites on the border and $E(\varphi,\eta)$ an offset energy resulting from the cluster mean field approximation. The site dependent mean fields $\varphi_i$ and $\eta_i$ are determined by the neighboring clusters wave functions. Finding the ground state of the system requires a self-consistent iterative procedure which involves solving of the cluster Schrödinger equation and calculation of the surrounding mean fields until convergence.\\
Compared to the single-site Gutzwiller mean field approximation, non-local quantum fluctuations are included up to a certain degree dependent on the cluster size in the CGA. Numerous theoretical studies have shown that the negligence of quantum fluctuations might affect phase boundaries or even lead to inaccurate predictions of quantum states ~\cite{NN2,NNmeanfield,ClusterLuhmann}. We therefore choose the CGA and use sufficiently big cluster sizes in order to capture quantum fluctuations (for a discussion of the cluster size see Appendix D).\\
Furthermore we study the effect of dissipation and dephasing with the Cluster Gutzwiller Lindblad master equation (CGLE), a CGA version of the master equation in Lindblad form. We define the cluster density operator $\hat \rho_\mathcal{C} = |\Psi\rangle_\mathcal{C}  \langle \Psi|_\mathcal{C}$ for which the CGLE reads
\begin{equation}
\frac{d\hat \rho_\mathcal{C}}{dt} = -\frac{i}{\hbar} [\hat{H}_\mathcal{C},\hat \rho_\mathcal{C}] + \mathcal{L}(\hat \rho_\mathcal{C}),
\end{equation}
with the Lindblad superoperator $\mathcal{L}(\cdot) = \sum_k (\hat L_k (\cdot) \hat L^\dag_k - \frac{1}{2} \{\hat L^\dag_k \hat L_k,(\cdot)\})$, which describes non-unitary processes of the system. We consider both single particle loss $\hat L_\text{Ryd} = \sum_i \sqrt{\Gamma}_\text{Ryd} \hat b_i$ and macrodimer loss $\hat L_\text{mol} = \sum_{ij} \sqrt{\Gamma}_\text{mol} \hat b_i \hat b_j \delta_{R_{ij},R_e}$ with the previously discussed scattering rates $\Gamma_\text{Ryd}$ and $\Gamma_\text{mol}$. Given an initial state density operator we are able to compute the time evolution via the CGLE and identify possible phase transitions through calculation of local observables.

\section{Equilibrium states and time evolution simulations}
We first investigate the equilibrium ground state phase diagram for realistic values of the on-site interaction, hopping rate and the long-range interaction. We set the on-site interaction to be $U = 2\pi \times 0.5$ kHz, the range of the hopping rate to $J/2\pi \in [0,100]$ Hz and the range of the dressed interaction strength $V/2\pi \in [0, 250]$ Hz. We vary the chemical potential $\mu$ in such a way that the average filling $\bar{\rho} = 1/N \sum_{i \in \mathcal{C}} \langle \Psi| \hat{b}^\dag_i \hat{b}_i | \Psi \rangle_\mathcal{C}$ of a cluster $\mathcal{C}$ with $N$ sites is fixed (see Appendix E). We determine the different phases through calculation of the condensate order parameter $|\phi|$ and occupation number $n$ as well as the staggered observable $|\phi|_\text{stag}$.\\
We first fix the average filling to $\bar{\rho} = 0.5$. We vary the hopping rate $J$ and the long-range interaction $V$, and plot the various regimes of the equilibrium phase diagrams for both types of interaction, NN and NNN (see FIG. \ref{fig:equilibrium}. (a)). At low hopping amplitudes we obtain density wave phases (DW) whereas for low long-range interaction strengths the system becomes superfluid (SF). These two regimes are separated by supersolid phases (SS) which exhibit a staggered condensate order parameter and occupation number. Note that the NN interaction leads to checkerboard ordering while NNN interaction favors striped ordering (see FIG. \ref{fig:equilibrium}. (b)). We clearly see that the SS regime is broader in the NNN case compared to the NN case. We believe that this phenomenon arises from the competition between long-range interaction and particle hopping which differs in both cases. For NN interaction, the long-range interaction and particle hopping mechanism couple the same pair of sites. While the repulsive long-range interaction favors a staggered occupation number, the superfluid hopping profits from both sites being occupied which results in a direct competition between these processes. In the NNN case the long-range interaction couples sites not coupled by the hopping mechanism, leading to a weaker competition and therefore allows to have ordered states at lower long-range interaction strengths for finite hopping rate. We conclude that NNN interaction is favorable for supersolidity.
\par
\noindent
\begin{figure*}
\begin{minipage}{0.325\linewidth}
\begin{overpic}[width = 1\textwidth, trim = {0 0 0 0}, clip]{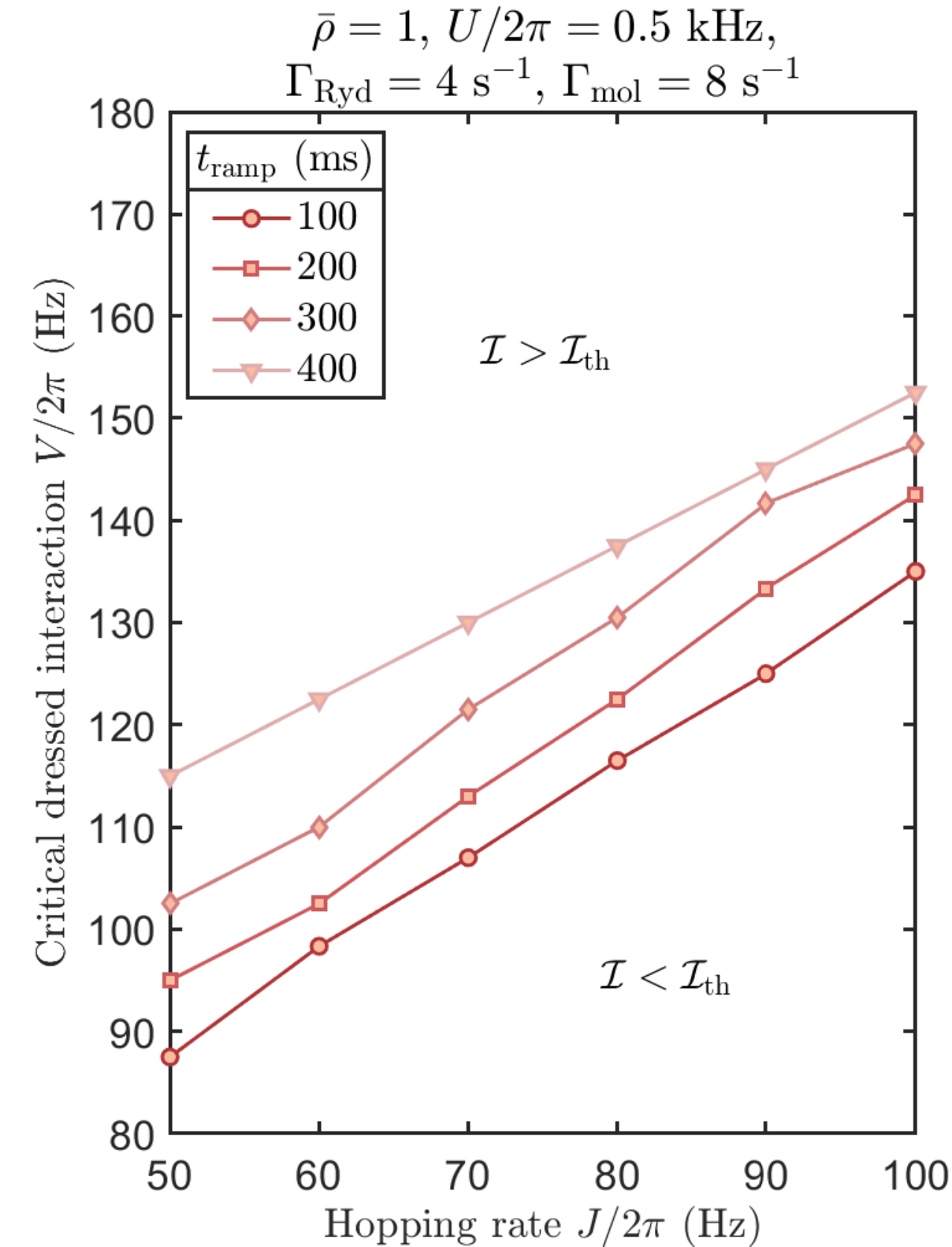}
\put (0,97) {\small(a)}
\end{overpic}
\end{minipage}
\begin{minipage}{.325\linewidth}
\begin{overpic}[width = 1\textwidth, trim = {0 0 0 0}, clip]{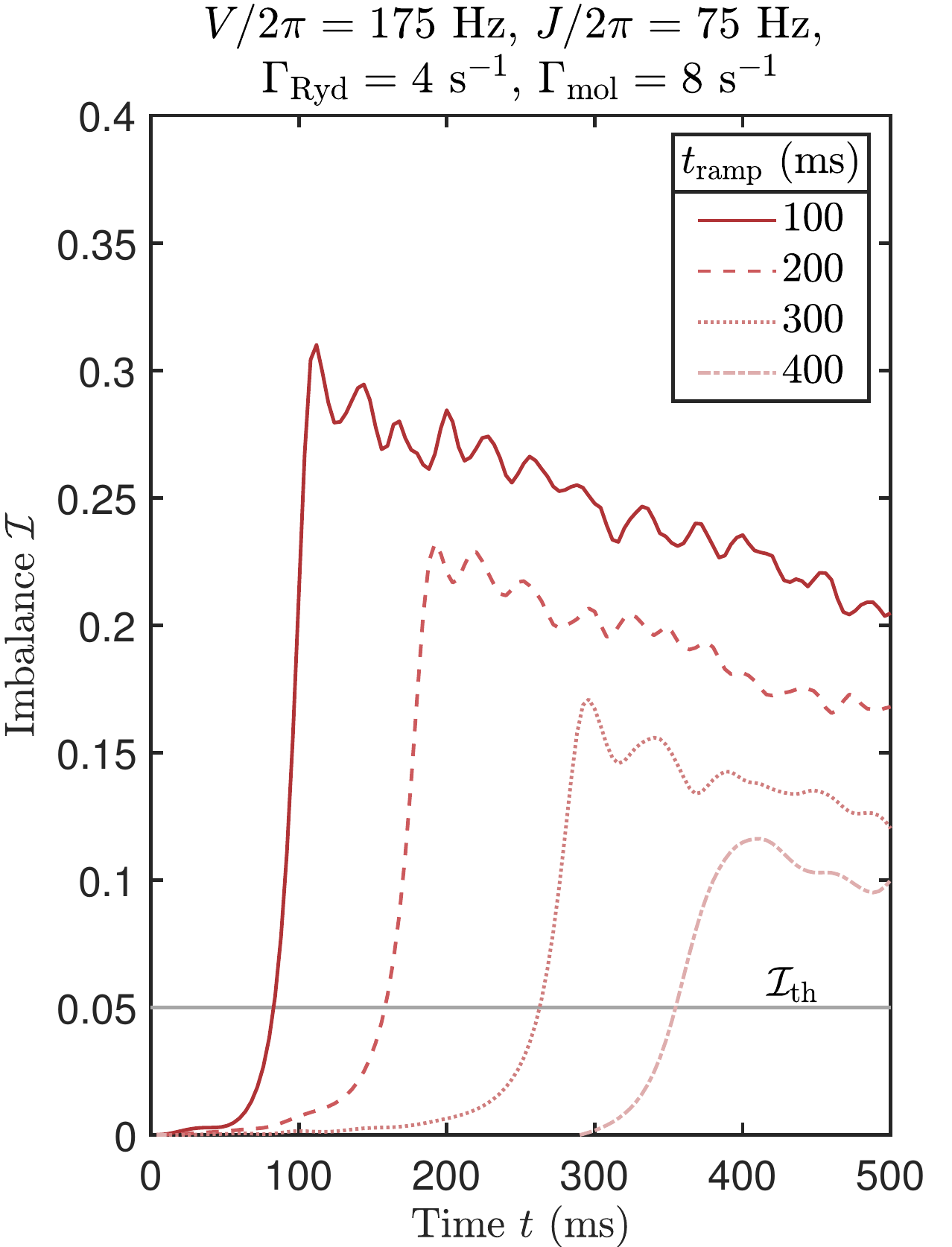}
\put (0,97) {\small(b)}
\end{overpic}
\end{minipage}
\begin{minipage}{.28\linewidth}
\begin{overpic}[width = 1.01\textwidth, trim = {0 0 0 -27}, clip]{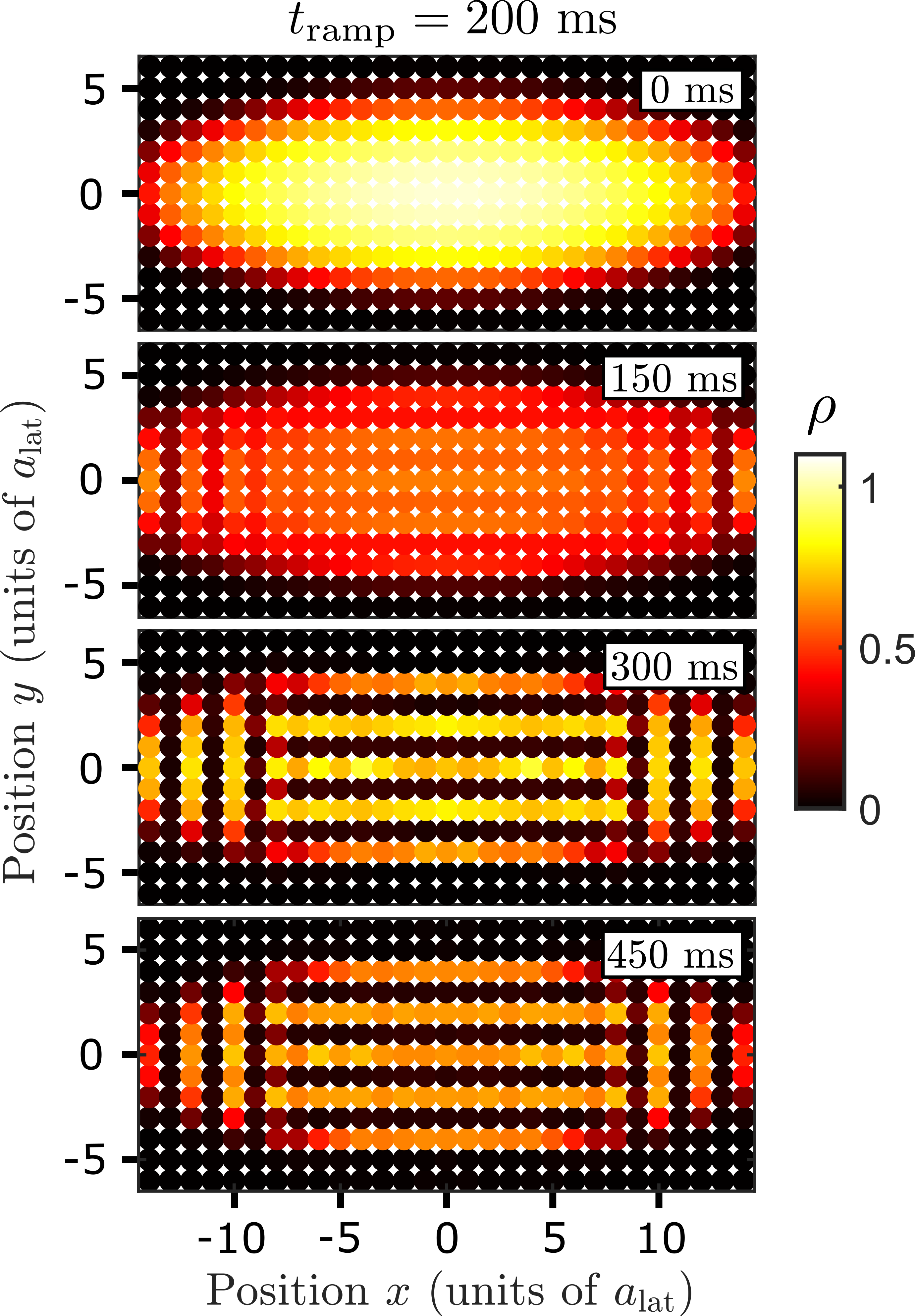}
\put (0,97) {\small(c)}
\end{overpic}
\end{minipage}
\caption{(a) Parameter regime of spontaneous symmetry broken time evolution ($\mathcal{I} > \mathcal{I}_\mathrm{th}$) and homogeneous time evolution ($\mathcal{I} < \mathcal{I}_\mathrm{th}$) given by the critical dressed interaction $V_c$ versus the hopping rate $J$ for different values of the ramping times with $\mathcal{I}_\mathrm{th} = 0.05$. Longer ramping times require larger dressed interaction for the imbalance $\mathcal{I}$ to become finite within the evolution times considered here. (b) Imbalance $\mathcal{I}$ versus time $t$ for an initial SF state for fixed parameters of the extended Bose-Hubbard model $U = 2\pi \times 0.5$ kHz, $J = 2\pi \times 75$ Hz, $V = 2\pi \times 175$ Hz, a bare Rydberg scattering rate $\Gamma_\mathrm{Ryd} = 4$ s$^\text{-1}$ and a molecular scattering rate $\Gamma_\mathrm{mol} = 8$ s$^\text{-1}$ for different values of the ramping times $t_\text{ramp} \in [100,400]$ ms. We find a finite imbalance emerging after ramping up the coupling to the macrodimer state. The later onset of a finite imbalance occurs for longer ramping times and its value decays due to the single particle and macrodimer losses. (c) Exemplary depiction of the time evolution of (b) with ramping time $t_\text{ramp} = 200$ ms at times $t \in [0,150,300,450]$ ms where the occupation number $\rho$ at each site is shown. The average filling $\bar \rho = 1$ is determined at the center of the trap. Due to the anisotropic harmonic confinement, striped order in $y$-direction becomes more favorable.}
\label{fig:time}
\end{figure*}
We additionally vary the filling $\bar{\rho}$ and study its effect on the phase boundary between the SF and the SS regime (see FIG. \ref{fig:equilibrium}. (c)). By increasing the filling up to $\bar{\rho} = 1$, we are able to lower the critical long-range interaction strength by a significant amount in the regime of high hopping amplitudes. Since striped phases are less susceptible to the dressed interaction in comparison to homogeneous phases, they are energetically more favorable for greater densities. For $\bar{\rho} = 1$, we perform second order perturbation theory in order to determine the phase transition between the homogeneous SF and the SS with long-range density wave order \cite{Perturbation}. We obtain the analytic value for the critical interaction $V_c = U/4 + J^2/U$, which coincides well with the numerically obtained phase boundary. We conclude that the dressed long-range interaction resulting from the avoided crossing of the chosen asymptotic pair states renders the task of observing density wave ordered states feasible. For lower hopping amplitudes we choose a lower filling, while higher fillings are helpful in the high hopping rate regime. Since higher hopping rates lead to faster dynamics during the time evolution, we focus on the $\bar{\rho} = 1$ case hereafter, even though higher densities also increase collective loss rates.\\
We now present possibilities for the preparation of itinerant states with density-wave order via NNN interaction. We start with an initial SF state and switch on the macrodimer dressing adiabatically, by linear ramping up of the dressed interaction. Since striped phases are fourfold degenerate with respect to rotation and translation, we impose an external, anisotropic harmonic confinement in order to lift the degeneracy and hereby enables the possibility of adiabatic time evolution. In order to determine the emergence of a striped phase, we define the imbalance $\mathcal{I} = |\rho_\text{odd} - \rho_\text{even}|/N$ as an order parameter where $N$ is the number of sites. We consider odd and even sites along the more strongly confined direction as we expect the stripes to form along the less strongly confined direction (see FIG. \ref{fig:time}. (c)). We determine whether the imbalance is finite or not based on a numerical threshold $\mathcal{I}_\mathrm{th} = 0.05$ (see Appendix E).\\
In the following simulation, we set the initial average occupation at the center of the harmonic confinement to $\bar{\rho} \approx 1$ and the on-site interaction $U = 2\pi \times 0.5$ kHz. For Rubidium at the considered principal quantum numbers, we obtain a bare Rydberg scattering rate around $\Gamma_\text{Ryd} = 4$ s$^\text{-1}$ and a macrodimer scattering rate of $\Gamma_\text{mol} = 8$ s$^\text{-1}$ (see Appendix C), assuming no collective loss processes. The dressed interaction is ramped up linearly with time $t_\text{ramp}$. We perform time evolution simulations of an initial SF state (i.e., at time $t = 0$ we start with the equilibrium state for $V = 0$) up to a time $t = 500$ ms. We perform these time evolution simulations for different values of the ramping time $t_\text{ramp}$, dressed interaction $V$ and hopping rate $J$ and determine a parameter regime which yields finite imbalance during the time evolution. With respect to the quantum adiabatic theorem, a ramping time of $t_\mathrm{ramp} > 20$ ms for any dressed interaction $V$ considered ensures the adiabaticity of the time evolution \cite{Adiabatic1,Adiabatic2}.\\
We see that the critical dressed interaction becoming larger as the ramping time becomes longer (see FIG. \ref{fig:time}. (a)). Similar to the phase transition of the equilibrium ground state calculations the critical interaction strength for striped order increases for larger hopping rates.\\
We depict the time evolution for an initial SF state ($\mathcal{I} = 0$) for fixed hopping rate and dressed interaction, while varying the ramping time $t_\text{ramp} \in [100,400]$ ms. We find the emergence of finite imbalance $\mathcal{I}$ at later times for higher ramping times. The imbalance also becomes smaller due to the losses as the ramping time grows, although it is far from zero.\\
We conclude that the considered scattering process does not inhibit the emergence of spontaneous striped density-wave order in an initial SF for reasonably slow ramping of the coupling to the macrodimer state. The depletion of the system due to the finite scattering rate indicates an upper limit of the possible ramping time, but achieving spontaneous symmetry breaking seems possible for realistic parameter values. The determined dressed interaction strengths appear to be significantly larger than the critical value necessary for the observation of symmetry breaking.\\

\section{Conclusion}
In conclusion, we study coupling rates to various macrodimer potentials over a wide range of experimental tuning parameters and provide the relevant scaling laws.
In addition to the single-color coupling scheme we propose a tunable two-color scheme, which allows to enhance the dressed interactions and cancel the overall AC Stark shift.
Using experimentally feasible values for the interactions strength, we obtain spatially modulated equilibrium phases, which can be realized by ramping up the coupling to the macrodimer state.
Thanks to its wide range of tunable parameters, we hope that one can find a parameter regime, where previously limiting avalanche losses, which were observed near-resonant to the Rydberg transition, can be avoided.\\


\section*{Acknowledgements}
We thank Jaromir Panas for his contribution to the computational aspect of this work, Jun Rui for the idea of the two-color coupling scheme and Johannes Zeiher for the idea of canceling out the AC Stark shift. Furthermore we thank all contributors to the open-source programs \texttt{pair interaction} and \texttt{ARC}. Support by the Deutsche Forschungsgemeinschaft via DFG SPP 1929 GiRyd, DFG HO 2407/8-1 and the high-performance computing center LOEWE-CSC is gratefully acknowledged. 

\bibliographystyle{apsrev4-1}


\begin{thebibliography}{72}%
\makeatletter
\providecommand \@ifxundefined [1]{%
 \@ifx{#1\undefined}
}%
\providecommand \@ifnum [1]{%
 \ifnum #1\expandafter \@firstoftwo
 \else \expandafter \@secondoftwo
 \fi
}%
\providecommand \@ifx [1]{%
 \ifx #1\expandafter \@firstoftwo
 \else \expandafter \@secondoftwo
 \fi
}%
\providecommand \natexlab [1]{#1}%
\providecommand \enquote  [1]{``#1''}%
\providecommand \bibnamefont  [1]{#1}%
\providecommand \bibfnamefont [1]{#1}%
\providecommand \citenamefont [1]{#1}%
\providecommand \href@noop [0]{\@secondoftwo}%
\providecommand \href [0]{\begingroup \@sanitize@url \@href}%
\providecommand \@href[1]{\@@startlink{#1}\@@href}%
\providecommand \@@href[1]{\endgroup#1\@@endlink}%
\providecommand \@sanitize@url [0]{\catcode `\\12\catcode `\$12\catcode
  `\&12\catcode `\#12\catcode `\^12\catcode `\_12\catcode `\%12\relax}%
\providecommand \@@startlink[1]{}%
\providecommand \@@endlink[0]{}%
\providecommand \url  [0]{\begingroup\@sanitize@url \@url }%
\providecommand \@url [1]{\endgroup\@href {#1}{\urlprefix }}%
\providecommand \urlprefix  [0]{URL }%
\providecommand \Eprint [0]{\href }%
\providecommand \doibase [0]{http://dx.doi.org/}%
\providecommand \selectlanguage [0]{\@gobble}%
\providecommand \bibinfo  [0]{\@secondoftwo}%
\providecommand \bibfield  [0]{\@secondoftwo}%
\providecommand \translation [1]{[#1]}%
\providecommand \BibitemOpen [0]{}%
\providecommand \bibitemStop [0]{}%
\providecommand \bibitemNoStop [0]{.\EOS\space}%
\providecommand \EOS [0]{\spacefactor3000\relax}%
\providecommand \BibitemShut  [1]{\csname bibitem#1\endcsname}%
\let\auto@bib@innerbib\@empty
\bibitem [{\citenamefont {Zeiher}\ \emph {et~al.}(2016)\citenamefont {Zeiher},
  \citenamefont {Bijnen}, \citenamefont {Schauß}, \citenamefont {Hild},
  \citenamefont {Choi}, \citenamefont {Pohl}, \citenamefont {Bloch},\ and\
  \citenamefont {Gross}}]{SpinLattice1}%
  \BibitemOpen
  \bibfield  {author} {\bibinfo {author} {\bibfnamefont {J.}~\bibnamefont
  {Zeiher}}, \bibinfo {author} {\bibfnamefont {R.}~\bibnamefont {Bijnen}},
  \bibinfo {author} {\bibfnamefont {P.}~\bibnamefont {Schauß}}, \bibinfo
  {author} {\bibfnamefont {S.}~\bibnamefont {Hild}}, \bibinfo {author}
  {\bibfnamefont {J.-y.}\ \bibnamefont {Choi}}, \bibinfo {author}
  {\bibfnamefont {T.}~\bibnamefont {Pohl}}, \bibinfo {author} {\bibfnamefont
  {I.}~\bibnamefont {Bloch}}, \ and\ \bibinfo {author} {\bibfnamefont
  {C.}~\bibnamefont {Gross}},\ }\href
  {https://www.nature.com/articles/nphys3835} {\bibfield  {journal} {\bibinfo
  {journal} {Nature Physics}\ }\textbf {\bibinfo {volume} {12}} (\bibinfo
  {year} {2016})}\BibitemShut {NoStop}%
\bibitem [{\citenamefont {Jau}\ \emph {et~al.}(2015)\citenamefont {Jau},
  \citenamefont {Hankin}, \citenamefont {Keating}, \citenamefont {Deutsch},\
  and\ \citenamefont {Biedermann}}]{SpinLattice2}%
  \BibitemOpen
  \bibfield  {author} {\bibinfo {author} {\bibfnamefont {Y.-Y.}\ \bibnamefont
  {Jau}}, \bibinfo {author} {\bibfnamefont {A.}~\bibnamefont {Hankin}},
  \bibinfo {author} {\bibfnamefont {T.}~\bibnamefont {Keating}}, \bibinfo
  {author} {\bibfnamefont {I.}~\bibnamefont {Deutsch}}, \ and\ \bibinfo
  {author} {\bibfnamefont {G.}~\bibnamefont {Biedermann}},\ }\href
  {https://www.nature.com/articles/nphys3487} {\bibfield  {journal} {\bibinfo
  {journal} {Nature Physics}\ }\textbf {\bibinfo {volume} {12}} (\bibinfo
  {year} {2015})}\BibitemShut {NoStop}%
\bibitem [{\citenamefont {Browaeys}\ and\ \citenamefont
  {Lahaye}(2020)}]{DressingTweezer}%
  \BibitemOpen
  \bibfield  {author} {\bibinfo {author} {\bibfnamefont {A.}~\bibnamefont
  {Browaeys}}\ and\ \bibinfo {author} {\bibfnamefont {T.}~\bibnamefont
  {Lahaye}},\ }\href {\doibase 10.1038/s41567-019-0733-z} {\bibfield  {journal}
  {\bibinfo  {journal} {Nature Physics}\ }\textbf {\bibinfo {volume} {16}},\
  \bibinfo {pages} {1} (\bibinfo {year} {2020})}\BibitemShut {NoStop}%
\bibitem [{\citenamefont {Boulier}\ \emph
  {et~al.}(2017{\natexlab{a}})\citenamefont {Boulier}, \citenamefont {Magnan},
  \citenamefont {Bracamontes}, \citenamefont {Maslek}, \citenamefont
  {Goldschmidt}, \citenamefont {Young}, \citenamefont {Gorshkov}, \citenamefont
  {Rolston},\ and\ \citenamefont {Porto}}]{Cryogenic}%
  \BibitemOpen
  \bibfield  {author} {\bibinfo {author} {\bibfnamefont {T.}~\bibnamefont
  {Boulier}}, \bibinfo {author} {\bibfnamefont {E.}~\bibnamefont {Magnan}},
  \bibinfo {author} {\bibfnamefont {C.}~\bibnamefont {Bracamontes}}, \bibinfo
  {author} {\bibfnamefont {J.}~\bibnamefont {Maslek}}, \bibinfo {author}
  {\bibfnamefont {E.~A.}\ \bibnamefont {Goldschmidt}}, \bibinfo {author}
  {\bibfnamefont {J.~T.}\ \bibnamefont {Young}}, \bibinfo {author}
  {\bibfnamefont {A.~V.}\ \bibnamefont {Gorshkov}}, \bibinfo {author}
  {\bibfnamefont {S.~L.}\ \bibnamefont {Rolston}}, \ and\ \bibinfo {author}
  {\bibfnamefont {J.~V.}\ \bibnamefont {Porto}},\ }\href {\doibase
  10.1103/PhysRevA.96.053409} {\bibfield  {journal} {\bibinfo  {journal} {Phys.
  Rev. A}\ }\textbf {\bibinfo {volume} {96}},\ \bibinfo {pages} {053409}
  (\bibinfo {year} {2017}{\natexlab{a}})}\BibitemShut {NoStop}%
\bibitem [{\citenamefont {Li}\ \emph {et~al.}(2018)\citenamefont {Li},
  \citenamefont {Gei\ss{}ler}, \citenamefont {Hofstetter},\ and\ \citenamefont
  {Li}}]{RydbergPhasediagram1}%
  \BibitemOpen
  \bibfield  {author} {\bibinfo {author} {\bibfnamefont {Y.}~\bibnamefont
  {Li}}, \bibinfo {author} {\bibfnamefont {A.}~\bibnamefont {Gei\ss{}ler}},
  \bibinfo {author} {\bibfnamefont {W.}~\bibnamefont {Hofstetter}}, \ and\
  \bibinfo {author} {\bibfnamefont {W.}~\bibnamefont {Li}},\ }\href {\doibase
  10.1103/PhysRevA.97.023619} {\bibfield  {journal} {\bibinfo  {journal} {Phys.
  Rev. A}\ }\textbf {\bibinfo {volume} {97}},\ \bibinfo {pages} {023619}
  (\bibinfo {year} {2018})}\BibitemShut {NoStop}%
\bibitem [{\citenamefont {Caballero-Benitez}\ \emph {et~al.}(2016)\citenamefont
  {Caballero-Benitez}, \citenamefont {Mazzucchi},\ and\ \citenamefont
  {Mekhov}}]{RydbergSupersolidsIII}%
  \BibitemOpen
  \bibfield  {author} {\bibinfo {author} {\bibfnamefont {S.~F.}\ \bibnamefont
  {Caballero-Benitez}}, \bibinfo {author} {\bibfnamefont {G.}~\bibnamefont
  {Mazzucchi}}, \ and\ \bibinfo {author} {\bibfnamefont {I.~B.}\ \bibnamefont
  {Mekhov}},\ }\href {\doibase 10.1103/PhysRevA.93.063632} {\bibfield
  {journal} {\bibinfo  {journal} {Phys. Rev. A}\ }\textbf {\bibinfo {volume}
  {93}},\ \bibinfo {pages} {063632} (\bibinfo {year} {2016})}\BibitemShut
  {NoStop}%
\bibitem [{\citenamefont {Pupillo}\ \emph {et~al.}(2010)\citenamefont
  {Pupillo}, \citenamefont {Micheli}, \citenamefont {Boninsegni}, \citenamefont
  {Lesanovsky},\ and\ \citenamefont {Zoller}}]{RydbergSupersolidsV}%
  \BibitemOpen
  \bibfield  {author} {\bibinfo {author} {\bibfnamefont {G.}~\bibnamefont
  {Pupillo}}, \bibinfo {author} {\bibfnamefont {A.}~\bibnamefont {Micheli}},
  \bibinfo {author} {\bibfnamefont {M.}~\bibnamefont {Boninsegni}}, \bibinfo
  {author} {\bibfnamefont {I.}~\bibnamefont {Lesanovsky}}, \ and\ \bibinfo
  {author} {\bibfnamefont {P.}~\bibnamefont {Zoller}},\ }\href {\doibase
  10.1103/PhysRevLett.104.223002} {\bibfield  {journal} {\bibinfo  {journal}
  {Phys. Rev. Lett.}\ }\textbf {\bibinfo {volume} {104}},\ \bibinfo {pages}
  {223002} (\bibinfo {year} {2010})}\BibitemShut {NoStop}%
\bibitem [{\citenamefont {Gei\ss{}ler}\ \emph {et~al.}(2017)\citenamefont
  {Gei\ss{}ler}, \citenamefont {Vasi\ifmmode~\acute{c}\else \'{c}\fi{}},\ and\
  \citenamefont {Hofstetter}}]{RydbergPhasediagram3}%
  \BibitemOpen
  \bibfield  {author} {\bibinfo {author} {\bibfnamefont {A.}~\bibnamefont
  {Gei\ss{}ler}}, \bibinfo {author} {\bibfnamefont {I.}~\bibnamefont
  {Vasi\ifmmode~\acute{c}\else \'{c}\fi{}}}, \ and\ \bibinfo {author}
  {\bibfnamefont {W.}~\bibnamefont {Hofstetter}},\ }\href {\doibase
  10.1103/PhysRevA.95.063608} {\bibfield  {journal} {\bibinfo  {journal} {Phys.
  Rev. A}\ }\textbf {\bibinfo {volume} {95}},\ \bibinfo {pages} {063608}
  (\bibinfo {year} {2017})}\BibitemShut {NoStop}%
\bibitem [{\citenamefont {Weimer}\ \emph {et~al.}(2008)\citenamefont {Weimer},
  \citenamefont {L\"ow}, \citenamefont {Pfau},\ and\ \citenamefont
  {B\"uchler}}]{RydbergPhasediagram4}%
  \BibitemOpen
  \bibfield  {author} {\bibinfo {author} {\bibfnamefont {H.}~\bibnamefont
  {Weimer}}, \bibinfo {author} {\bibfnamefont {R.}~\bibnamefont {L\"ow}},
  \bibinfo {author} {\bibfnamefont {T.}~\bibnamefont {Pfau}}, \ and\ \bibinfo
  {author} {\bibfnamefont {H.~P.}\ \bibnamefont {B\"uchler}},\ }\href {\doibase
  10.1103/PhysRevLett.101.250601} {\bibfield  {journal} {\bibinfo  {journal}
  {Phys. Rev. Lett.}\ }\textbf {\bibinfo {volume} {101}},\ \bibinfo {pages}
  {250601} (\bibinfo {year} {2008})}\BibitemShut {NoStop}%
\bibitem [{\citenamefont {Ji}\ \emph {et~al.}(2011)\citenamefont {Ji},
  \citenamefont {Ates},\ and\ \citenamefont
  {Lesanovsky}}]{RydbergPhasediagram5}%
  \BibitemOpen
  \bibfield  {author} {\bibinfo {author} {\bibfnamefont {S.}~\bibnamefont
  {Ji}}, \bibinfo {author} {\bibfnamefont {C.}~\bibnamefont {Ates}}, \ and\
  \bibinfo {author} {\bibfnamefont {I.}~\bibnamefont {Lesanovsky}},\ }\href
  {\doibase 10.1103/PhysRevLett.107.060406} {\bibfield  {journal} {\bibinfo
  {journal} {Phys. Rev. Lett.}\ }\textbf {\bibinfo {volume} {107}},\ \bibinfo
  {pages} {060406} (\bibinfo {year} {2011})}\BibitemShut {NoStop}%
\bibitem [{\citenamefont {Lesanovsky}(2011)}]{RydbergPhasediagram6}%
  \BibitemOpen
  \bibfield  {author} {\bibinfo {author} {\bibfnamefont {I.}~\bibnamefont
  {Lesanovsky}},\ }\href {\doibase 10.1103/PhysRevLett.106.025301} {\bibfield
  {journal} {\bibinfo  {journal} {Phys. Rev. Lett.}\ }\textbf {\bibinfo
  {volume} {106}},\ \bibinfo {pages} {025301} (\bibinfo {year}
  {2011})}\BibitemShut {NoStop}%
\bibitem [{\citenamefont {Goldschmidt}\ \emph {et~al.}(2016)\citenamefont
  {Goldschmidt}, \citenamefont {Boulier}, \citenamefont {Brown}, \citenamefont
  {Koller}, \citenamefont {Young}, \citenamefont {Gorshkov}, \citenamefont
  {Rolston},\ and\ \citenamefont {Porto}}]{AvalancheII}%
  \BibitemOpen
  \bibfield  {author} {\bibinfo {author} {\bibfnamefont {E.~A.}\ \bibnamefont
  {Goldschmidt}}, \bibinfo {author} {\bibfnamefont {T.}~\bibnamefont
  {Boulier}}, \bibinfo {author} {\bibfnamefont {R.~C.}\ \bibnamefont {Brown}},
  \bibinfo {author} {\bibfnamefont {S.~B.}\ \bibnamefont {Koller}}, \bibinfo
  {author} {\bibfnamefont {J.~T.}\ \bibnamefont {Young}}, \bibinfo {author}
  {\bibfnamefont {A.~V.}\ \bibnamefont {Gorshkov}}, \bibinfo {author}
  {\bibfnamefont {S.~L.}\ \bibnamefont {Rolston}}, \ and\ \bibinfo {author}
  {\bibfnamefont {J.~V.}\ \bibnamefont {Porto}},\ }\href {\doibase
  10.1103/PhysRevLett.116.113001} {\bibfield  {journal} {\bibinfo  {journal}
  {Phys. Rev. Lett.}\ }\textbf {\bibinfo {volume} {116}},\ \bibinfo {pages}
  {113001} (\bibinfo {year} {2016})}\BibitemShut {NoStop}%
\bibitem [{\citenamefont {Aman}\ \emph {et~al.}(2016)\citenamefont {Aman},
  \citenamefont {DeSalvo}, \citenamefont {Dunning}, \citenamefont {Killian},
  \citenamefont {Yoshida},\ and\ \citenamefont {Burgd{\"o}rfer}}]{AtomLoss2}%
  \BibitemOpen
  \bibfield  {author} {\bibinfo {author} {\bibfnamefont {J.~A.}\ \bibnamefont
  {Aman}}, \bibinfo {author} {\bibfnamefont {B.~J.}\ \bibnamefont {DeSalvo}},
  \bibinfo {author} {\bibfnamefont {F.~B.}\ \bibnamefont {Dunning}}, \bibinfo
  {author} {\bibfnamefont {T.~C.}\ \bibnamefont {Killian}}, \bibinfo {author}
  {\bibfnamefont {S.}~\bibnamefont {Yoshida}}, \ and\ \bibinfo {author}
  {\bibfnamefont {J.}~\bibnamefont {Burgd{\"o}rfer}},\ }\href {\doibase
  10.1103/PhysRevA.93.043425} {\bibfield  {journal} {\bibinfo  {journal} {Phys.
  Rev. A}\ }\textbf {\bibinfo {volume} {93}},\ \bibinfo {pages} {043425}
  (\bibinfo {year} {2016})}\BibitemShut {NoStop}%
\bibitem [{\citenamefont {Zeiher}\ \emph {et~al.}(2017)\citenamefont {Zeiher},
  \citenamefont {Choi}, \citenamefont {Rubio-Abadal}, \citenamefont {Pohl},
  \citenamefont {van Bijnen}, \citenamefont {Bloch},\ and\ \citenamefont
  {Gross}}]{AtomLoss1}%
  \BibitemOpen
  \bibfield  {author} {\bibinfo {author} {\bibfnamefont {J.}~\bibnamefont
  {Zeiher}}, \bibinfo {author} {\bibfnamefont {J.-y.}\ \bibnamefont {Choi}},
  \bibinfo {author} {\bibfnamefont {A.}~\bibnamefont {Rubio-Abadal}}, \bibinfo
  {author} {\bibfnamefont {T.}~\bibnamefont {Pohl}}, \bibinfo {author}
  {\bibfnamefont {R.}~\bibnamefont {van Bijnen}}, \bibinfo {author}
  {\bibfnamefont {I.}~\bibnamefont {Bloch}}, \ and\ \bibinfo {author}
  {\bibfnamefont {C.}~\bibnamefont {Gross}},\ }\href {\doibase
  10.1103/PhysRevX.7.041063} {\bibfield  {journal} {\bibinfo  {journal} {Phys.
  Rev. X}\ }\textbf {\bibinfo {volume} {7}},\ \bibinfo {pages} {041063}
  (\bibinfo {year} {2017})}\BibitemShut {NoStop}%
\bibitem [{\citenamefont {Robert-de Saint-Vincent}\ \emph
  {et~al.}(2013)\citenamefont {Robert-de Saint-Vincent}, \citenamefont
  {Hofmann}, \citenamefont {Schempp}, \citenamefont {G\"unter}, \citenamefont
  {Whitlock},\ and\ \citenamefont {Weidem\"uller}}]{AvalancheIV}%
  \BibitemOpen
  \bibfield  {author} {\bibinfo {author} {\bibfnamefont {M.}~\bibnamefont
  {Robert-de Saint-Vincent}}, \bibinfo {author} {\bibfnamefont {C.~S.}\
  \bibnamefont {Hofmann}}, \bibinfo {author} {\bibfnamefont {H.}~\bibnamefont
  {Schempp}}, \bibinfo {author} {\bibfnamefont {G.}~\bibnamefont {G\"unter}},
  \bibinfo {author} {\bibfnamefont {S.}~\bibnamefont {Whitlock}}, \ and\
  \bibinfo {author} {\bibfnamefont {M.}~\bibnamefont {Weidem\"uller}},\ }\href
  {\doibase 10.1103/PhysRevLett.110.045004} {\bibfield  {journal} {\bibinfo
  {journal} {Phys. Rev. Lett.}\ }\textbf {\bibinfo {volume} {110}},\ \bibinfo
  {pages} {045004} (\bibinfo {year} {2013})}\BibitemShut {NoStop}%
\bibitem [{\citenamefont {Labuhn}\ \emph {et~al.}(2014)\citenamefont {Labuhn},
  \citenamefont {Ravets}, \citenamefont {Barredo}, \citenamefont {B\'eguin},
  \citenamefont {Nogrette}, \citenamefont {Lahaye},\ and\ \citenamefont
  {Browaeys}}]{ACStark1}%
  \BibitemOpen
  \bibfield  {author} {\bibinfo {author} {\bibfnamefont {H.}~\bibnamefont
  {Labuhn}}, \bibinfo {author} {\bibfnamefont {S.}~\bibnamefont {Ravets}},
  \bibinfo {author} {\bibfnamefont {D.}~\bibnamefont {Barredo}}, \bibinfo
  {author} {\bibfnamefont {L.}~\bibnamefont {B\'eguin}}, \bibinfo {author}
  {\bibfnamefont {F.}~\bibnamefont {Nogrette}}, \bibinfo {author}
  {\bibfnamefont {T.}~\bibnamefont {Lahaye}}, \ and\ \bibinfo {author}
  {\bibfnamefont {A.}~\bibnamefont {Browaeys}},\ }\href {\doibase
  10.1103/PhysRevA.90.023415} {\bibfield  {journal} {\bibinfo  {journal} {Phys.
  Rev. A}\ }\textbf {\bibinfo {volume} {90}},\ \bibinfo {pages} {023415}
  (\bibinfo {year} {2014})}\BibitemShut {NoStop}%
\bibitem [{\citenamefont {Wilson}\ \emph {et~al.}(2019)\citenamefont {Wilson},
  \citenamefont {Saskin}, \citenamefont {Meng}, \citenamefont {Ma},
  \citenamefont {Dilip}, \citenamefont {Burgers},\ and\ \citenamefont
  {Thompson}}]{ACStark2}%
  \BibitemOpen
  \bibfield  {author} {\bibinfo {author} {\bibfnamefont {J.}~\bibnamefont
  {Wilson}}, \bibinfo {author} {\bibfnamefont {S.}~\bibnamefont {Saskin}},
  \bibinfo {author} {\bibfnamefont {Y.}~\bibnamefont {Meng}}, \bibinfo {author}
  {\bibfnamefont {S.}~\bibnamefont {Ma}}, \bibinfo {author} {\bibfnamefont
  {R.}~\bibnamefont {Dilip}}, \bibinfo {author} {\bibfnamefont
  {A.}~\bibnamefont {Burgers}}, \ and\ \bibinfo {author} {\bibfnamefont
  {J.}~\bibnamefont {Thompson}},\ }\href@noop {} {} (\bibinfo {year} {2019}),\
  \Eprint {http://arxiv.org/abs/1912.08754} {arXiv:1912.08754 [quant-ph]}
  \BibitemShut {NoStop}%
\bibitem [{\citenamefont {Guardado-Sanchez}\ \emph {et~al.}(2021)\citenamefont
  {Guardado-Sanchez}, \citenamefont {Spar}, \citenamefont {Schauss},
  \citenamefont {Belyansky}, \citenamefont {Young}, \citenamefont {Bienias},
  \citenamefont {Gorshkov}, \citenamefont {Iadecola},\ and\ \citenamefont
  {Bakr}}]{ItinerantRydberg}%
  \BibitemOpen
  \bibfield  {author} {\bibinfo {author} {\bibfnamefont {E.}~\bibnamefont
  {Guardado-Sanchez}}, \bibinfo {author} {\bibfnamefont {B.~M.}\ \bibnamefont
  {Spar}}, \bibinfo {author} {\bibfnamefont {P.}~\bibnamefont {Schauss}},
  \bibinfo {author} {\bibfnamefont {R.}~\bibnamefont {Belyansky}}, \bibinfo
  {author} {\bibfnamefont {J.~T.}\ \bibnamefont {Young}}, \bibinfo {author}
  {\bibfnamefont {P.}~\bibnamefont {Bienias}}, \bibinfo {author} {\bibfnamefont
  {A.~V.}\ \bibnamefont {Gorshkov}}, \bibinfo {author} {\bibfnamefont
  {T.}~\bibnamefont {Iadecola}}, \ and\ \bibinfo {author} {\bibfnamefont
  {W.~S.}\ \bibnamefont {Bakr}},\ }\href {\doibase 10.1103/PhysRevX.11.021036}
  {\bibfield  {journal} {\bibinfo  {journal} {Phys. Rev. X}\ }\textbf {\bibinfo
  {volume} {11}},\ \bibinfo {pages} {021036} (\bibinfo {year}
  {2021})}\BibitemShut {NoStop}%
\bibitem [{\citenamefont {van Bijnen}\ and\ \citenamefont
  {Pohl}(2015)}]{MacrodimerDressing1}%
  \BibitemOpen
  \bibfield  {author} {\bibinfo {author} {\bibfnamefont {R.~M.~W.}\
  \bibnamefont {van Bijnen}}\ and\ \bibinfo {author} {\bibfnamefont
  {T.}~\bibnamefont {Pohl}},\ }\href {\doibase 10.1103/PhysRevLett.114.243002}
  {\bibfield  {journal} {\bibinfo  {journal} {Phys. Rev. Lett.}\ }\textbf
  {\bibinfo {volume} {114}},\ \bibinfo {pages} {243002} (\bibinfo {year}
  {2015})}\BibitemShut {NoStop}%
\bibitem [{\citenamefont {Samboy}\ and\ \citenamefont
  {C{\^{o}}t{\'{e}}}(2011)}]{MacrodimerScaling}%
  \BibitemOpen
  \bibfield  {author} {\bibinfo {author} {\bibfnamefont {N.}~\bibnamefont
  {Samboy}}\ and\ \bibinfo {author} {\bibfnamefont {R.}~\bibnamefont
  {C{\^{o}}t{\'{e}}}},\ }\href {\doibase 10.1088/0953-4075/44/18/184006}
  {\bibfield  {journal} {\bibinfo  {journal} {Journal of Physics B: Atomic,
  Molecular and Optical Physics}\ }\textbf {\bibinfo {volume} {44}},\ \bibinfo
  {pages} {184006} (\bibinfo {year} {2011})}\BibitemShut {NoStop}%
\bibitem [{\citenamefont {Samboy}\ \emph {et~al.}(2011)\citenamefont {Samboy},
  \citenamefont {Stanojevic},\ and\ \citenamefont
  {C\^ot\'e}}]{MacrodimerPrediction2}%
  \BibitemOpen
  \bibfield  {author} {\bibinfo {author} {\bibfnamefont {N.}~\bibnamefont
  {Samboy}}, \bibinfo {author} {\bibfnamefont {J.}~\bibnamefont {Stanojevic}},
  \ and\ \bibinfo {author} {\bibfnamefont {R.}~\bibnamefont {C\^ot\'e}},\
  }\href {\doibase 10.1103/PhysRevA.83.050501} {\bibfield  {journal} {\bibinfo
  {journal} {Phys. Rev. A}\ }\textbf {\bibinfo {volume} {83}},\ \bibinfo
  {pages} {050501} (\bibinfo {year} {2011})}\BibitemShut {NoStop}%
\bibitem [{\citenamefont {Schwettmann}\ \emph {et~al.}(2007)\citenamefont
  {Schwettmann}, \citenamefont {Overstreet}, \citenamefont {Tallant},\ and\
  \citenamefont {Shaffer}}]{MacrodimerPrediction3}%
  \BibitemOpen
  \bibfield  {author} {\bibinfo {author} {\bibfnamefont {A.}~\bibnamefont
  {Schwettmann}}, \bibinfo {author} {\bibfnamefont {K.}~\bibnamefont
  {Overstreet}}, \bibinfo {author} {\bibfnamefont {J.}~\bibnamefont {Tallant}},
  \ and\ \bibinfo {author} {\bibfnamefont {J.}~\bibnamefont {Shaffer}},\ }\href
  {https://www.tandfonline.com/doi/abs/10.1080/09500340701584076} {\bibfield
  {journal} {\bibinfo  {journal} {Journal of Modern Optics}\ }\textbf {\bibinfo
  {volume} {54}},\ \bibinfo {pages} {2551} (\bibinfo {year}
  {2007})}\BibitemShut {NoStop}%
\bibitem [{\citenamefont {Overstreet}\ \emph {et~al.}(2009)\citenamefont
  {Overstreet}, \citenamefont {Schwettmann}, \citenamefont {Tallant},
  \citenamefont {Booth},\ and\ \citenamefont {Shaffer}}]{MacrodimerCs1}%
  \BibitemOpen
  \bibfield  {author} {\bibinfo {author} {\bibfnamefont {K.}~\bibnamefont
  {Overstreet}}, \bibinfo {author} {\bibfnamefont {A.}~\bibnamefont
  {Schwettmann}}, \bibinfo {author} {\bibfnamefont {J.}~\bibnamefont
  {Tallant}}, \bibinfo {author} {\bibfnamefont {D.}~\bibnamefont {Booth}}, \
  and\ \bibinfo {author} {\bibfnamefont {J.}~\bibnamefont {Shaffer}},\ }\href
  {https://www.nature.com/articles/nphys1307} {\bibfield  {journal} {\bibinfo
  {journal} {Nature Physics}\ }\textbf {\bibinfo {volume} {5}} (\bibinfo {year}
  {2009})}\BibitemShut {NoStop}%
\bibitem [{\citenamefont {Shaffer}\ \emph {et~al.}(2018)\citenamefont
  {Shaffer}, \citenamefont {Rittenhouse},\ and\ \citenamefont
  {Sadeghpour}}]{MacrodimerCs2}%
  \BibitemOpen
  \bibfield  {author} {\bibinfo {author} {\bibfnamefont {J.}~\bibnamefont
  {Shaffer}}, \bibinfo {author} {\bibfnamefont {S.}~\bibnamefont
  {Rittenhouse}}, \ and\ \bibinfo {author} {\bibfnamefont {H.}~\bibnamefont
  {Sadeghpour}},\ }\href {https://www.nature.com/articles/s41467-018-04135-6}
  {\bibfield  {journal} {\bibinfo  {journal} {Nature Communications}\ }\textbf
  {\bibinfo {volume} {9}} (\bibinfo {year} {2018})}\BibitemShut {NoStop}%
\bibitem [{\citenamefont {Sa\ss{}mannshausen}\ and\ \citenamefont
  {Deiglmayr}(2016)}]{MacrodimerCs3}%
  \BibitemOpen
  \bibfield  {author} {\bibinfo {author} {\bibfnamefont {H.}~\bibnamefont
  {Sa\ss{}mannshausen}}\ and\ \bibinfo {author} {\bibfnamefont
  {J.}~\bibnamefont {Deiglmayr}},\ }\href {\doibase
  10.1103/PhysRevLett.117.083401} {\bibfield  {journal} {\bibinfo  {journal}
  {Phys. Rev. Lett.}\ }\textbf {\bibinfo {volume} {117}},\ \bibinfo {pages}
  {083401} (\bibinfo {year} {2016})}\BibitemShut {NoStop}%
\bibitem [{\citenamefont {Šibalić}\ \emph {et~al.}(2017)\citenamefont
  {Šibalić}, \citenamefont {Pritchard}, \citenamefont {Adams},\ and\
  \citenamefont {Weatherill}}]{ARC}%
  \BibitemOpen
  \bibfield  {author} {\bibinfo {author} {\bibfnamefont {N.}~\bibnamefont
  {Šibalić}}, \bibinfo {author} {\bibfnamefont {J.}~\bibnamefont
  {Pritchard}}, \bibinfo {author} {\bibfnamefont {C.}~\bibnamefont {Adams}}, \
  and\ \bibinfo {author} {\bibfnamefont {K.}~\bibnamefont {Weatherill}},\
  }\href {\doibase https://doi.org/10.1016/j.cpc.2017.06.015} {\bibfield
  {journal} {\bibinfo  {journal} {Computer Physics Communications}\ }\textbf
  {\bibinfo {volume} {220}},\ \bibinfo {pages} {319} (\bibinfo {year}
  {2017})}\BibitemShut {NoStop}%
\bibitem [{\citenamefont {Weber}\ \emph {et~al.}(2017)\citenamefont {Weber},
  \citenamefont {Tresp}, \citenamefont {Menke}, \citenamefont {Urvoy},
  \citenamefont {Firstenberg}, \citenamefont {Büchler},\ and\ \citenamefont
  {Hofferberth}}]{PI}%
  \BibitemOpen
  \bibfield  {author} {\bibinfo {author} {\bibfnamefont {S.}~\bibnamefont
  {Weber}}, \bibinfo {author} {\bibfnamefont {C.}~\bibnamefont {Tresp}},
  \bibinfo {author} {\bibfnamefont {H.}~\bibnamefont {Menke}}, \bibinfo
  {author} {\bibfnamefont {A.}~\bibnamefont {Urvoy}}, \bibinfo {author}
  {\bibfnamefont {O.}~\bibnamefont {Firstenberg}}, \bibinfo {author}
  {\bibfnamefont {H.~P.}\ \bibnamefont {Büchler}}, \ and\ \bibinfo {author}
  {\bibfnamefont {S.}~\bibnamefont {Hofferberth}},\ }\href {\doibase
  10.1088/1361-6455/aa743a} {\bibfield  {journal} {\bibinfo  {journal} {Journal
  of Physics B: Atomic, Molecular and Optical Physics}\ }\textbf {\bibinfo
  {volume} {50}},\ \bibinfo {pages} {133001} (\bibinfo {year}
  {2017})}\BibitemShut {NoStop}%
\bibitem [{\citenamefont {Henkel}\ \emph {et~al.}(2010)\citenamefont {Henkel},
  \citenamefont {Nath},\ and\ \citenamefont {Pohl}}]{DressingDephasing2}%
  \BibitemOpen
  \bibfield  {author} {\bibinfo {author} {\bibfnamefont {N.}~\bibnamefont
  {Henkel}}, \bibinfo {author} {\bibfnamefont {R.}~\bibnamefont {Nath}}, \ and\
  \bibinfo {author} {\bibfnamefont {T.}~\bibnamefont {Pohl}},\ }\href {\doibase
  10.1103/PhysRevLett.104.195302} {\bibfield  {journal} {\bibinfo  {journal}
  {Phys. Rev. Lett.}\ }\textbf {\bibinfo {volume} {104}},\ \bibinfo {pages}
  {195302} (\bibinfo {year} {2010})}\BibitemShut {NoStop}%
\bibitem [{\citenamefont {Honer}\ \emph {et~al.}(2010)\citenamefont {Honer},
  \citenamefont {Weimer}, \citenamefont {Pfau},\ and\ \citenamefont
  {B\"uchler}}]{RydbergDressing}%
  \BibitemOpen
  \bibfield  {author} {\bibinfo {author} {\bibfnamefont {J.}~\bibnamefont
  {Honer}}, \bibinfo {author} {\bibfnamefont {H.}~\bibnamefont {Weimer}},
  \bibinfo {author} {\bibfnamefont {T.}~\bibnamefont {Pfau}}, \ and\ \bibinfo
  {author} {\bibfnamefont {H.~P.}\ \bibnamefont {B\"uchler}},\ }\href {\doibase
  10.1103/PhysRevLett.105.160404} {\bibfield  {journal} {\bibinfo  {journal}
  {Phys. Rev. Lett.}\ }\textbf {\bibinfo {volume} {105}},\ \bibinfo {pages}
  {160404} (\bibinfo {year} {2010})}\BibitemShut {NoStop}%
\bibitem [{\citenamefont {Johnson}\ and\ \citenamefont
  {Rolston}(2010)}]{Softcore}%
  \BibitemOpen
  \bibfield  {author} {\bibinfo {author} {\bibfnamefont {J.~E.}\ \bibnamefont
  {Johnson}}\ and\ \bibinfo {author} {\bibfnamefont {S.~L.}\ \bibnamefont
  {Rolston}},\ }\href {\doibase 10.1103/PhysRevA.82.033412} {\bibfield
  {journal} {\bibinfo  {journal} {Phys. Rev. A}\ }\textbf {\bibinfo {volume}
  {82}},\ \bibinfo {pages} {033412} (\bibinfo {year} {2010})}\BibitemShut
  {NoStop}%
\bibitem [{\citenamefont {Helmrich}\ \emph {et~al.}(2016)\citenamefont
  {Helmrich}, \citenamefont {Arias}, \citenamefont {Pehoviak},\ and\
  \citenamefont {Whitlock}}]{Softcore2}%
  \BibitemOpen
  \bibfield  {author} {\bibinfo {author} {\bibfnamefont {S.}~\bibnamefont
  {Helmrich}}, \bibinfo {author} {\bibfnamefont {A.}~\bibnamefont {Arias}},
  \bibinfo {author} {\bibfnamefont {N.}~\bibnamefont {Pehoviak}}, \ and\
  \bibinfo {author} {\bibfnamefont {S.}~\bibnamefont {Whitlock}},\ }\href
  {https://doi.org/10.1088/0953-4075/49/3/03lt02} {\bibfield  {journal}
  {\bibinfo  {journal} {Journal of Physics B: Atomic, Molecular and Optical
  Physics}\ }\textbf {\bibinfo {volume} {49}} (\bibinfo {year}
  {2016})}\BibitemShut {NoStop}%
\bibitem [{\citenamefont {Balewski}\ \emph {et~al.}(2014)\citenamefont
  {Balewski}, \citenamefont {Krupp}, \citenamefont {Gaj}, \citenamefont
  {Hofferberth}, \citenamefont {Löw},\ and\ \citenamefont {Pfau}}]{Softcore3}%
  \BibitemOpen
  \bibfield  {author} {\bibinfo {author} {\bibfnamefont {J.~B.}\ \bibnamefont
  {Balewski}}, \bibinfo {author} {\bibfnamefont {A.~T.}\ \bibnamefont {Krupp}},
  \bibinfo {author} {\bibfnamefont {A.}~\bibnamefont {Gaj}}, \bibinfo {author}
  {\bibfnamefont {S.}~\bibnamefont {Hofferberth}}, \bibinfo {author}
  {\bibfnamefont {R.}~\bibnamefont {Löw}}, \ and\ \bibinfo {author}
  {\bibfnamefont {T.}~\bibnamefont {Pfau}},\ }\href {\doibase
  10.1088/1367-2630/16/6/063012} {\bibfield  {journal} {\bibinfo  {journal}
  {New Journal of Physics}\ }\textbf {\bibinfo {volume} {16}},\ \bibinfo
  {pages} {063012} (\bibinfo {year} {2014})}\BibitemShut {NoStop}%
\bibitem [{\citenamefont {Hollerith}\ \emph {et~al.}(2021)\citenamefont
  {Hollerith}, \citenamefont {Rui}, \citenamefont {Rubio-Abadal}, \citenamefont
  {Srakaew}, \citenamefont {Wei}, \citenamefont {Zeiher}, \citenamefont
  {Gross},\ and\ \citenamefont {Bloch}}]{Alpha}%
  \BibitemOpen
  \bibfield  {author} {\bibinfo {author} {\bibfnamefont {S.}~\bibnamefont
  {Hollerith}}, \bibinfo {author} {\bibfnamefont {J.}~\bibnamefont {Rui}},
  \bibinfo {author} {\bibfnamefont {A.}~\bibnamefont {Rubio-Abadal}}, \bibinfo
  {author} {\bibfnamefont {K.}~\bibnamefont {Srakaew}}, \bibinfo {author}
  {\bibfnamefont {D.}~\bibnamefont {Wei}}, \bibinfo {author} {\bibfnamefont
  {J.}~\bibnamefont {Zeiher}}, \bibinfo {author} {\bibfnamefont
  {C.}~\bibnamefont {Gross}}, \ and\ \bibinfo {author} {\bibfnamefont
  {I.}~\bibnamefont {Bloch}},\ }\href {\doibase
  10.1103/PhysRevResearch.3.013252} {\bibfield  {journal} {\bibinfo  {journal}
  {Phys. Rev. Research}\ }\textbf {\bibinfo {volume} {3}},\ \bibinfo {pages}
  {013252} (\bibinfo {year} {2021})}\BibitemShut {NoStop}%
\bibitem [{\citenamefont {Bloch}\ \emph {et~al.}(2008)\citenamefont {Bloch},
  \citenamefont {Dalibard},\ and\ \citenamefont {Zwerger}}]{Wannier}%
  \BibitemOpen
  \bibfield  {author} {\bibinfo {author} {\bibfnamefont {I.}~\bibnamefont
  {Bloch}}, \bibinfo {author} {\bibfnamefont {J.}~\bibnamefont {Dalibard}}, \
  and\ \bibinfo {author} {\bibfnamefont {W.}~\bibnamefont {Zwerger}},\ }\href
  {\doibase 10.1103/RevModPhys.80.885} {\bibfield  {journal} {\bibinfo
  {journal} {Rev. Mod. Phys.}\ }\textbf {\bibinfo {volume} {80}},\ \bibinfo
  {pages} {885} (\bibinfo {year} {2008})}\BibitemShut {NoStop}%
\bibitem [{\citenamefont {Beterov}\ \emph {et~al.}(2009)\citenamefont
  {Beterov}, \citenamefont {Ryabtsev}, \citenamefont {Tretyakov},\ and\
  \citenamefont {Entin}}]{DissipationI}%
  \BibitemOpen
  \bibfield  {author} {\bibinfo {author} {\bibfnamefont {I.~I.}\ \bibnamefont
  {Beterov}}, \bibinfo {author} {\bibfnamefont {I.~I.}\ \bibnamefont
  {Ryabtsev}}, \bibinfo {author} {\bibfnamefont {D.~B.}\ \bibnamefont
  {Tretyakov}}, \ and\ \bibinfo {author} {\bibfnamefont {V.~M.}\ \bibnamefont
  {Entin}},\ }\href {\doibase 10.1103/PhysRevA.79.052504} {\bibfield  {journal}
  {\bibinfo  {journal} {Phys. Rev. A}\ }\textbf {\bibinfo {volume} {79}},\
  \bibinfo {pages} {052504} (\bibinfo {year} {2009})}\BibitemShut {NoStop}%
\bibitem [{\citenamefont {Seaton}(1983)}]{QuantumDefect1}%
  \BibitemOpen
  \bibfield  {author} {\bibinfo {author} {\bibfnamefont {M.~J.}\ \bibnamefont
  {Seaton}},\ }\href {\doibase 10.1088/0034-4885/46/2/002} {\bibfield
  {journal} {\bibinfo  {journal} {Reports on Progress in Physics}\ }\textbf
  {\bibinfo {volume} {46}},\ \bibinfo {pages} {167} (\bibinfo {year}
  {1983})}\BibitemShut {NoStop}%
\bibitem [{\citenamefont {Gruninger}\ \emph {et~al.}(1973)\citenamefont
  {Gruninger}, \citenamefont {Clements},\ and\ \citenamefont
  {Jaworowicz}}]{QuantumDefect2}%
  \BibitemOpen
  \bibfield  {author} {\bibinfo {author} {\bibfnamefont {J.}~\bibnamefont
  {Gruninger}}, \bibinfo {author} {\bibfnamefont {W.}~\bibnamefont {Clements}},
  \ and\ \bibinfo {author} {\bibfnamefont {S.}~\bibnamefont {Jaworowicz}},\
  }\href {\doibase https://doi.org/10.1002/qua.560070714} {\bibfield  {journal}
  {\bibinfo  {journal} {International Journal of Quantum Chemistry}\ }\textbf
  {\bibinfo {volume} {7}},\ \bibinfo {pages} {103} (\bibinfo {year} {1973})},\
  \Eprint
  {http://arxiv.org/abs/https://onlinelibrary.wiley.com/doi/pdf/10.1002/qua.560070714}
  {https://onlinelibrary.wiley.com/doi/pdf/10.1002/qua.560070714} \BibitemShut
  {NoStop}%
\bibitem [{\citenamefont {Cooper}\ \emph {et~al.}(2018)\citenamefont {Cooper},
  \citenamefont {Covey}, \citenamefont {Madjarov}, \citenamefont {Porsev},
  \citenamefont {Safronova},\ and\ \citenamefont {Endres}}]{RydbergSr1}%
  \BibitemOpen
  \bibfield  {author} {\bibinfo {author} {\bibfnamefont {A.}~\bibnamefont
  {Cooper}}, \bibinfo {author} {\bibfnamefont {J.~P.}\ \bibnamefont {Covey}},
  \bibinfo {author} {\bibfnamefont {I.~S.}\ \bibnamefont {Madjarov}}, \bibinfo
  {author} {\bibfnamefont {S.~G.}\ \bibnamefont {Porsev}}, \bibinfo {author}
  {\bibfnamefont {M.~S.}\ \bibnamefont {Safronova}}, \ and\ \bibinfo {author}
  {\bibfnamefont {M.}~\bibnamefont {Endres}},\ }\href {\doibase
  10.1103/PhysRevX.8.041055} {\bibfield  {journal} {\bibinfo  {journal} {Phys.
  Rev. X}\ }\textbf {\bibinfo {volume} {8}},\ \bibinfo {pages} {041055}
  (\bibinfo {year} {2018})}\BibitemShut {NoStop}%
\bibitem [{\citenamefont {Gil}\ \emph {et~al.}(2014)\citenamefont {Gil},
  \citenamefont {Mukherjee}, \citenamefont {Bridge}, \citenamefont {Jones},\
  and\ \citenamefont {Pohl}}]{RydbergSr2}%
  \BibitemOpen
  \bibfield  {author} {\bibinfo {author} {\bibfnamefont {L.~I.~R.}\
  \bibnamefont {Gil}}, \bibinfo {author} {\bibfnamefont {R.}~\bibnamefont
  {Mukherjee}}, \bibinfo {author} {\bibfnamefont {E.~M.}\ \bibnamefont
  {Bridge}}, \bibinfo {author} {\bibfnamefont {M.~P.~A.}\ \bibnamefont
  {Jones}}, \ and\ \bibinfo {author} {\bibfnamefont {T.}~\bibnamefont {Pohl}},\
  }\href {\doibase 10.1103/PhysRevLett.112.103601} {\bibfield  {journal}
  {\bibinfo  {journal} {Phys. Rev. Lett.}\ }\textbf {\bibinfo {volume} {112}},\
  \bibinfo {pages} {103601} (\bibinfo {year} {2014})}\BibitemShut {NoStop}%
\bibitem [{\citenamefont {Bridge}\ \emph {et~al.}(2016)\citenamefont {Bridge},
  \citenamefont {Keegan}, \citenamefont {Bounds}, \citenamefont {Boddy},
  \citenamefont {Sadler},\ and\ \citenamefont {Jones}}]{RydbergSr3}%
  \BibitemOpen
  \bibfield  {author} {\bibinfo {author} {\bibfnamefont {E.~M.}\ \bibnamefont
  {Bridge}}, \bibinfo {author} {\bibfnamefont {N.~C.}\ \bibnamefont {Keegan}},
  \bibinfo {author} {\bibfnamefont {A.~D.}\ \bibnamefont {Bounds}}, \bibinfo
  {author} {\bibfnamefont {D.}~\bibnamefont {Boddy}}, \bibinfo {author}
  {\bibfnamefont {D.~P.}\ \bibnamefont {Sadler}}, \ and\ \bibinfo {author}
  {\bibfnamefont {M.~P.~A.}\ \bibnamefont {Jones}},\ }\href {\doibase
  10.1364/OE.24.002281} {\bibfield  {journal} {\bibinfo  {journal} {Opt.
  Express}\ }\textbf {\bibinfo {volume} {24}},\ \bibinfo {pages} {2281}
  (\bibinfo {year} {2016})}\BibitemShut {NoStop}%
\bibitem [{\citenamefont {wei Fang}\ \emph {et~al.}(2001)\citenamefont {wei
  Fang}, \citenamefont {jun Xie}, \citenamefont {Zhang}, \citenamefont {Hu},\
  and\ \citenamefont {yan Liu}}]{RydbergYt}%
  \BibitemOpen
  \bibfield  {author} {\bibinfo {author} {\bibfnamefont {D.}~\bibnamefont {wei
  Fang}}, \bibinfo {author} {\bibfnamefont {W.}~\bibnamefont {jun Xie}},
  \bibinfo {author} {\bibfnamefont {Y.}~\bibnamefont {Zhang}}, \bibinfo
  {author} {\bibfnamefont {X.}~\bibnamefont {Hu}}, \ and\ \bibinfo {author}
  {\bibfnamefont {Y.}~\bibnamefont {yan Liu}},\ }\href {\doibase
  https://doi.org/10.1016/S0022-4073(00)00096-0} {\bibfield  {journal}
  {\bibinfo  {journal} {Journal of Quantitative Spectroscopy and Radiative
  Transfer}\ }\textbf {\bibinfo {volume} {69}},\ \bibinfo {pages} {469}
  (\bibinfo {year} {2001})}\BibitemShut {NoStop}%
\bibitem [{\citenamefont {Greiner}\ \emph {et~al.}(2002)\citenamefont
  {Greiner}, \citenamefont {Mandel}, \citenamefont {Esslinger}, \citenamefont
  {Haensch},\ and\ \citenamefont {Bloch}}]{SuperfluidMott}%
  \BibitemOpen
  \bibfield  {author} {\bibinfo {author} {\bibfnamefont {M.}~\bibnamefont
  {Greiner}}, \bibinfo {author} {\bibfnamefont {O.}~\bibnamefont {Mandel}},
  \bibinfo {author} {\bibfnamefont {T.}~\bibnamefont {Esslinger}}, \bibinfo
  {author} {\bibfnamefont {T.}~\bibnamefont {Haensch}}, \ and\ \bibinfo
  {author} {\bibfnamefont {I.}~\bibnamefont {Bloch}},\ }\href {\doibase
  10.1038/415039a} {\bibfield  {journal} {\bibinfo  {journal} {Nature}\
  }\textbf {\bibinfo {volume} {415}},\ \bibinfo {pages} {39} (\bibinfo {year}
  {2002})}\BibitemShut {NoStop}%
\bibitem [{\citenamefont {Barredo}\ \emph {et~al.}(2020)\citenamefont
  {Barredo}, \citenamefont {Lienhard}, \citenamefont {Scholl}, \citenamefont
  {de~L\'es\'eleuc}, \citenamefont {Boulier}, \citenamefont {Browaeys},\ and\
  \citenamefont {Lahaye}}]{OpticalTweezers1}%
  \BibitemOpen
  \bibfield  {author} {\bibinfo {author} {\bibfnamefont {D.}~\bibnamefont
  {Barredo}}, \bibinfo {author} {\bibfnamefont {V.}~\bibnamefont {Lienhard}},
  \bibinfo {author} {\bibfnamefont {P.}~\bibnamefont {Scholl}}, \bibinfo
  {author} {\bibfnamefont {S.}~\bibnamefont {de~L\'es\'eleuc}}, \bibinfo
  {author} {\bibfnamefont {T.}~\bibnamefont {Boulier}}, \bibinfo {author}
  {\bibfnamefont {A.}~\bibnamefont {Browaeys}}, \ and\ \bibinfo {author}
  {\bibfnamefont {T.}~\bibnamefont {Lahaye}},\ }\href {\doibase
  10.1103/PhysRevLett.124.023201} {\bibfield  {journal} {\bibinfo  {journal}
  {Phys. Rev. Lett.}\ }\textbf {\bibinfo {volume} {124}},\ \bibinfo {pages}
  {023201} (\bibinfo {year} {2020})}\BibitemShut {NoStop}%
\bibitem [{\citenamefont {Kim}\ \emph {et~al.}(2018)\citenamefont {Kim},
  \citenamefont {Park}, \citenamefont {Kim}, \citenamefont {Sim},\ and\
  \citenamefont {Ahn}}]{OpticalTweezers2}%
  \BibitemOpen
  \bibfield  {author} {\bibinfo {author} {\bibfnamefont {H.}~\bibnamefont
  {Kim}}, \bibinfo {author} {\bibfnamefont {Y.}~\bibnamefont {Park}}, \bibinfo
  {author} {\bibfnamefont {K.}~\bibnamefont {Kim}}, \bibinfo {author}
  {\bibfnamefont {H.-S.}\ \bibnamefont {Sim}}, \ and\ \bibinfo {author}
  {\bibfnamefont {J.}~\bibnamefont {Ahn}},\ }\href {\doibase
  10.1103/PhysRevLett.120.180502} {\bibfield  {journal} {\bibinfo  {journal}
  {Phys. Rev. Lett.}\ }\textbf {\bibinfo {volume} {120}},\ \bibinfo {pages}
  {180502} (\bibinfo {year} {2018})}\BibitemShut {NoStop}%
\bibitem [{\citenamefont {Bernien}\ \emph {et~al.}(2017)\citenamefont
  {Bernien}, \citenamefont {Schwartz}, \citenamefont {Keesling}, \citenamefont
  {Levine}, \citenamefont {Omran}, \citenamefont {Pichler}, \citenamefont
  {Choi}, \citenamefont {Zibrov}, \citenamefont {Endres}, \citenamefont
  {Greiner}, \citenamefont {Vuletic},\ and\ \citenamefont
  {Lukin}}]{OpticalTweezers3}%
  \BibitemOpen
  \bibfield  {author} {\bibinfo {author} {\bibfnamefont {H.}~\bibnamefont
  {Bernien}}, \bibinfo {author} {\bibfnamefont {S.}~\bibnamefont {Schwartz}},
  \bibinfo {author} {\bibfnamefont {A.}~\bibnamefont {Keesling}}, \bibinfo
  {author} {\bibfnamefont {H.}~\bibnamefont {Levine}}, \bibinfo {author}
  {\bibfnamefont {A.}~\bibnamefont {Omran}}, \bibinfo {author} {\bibfnamefont
  {H.}~\bibnamefont {Pichler}}, \bibinfo {author} {\bibfnamefont
  {S.}~\bibnamefont {Choi}}, \bibinfo {author} {\bibfnamefont {A.}~\bibnamefont
  {Zibrov}}, \bibinfo {author} {\bibfnamefont {M.}~\bibnamefont {Endres}},
  \bibinfo {author} {\bibfnamefont {M.}~\bibnamefont {Greiner}}, \bibinfo
  {author} {\bibfnamefont {V.}~\bibnamefont {Vuletic}}, \ and\ \bibinfo
  {author} {\bibfnamefont {M.}~\bibnamefont {Lukin}},\ }\href {\doibase
  10.1038/nature24622} {\bibfield  {journal} {\bibinfo  {journal} {Nature}\
  }\textbf {\bibinfo {volume} {551}} (\bibinfo {year} {2017}),\
  10.1038/nature24622}\BibitemShut {NoStop}%
\bibitem [{\citenamefont {Labuhn}\ \emph {et~al.}(2016)\citenamefont {Labuhn},
  \citenamefont {Barredo}, \citenamefont {Ravets}, \citenamefont {Léséleuc},
  \citenamefont {Macri}, \citenamefont {Lahaye},\ and\ \citenamefont
  {Browaeys}}]{OpticalTweezers4}%
  \BibitemOpen
  \bibfield  {author} {\bibinfo {author} {\bibfnamefont {H.}~\bibnamefont
  {Labuhn}}, \bibinfo {author} {\bibfnamefont {D.}~\bibnamefont {Barredo}},
  \bibinfo {author} {\bibfnamefont {S.}~\bibnamefont {Ravets}}, \bibinfo
  {author} {\bibfnamefont {S.}~\bibnamefont {Léséleuc}}, \bibinfo {author}
  {\bibfnamefont {T.}~\bibnamefont {Macri}}, \bibinfo {author} {\bibfnamefont
  {T.}~\bibnamefont {Lahaye}}, \ and\ \bibinfo {author} {\bibfnamefont
  {A.}~\bibnamefont {Browaeys}},\ }\href {\doibase 10.1038/nature18274}
  {\bibfield  {journal} {\bibinfo  {journal} {Nature}\ }\textbf {\bibinfo
  {volume} {534}} (\bibinfo {year} {2016}),\ 10.1038/nature18274}\BibitemShut
  {NoStop}%
\bibitem [{\citenamefont {Theodosiou}(1984)}]{DissipationIII}%
  \BibitemOpen
  \bibfield  {author} {\bibinfo {author} {\bibfnamefont {C.~E.}\ \bibnamefont
  {Theodosiou}},\ }\href {\doibase 10.1103/PhysRevA.30.2881} {\bibfield
  {journal} {\bibinfo  {journal} {Phys. Rev. A}\ }\textbf {\bibinfo {volume}
  {30}},\ \bibinfo {pages} {2881} (\bibinfo {year} {1984})}\BibitemShut
  {NoStop}%
\bibitem [{\citenamefont {Löw}\ \emph {et~al.}(2012)\citenamefont {Löw},
  \citenamefont {Weimer}, \citenamefont {Nipper}, \citenamefont {Balewski},
  \citenamefont {Butscher}, \citenamefont {Büchler},\ and\ \citenamefont
  {Pfau}}]{Polarizability}%
  \BibitemOpen
  \bibfield  {author} {\bibinfo {author} {\bibfnamefont {R.}~\bibnamefont
  {Löw}}, \bibinfo {author} {\bibfnamefont {H.}~\bibnamefont {Weimer}},
  \bibinfo {author} {\bibfnamefont {J.}~\bibnamefont {Nipper}}, \bibinfo
  {author} {\bibfnamefont {J.~B.}\ \bibnamefont {Balewski}}, \bibinfo {author}
  {\bibfnamefont {B.}~\bibnamefont {Butscher}}, \bibinfo {author}
  {\bibfnamefont {H.~P.}\ \bibnamefont {Büchler}}, \ and\ \bibinfo {author}
  {\bibfnamefont {T.}~\bibnamefont {Pfau}},\ }\href
  {http://stacks.iop.org/0953-4075/45/i=11/a=113001} {\bibfield  {journal}
  {\bibinfo  {journal} {Journal of Physics B: Atomic, Molecular and Optical
  Physics}\ }\textbf {\bibinfo {volume} {45}},\ \bibinfo {pages} {113001}
  (\bibinfo {year} {2012})}\BibitemShut {NoStop}%
\bibitem [{\citenamefont {Cooke}\ and\ \citenamefont
  {Gallagher}(1980)}]{DissipationII}%
  \BibitemOpen
  \bibfield  {author} {\bibinfo {author} {\bibfnamefont {W.~E.}\ \bibnamefont
  {Cooke}}\ and\ \bibinfo {author} {\bibfnamefont {T.~F.}\ \bibnamefont
  {Gallagher}},\ }\href {\doibase 10.1103/PhysRevA.21.588} {\bibfield
  {journal} {\bibinfo  {journal} {Phys. Rev. A}\ }\textbf {\bibinfo {volume}
  {21}},\ \bibinfo {pages} {588} (\bibinfo {year} {1980})}\BibitemShut
  {NoStop}%
\bibitem [{\citenamefont {Mukherjee}\ \emph {et~al.}(2011)\citenamefont
  {Mukherjee}, \citenamefont {Millen}, \citenamefont {Nath}, \citenamefont
  {Jones},\ and\ \citenamefont {Pohl}}]{Lifetimes}%
  \BibitemOpen
  \bibfield  {author} {\bibinfo {author} {\bibfnamefont {R.}~\bibnamefont
  {Mukherjee}}, \bibinfo {author} {\bibfnamefont {J.}~\bibnamefont {Millen}},
  \bibinfo {author} {\bibfnamefont {R.}~\bibnamefont {Nath}}, \bibinfo {author}
  {\bibfnamefont {M.~P.~A.}\ \bibnamefont {Jones}}, \ and\ \bibinfo {author}
  {\bibfnamefont {T.}~\bibnamefont {Pohl}},\ }\href {\doibase
  10.1088/0953-4075/44/18/184010} {\bibfield  {journal} {\bibinfo  {journal}
  {Journal of Physics B: Atomic, Molecular and Optical Physics}\ }\textbf
  {\bibinfo {volume} {44}},\ \bibinfo {pages} {184010} (\bibinfo {year}
  {2011})}\BibitemShut {NoStop}%
\bibitem [{\citenamefont {Kovrizhin}\ \emph {et~al.}(2007)\citenamefont
  {Kovrizhin}, \citenamefont {Pai},\ and\ \citenamefont {Sinha}}]{NN1}%
  \BibitemOpen
  \bibfield  {author} {\bibinfo {author} {\bibfnamefont {D.}~\bibnamefont
  {Kovrizhin}}, \bibinfo {author} {\bibfnamefont {G.}~\bibnamefont {Pai}}, \
  and\ \bibinfo {author} {\bibfnamefont {S.}~\bibnamefont {Sinha}},\ }\href
  {\doibase 10.1209/epl/i2005-10231-y} {\bibfield  {journal} {\bibinfo
  {journal} {EPL (Europhysics Letters)}\ }\textbf {\bibinfo {volume} {72}},\
  \bibinfo {pages} {162} (\bibinfo {year} {2007})}\BibitemShut {NoStop}%
\bibitem [{\citenamefont {Batrouni}\ and\ \citenamefont
  {Scalettar}(2000)}]{NN3}%
  \BibitemOpen
  \bibfield  {author} {\bibinfo {author} {\bibfnamefont {G.}~\bibnamefont
  {Batrouni}}\ and\ \bibinfo {author} {\bibfnamefont {R.}~\bibnamefont
  {Scalettar}},\ }\href {\doibase 10.1103/PhysRevLett.84.1599} {\bibfield
  {journal} {\bibinfo  {journal} {Phys. Rev. Lett.}\ }\textbf {\bibinfo
  {volume} {84}},\ \bibinfo {pages} {1599} (\bibinfo {year}
  {2000})}\BibitemShut {NoStop}%
\bibitem [{\citenamefont {van Otterlo}\ \emph {et~al.}(1995)\citenamefont {van
  Otterlo}, \citenamefont {Wagenblast}, \citenamefont {Baltin}, \citenamefont
  {Bruder}, \citenamefont {Fazio},\ and\ \citenamefont {Sch\"on}}]{NN4}%
  \BibitemOpen
  \bibfield  {author} {\bibinfo {author} {\bibfnamefont {A.}~\bibnamefont {van
  Otterlo}}, \bibinfo {author} {\bibfnamefont {K.-H.}\ \bibnamefont
  {Wagenblast}}, \bibinfo {author} {\bibfnamefont {R.}~\bibnamefont {Baltin}},
  \bibinfo {author} {\bibfnamefont {C.}~\bibnamefont {Bruder}}, \bibinfo
  {author} {\bibfnamefont {R.}~\bibnamefont {Fazio}}, \ and\ \bibinfo {author}
  {\bibfnamefont {G.}~\bibnamefont {Sch\"on}},\ }\href {\doibase
  10.1103/PhysRevB.52.16176} {\bibfield  {journal} {\bibinfo  {journal} {Phys.
  Rev. B}\ }\textbf {\bibinfo {volume} {52}},\ \bibinfo {pages} {16176}
  (\bibinfo {year} {1995})}\BibitemShut {NoStop}%
\bibitem [{\citenamefont {van Dongen}(1995)}]{NN5}%
  \BibitemOpen
  \bibfield  {author} {\bibinfo {author} {\bibfnamefont {P.~G.~J.}\
  \bibnamefont {van Dongen}},\ }\href {\doibase 10.1103/PhysRevLett.74.182}
  {\bibfield  {journal} {\bibinfo  {journal} {Phys. Rev. Lett.}\ }\textbf
  {\bibinfo {volume} {74}},\ \bibinfo {pages} {182} (\bibinfo {year}
  {1995})}\BibitemShut {NoStop}%
\bibitem [{\citenamefont {Wernsdorfer}\ \emph {et~al.}(2010)\citenamefont
  {Wernsdorfer}, \citenamefont {Snoek},\ and\ \citenamefont
  {Hofstetter}}]{SingleBand2}%
  \BibitemOpen
  \bibfield  {author} {\bibinfo {author} {\bibfnamefont {J.}~\bibnamefont
  {Wernsdorfer}}, \bibinfo {author} {\bibfnamefont {M.}~\bibnamefont {Snoek}},
  \ and\ \bibinfo {author} {\bibfnamefont {W.}~\bibnamefont {Hofstetter}},\
  }\href {\doibase 10.1103/PhysRevA.81.043620} {\bibfield  {journal} {\bibinfo
  {journal} {Phys. Rev. A}\ }\textbf {\bibinfo {volume} {81}},\ \bibinfo
  {pages} {043620} (\bibinfo {year} {2010})}\BibitemShut {NoStop}%
\bibitem [{\citenamefont {Zhang}\ \emph {et~al.}(1995)\citenamefont {Zhang},
  \citenamefont {Ma}, \citenamefont {Sun}, \citenamefont {Lee},\ and\
  \citenamefont {Park}}]{ClusterPairing}%
  \BibitemOpen
  \bibfield  {author} {\bibinfo {author} {\bibfnamefont {G.~P.}\ \bibnamefont
  {Zhang}}, \bibinfo {author} {\bibfnamefont {Y.~S.}\ \bibnamefont {Ma}},
  \bibinfo {author} {\bibfnamefont {X.}~\bibnamefont {Sun}}, \bibinfo {author}
  {\bibfnamefont {K.~H.}\ \bibnamefont {Lee}}, \ and\ \bibinfo {author}
  {\bibfnamefont {T.~Y.}\ \bibnamefont {Park}},\ }\href {\doibase
  10.1103/PhysRevB.52.6081} {\bibfield  {journal} {\bibinfo  {journal} {Phys.
  Rev. B}\ }\textbf {\bibinfo {volume} {52}},\ \bibinfo {pages} {6081}
  (\bibinfo {year} {1995})}\BibitemShut {NoStop}%
\bibitem [{\citenamefont {Pisarski}\ \emph {et~al.}(2011)\citenamefont
  {Pisarski}, \citenamefont {Jones},\ and\ \citenamefont
  {Gooding}}]{ClusterDisorder}%
  \BibitemOpen
  \bibfield  {author} {\bibinfo {author} {\bibfnamefont {P.}~\bibnamefont
  {Pisarski}}, \bibinfo {author} {\bibfnamefont {R.~M.}\ \bibnamefont {Jones}},
  \ and\ \bibinfo {author} {\bibfnamefont {R.~J.}\ \bibnamefont {Gooding}},\
  }\href {\doibase 10.1103/PhysRevA.83.053608} {\bibfield  {journal} {\bibinfo
  {journal} {Phys. Rev. A}\ }\textbf {\bibinfo {volume} {83}},\ \bibinfo
  {pages} {053608} (\bibinfo {year} {2011})}\BibitemShut {NoStop}%
\bibitem [{\citenamefont {Suthar}\ \emph
  {et~al.}(2020{\natexlab{a}})\citenamefont {Suthar}, \citenamefont {Kraus},
  \citenamefont {Sable}, \citenamefont {Angom}, \citenamefont {Morigi},\ and\
  \citenamefont {Zakrzewski}}]{ClusterDipolar}%
  \BibitemOpen
  \bibfield  {author} {\bibinfo {author} {\bibfnamefont {K.}~\bibnamefont
  {Suthar}}, \bibinfo {author} {\bibfnamefont {R.}~\bibnamefont {Kraus}},
  \bibinfo {author} {\bibfnamefont {H.}~\bibnamefont {Sable}}, \bibinfo
  {author} {\bibfnamefont {D.}~\bibnamefont {Angom}}, \bibinfo {author}
  {\bibfnamefont {G.}~\bibnamefont {Morigi}}, \ and\ \bibinfo {author}
  {\bibfnamefont {J.}~\bibnamefont {Zakrzewski}},\ }\href {\doibase
  10.1103/PhysRevB.102.214503} {\bibfield  {journal} {\bibinfo  {journal}
  {Phys. Rev. B}\ }\textbf {\bibinfo {volume} {102}},\ \bibinfo {pages}
  {214503} (\bibinfo {year} {2020}{\natexlab{a}})}\BibitemShut {NoStop}%
\bibitem [{\citenamefont {Suthar}\ \emph
  {et~al.}(2020{\natexlab{b}})\citenamefont {Suthar}, \citenamefont {Sable},
  \citenamefont {Bai}, \citenamefont {Bandyopadhyay}, \citenamefont {Pal},\
  and\ \citenamefont {Angom}}]{ClusterGauge}%
  \BibitemOpen
  \bibfield  {author} {\bibinfo {author} {\bibfnamefont {K.}~\bibnamefont
  {Suthar}}, \bibinfo {author} {\bibfnamefont {H.}~\bibnamefont {Sable}},
  \bibinfo {author} {\bibfnamefont {R.}~\bibnamefont {Bai}}, \bibinfo {author}
  {\bibfnamefont {S.}~\bibnamefont {Bandyopadhyay}}, \bibinfo {author}
  {\bibfnamefont {S.}~\bibnamefont {Pal}}, \ and\ \bibinfo {author}
  {\bibfnamefont {D.}~\bibnamefont {Angom}},\ }\href {\doibase
  10.1103/PhysRevA.102.013320} {\bibfield  {journal} {\bibinfo  {journal}
  {Phys. Rev. A}\ }\textbf {\bibinfo {volume} {102}},\ \bibinfo {pages}
  {013320} (\bibinfo {year} {2020}{\natexlab{b}})}\BibitemShut {NoStop}%
\bibitem [{\citenamefont {Yamamoto}\ \emph {et~al.}(2014)\citenamefont
  {Yamamoto}, \citenamefont {Marmorini},\ and\ \citenamefont
  {Danshita}}]{ClusterTriangular}%
  \BibitemOpen
  \bibfield  {author} {\bibinfo {author} {\bibfnamefont {D.}~\bibnamefont
  {Yamamoto}}, \bibinfo {author} {\bibfnamefont {G.}~\bibnamefont {Marmorini}},
  \ and\ \bibinfo {author} {\bibfnamefont {I.}~\bibnamefont {Danshita}},\
  }\href {\doibase 10.1103/PhysRevLett.112.127203} {\bibfield  {journal}
  {\bibinfo  {journal} {Phys. Rev. Lett.}\ }\textbf {\bibinfo {volume} {112}},\
  \bibinfo {pages} {127203} (\bibinfo {year} {2014})}\BibitemShut {NoStop}%
\bibitem [{\citenamefont {Natu}\ \emph {et~al.}(2016)\citenamefont {Natu},
  \citenamefont {Mueller},\ and\ \citenamefont
  {Das~Sarma}}]{ClusterHofstadter}%
  \BibitemOpen
  \bibfield  {author} {\bibinfo {author} {\bibfnamefont {S.~S.}\ \bibnamefont
  {Natu}}, \bibinfo {author} {\bibfnamefont {E.~J.}\ \bibnamefont {Mueller}}, \
  and\ \bibinfo {author} {\bibfnamefont {S.}~\bibnamefont {Das~Sarma}},\ }\href
  {\doibase 10.1103/PhysRevA.93.063610} {\bibfield  {journal} {\bibinfo
  {journal} {Phys. Rev. A}\ }\textbf {\bibinfo {volume} {93}},\ \bibinfo
  {pages} {063610} (\bibinfo {year} {2016})}\BibitemShut {NoStop}%
\bibitem [{\citenamefont {H\"ugel}\ \emph {et~al.}(2017)\citenamefont
  {H\"ugel}, \citenamefont {Strand}, \citenamefont {Werner},\ and\
  \citenamefont {Pollet}}]{ClusterAnisotropic}%
  \BibitemOpen
  \bibfield  {author} {\bibinfo {author} {\bibfnamefont {D.}~\bibnamefont
  {H\"ugel}}, \bibinfo {author} {\bibfnamefont {H.~U.~R.}\ \bibnamefont
  {Strand}}, \bibinfo {author} {\bibfnamefont {P.}~\bibnamefont {Werner}}, \
  and\ \bibinfo {author} {\bibfnamefont {L.}~\bibnamefont {Pollet}},\ }\href
  {\doibase 10.1103/PhysRevB.96.054431} {\bibfield  {journal} {\bibinfo
  {journal} {Phys. Rev. B}\ }\textbf {\bibinfo {volume} {96}},\ \bibinfo
  {pages} {054431} (\bibinfo {year} {2017})}\BibitemShut {NoStop}%
\bibitem [{\citenamefont {Chen}\ and\ \citenamefont
  {Yang}(2017)}]{ClusterTwoSpecies}%
  \BibitemOpen
  \bibfield  {author} {\bibinfo {author} {\bibfnamefont {Y.-C.}\ \bibnamefont
  {Chen}}\ and\ \bibinfo {author} {\bibfnamefont {M.-F.}\ \bibnamefont
  {Yang}},\ }\href {\doibase 10.1088/2399-6528/aa8bfb} {\bibfield  {journal}
  {\bibinfo  {journal} {Journal of Physics Communications}\ }\textbf {\bibinfo
  {volume} {1}},\ \bibinfo {pages} {035009} (\bibinfo {year}
  {2017})}\BibitemShut {NoStop}%
\bibitem [{\citenamefont {Niyaz}\ \emph {et~al.}(1991)\citenamefont {Niyaz},
  \citenamefont {Scalettar}, \citenamefont {Fong},\ and\ \citenamefont
  {Batrouni}}]{NN2}%
  \BibitemOpen
  \bibfield  {author} {\bibinfo {author} {\bibfnamefont {P.}~\bibnamefont
  {Niyaz}}, \bibinfo {author} {\bibfnamefont {R.}~\bibnamefont {Scalettar}},
  \bibinfo {author} {\bibfnamefont {C.}~\bibnamefont {Fong}}, \ and\ \bibinfo
  {author} {\bibfnamefont {G.}~\bibnamefont {Batrouni}},\ }\href {\doibase
  10.1103/PhysRevB.44.7143} {\bibfield  {journal} {\bibinfo  {journal}
  {Physical review. B, Condensed matter}\ }\textbf {\bibinfo {volume} {44}},\
  \bibinfo {pages} {7143} (\bibinfo {year} {1991})}\BibitemShut {NoStop}%
\bibitem [{\citenamefont {Ohgoe}\ \emph {et~al.}(2012)\citenamefont {Ohgoe},
  \citenamefont {Suzuki},\ and\ \citenamefont {Kawashima}}]{NNmeanfield}%
  \BibitemOpen
  \bibfield  {author} {\bibinfo {author} {\bibfnamefont {T.}~\bibnamefont
  {Ohgoe}}, \bibinfo {author} {\bibfnamefont {T.}~\bibnamefont {Suzuki}}, \
  and\ \bibinfo {author} {\bibfnamefont {N.}~\bibnamefont {Kawashima}},\ }\href
  {\doibase 10.1103/PhysRevB.86.054520} {\bibfield  {journal} {\bibinfo
  {journal} {Phys. Rev. B}\ }\textbf {\bibinfo {volume} {86}},\ \bibinfo
  {pages} {054520} (\bibinfo {year} {2012})}\BibitemShut {NoStop}%
\bibitem [{\citenamefont {L\"uhmann}(2013)}]{ClusterLuhmann}%
  \BibitemOpen
  \bibfield  {author} {\bibinfo {author} {\bibfnamefont {D.-S.}\ \bibnamefont
  {L\"uhmann}},\ }\href {\doibase 10.1103/PhysRevA.87.043619} {\bibfield
  {journal} {\bibinfo  {journal} {Phys. Rev. A}\ }\textbf {\bibinfo {volume}
  {87}},\ \bibinfo {pages} {043619} (\bibinfo {year} {2013})}\BibitemShut
  {NoStop}%
\bibitem [{\citenamefont {Iskin}\ and\ \citenamefont
  {Freericks}(2009)}]{Perturbation}%
  \BibitemOpen
  \bibfield  {author} {\bibinfo {author} {\bibfnamefont {M.}~\bibnamefont
  {Iskin}}\ and\ \bibinfo {author} {\bibfnamefont {J.~K.}\ \bibnamefont
  {Freericks}},\ }\href {\doibase 10.1103/PhysRevA.79.053634} {\bibfield
  {journal} {\bibinfo  {journal} {Phys. Rev. A}\ }\textbf {\bibinfo {volume}
  {79}},\ \bibinfo {pages} {053634} (\bibinfo {year} {2009})}\BibitemShut
  {NoStop}%
\bibitem [{\citenamefont {Messiah}(1960)}]{Adiabatic1}%
  \BibitemOpen
  \bibfield  {author} {\bibinfo {author} {\bibfnamefont {A.}~\bibnamefont
  {Messiah}},\ }\href {\doibase 10.1119/1.1935901} {\bibfield  {journal}
  {\bibinfo  {journal} {American Journal of Physics}\ }\textbf {\bibinfo
  {volume} {28}},\ \bibinfo {pages} {580} (\bibinfo {year} {1960})},\ \Eprint
  {http://arxiv.org/abs/https://doi.org/10.1119/1.1935901}
  {https://doi.org/10.1119/1.1935901} \BibitemShut {NoStop}%
\bibitem [{\citenamefont {Nenciu}(1980)}]{Adiabatic2}%
  \BibitemOpen
  \bibfield  {author} {\bibinfo {author} {\bibfnamefont {G.}~\bibnamefont
  {Nenciu}},\ }\href {\doibase 10.1088/0305-4470/13/2/002} {\bibfield
  {journal} {\bibinfo  {journal} {Journal of Physics A: Mathematical and
  General}\ }\textbf {\bibinfo {volume} {13}},\ \bibinfo {pages} {L15}
  (\bibinfo {year} {1980})}\BibitemShut {NoStop}%
\bibitem [{\citenamefont {Johnstone}\ \emph {et~al.}(2019)\citenamefont
  {Johnstone}, \citenamefont {Westerberg}, \citenamefont {Duncan},\ and\
  \citenamefont {\"Ohberg}}]{NN6}%
  \BibitemOpen
  \bibfield  {author} {\bibinfo {author} {\bibfnamefont {D.}~\bibnamefont
  {Johnstone}}, \bibinfo {author} {\bibfnamefont {N.}~\bibnamefont
  {Westerberg}}, \bibinfo {author} {\bibfnamefont {C.~W.}\ \bibnamefont
  {Duncan}}, \ and\ \bibinfo {author} {\bibfnamefont {P.}~\bibnamefont
  {\"Ohberg}},\ }\href {\doibase 10.1103/PhysRevA.100.043614} {\bibfield
  {journal} {\bibinfo  {journal} {Phys. Rev. A}\ }\textbf {\bibinfo {volume}
  {100}},\ \bibinfo {pages} {043614} (\bibinfo {year} {2019})}\BibitemShut
  {NoStop}%
\bibitem [{\citenamefont {Li}\ \emph {et~al.}(2013)\citenamefont {Li},
  \citenamefont {He},\ and\ \citenamefont {Hofstetter}}]{NN7}%
  \BibitemOpen
  \bibfield  {author} {\bibinfo {author} {\bibfnamefont {Y.}~\bibnamefont
  {Li}}, \bibinfo {author} {\bibfnamefont {L.}~\bibnamefont {He}}, \ and\
  \bibinfo {author} {\bibfnamefont {W.}~\bibnamefont {Hofstetter}},\ }\href
  {\doibase 10.1103/PhysRevA.87.051604} {\bibfield  {journal} {\bibinfo
  {journal} {Phys. Rev. A}\ }\textbf {\bibinfo {volume} {87}},\ \bibinfo
  {pages} {051604} (\bibinfo {year} {2013})}\BibitemShut {NoStop}%
\end{thebibliography}
%

\section*{Supplementary material}

\begin{figure}[b]
\includegraphics[width = 0.48\textwidth, trim = {0 0 0 0}, clip]{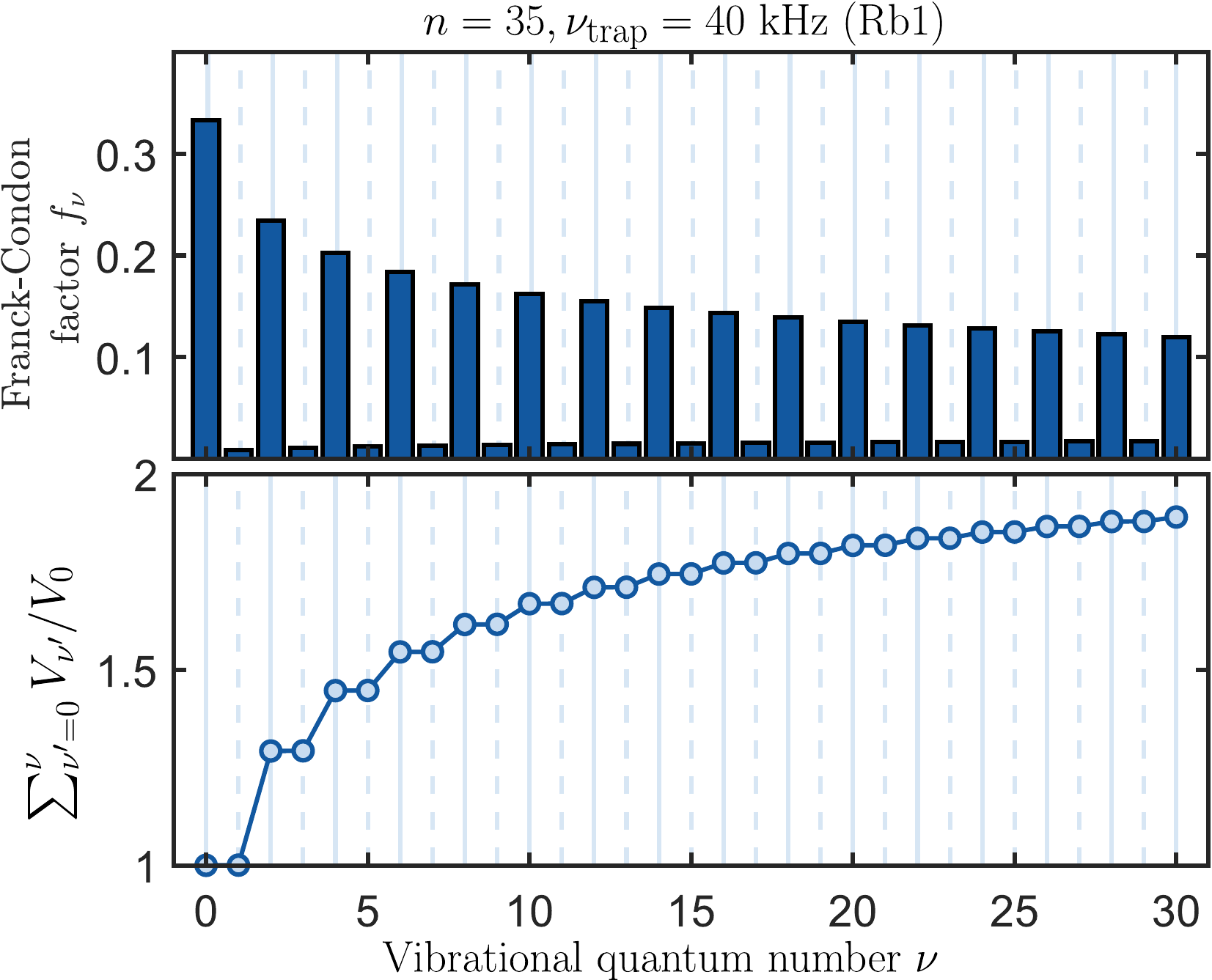}
\caption{Upper diagram depicts Franck-Condon factors $f_\nu$ versus vibrational quantum number $\nu$ for the interaction potential Rb1 and fixed $n = 55$, trapping frequency $\nu_\text{trap} = 40$ kHz. Lower diagram shows the accumulated contribution of vibrational bound states versus $\nu$ for $\delta = 2\pi \times 3$ MHz.}
\label{fig:f0}
\end{figure}
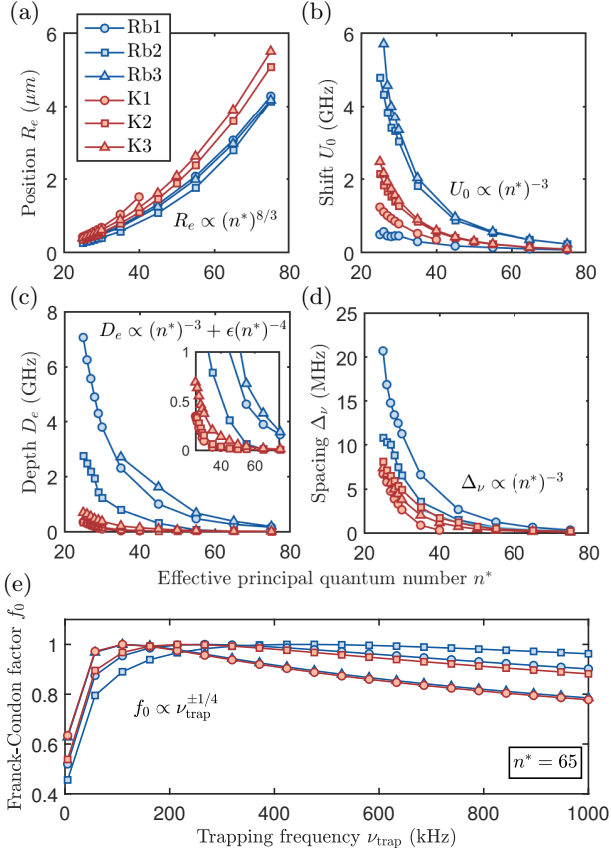
\begin{figure}[t]
\begin{overpic}[width = 0.48\textwidth, trim = {10 30 10 15}, clip]{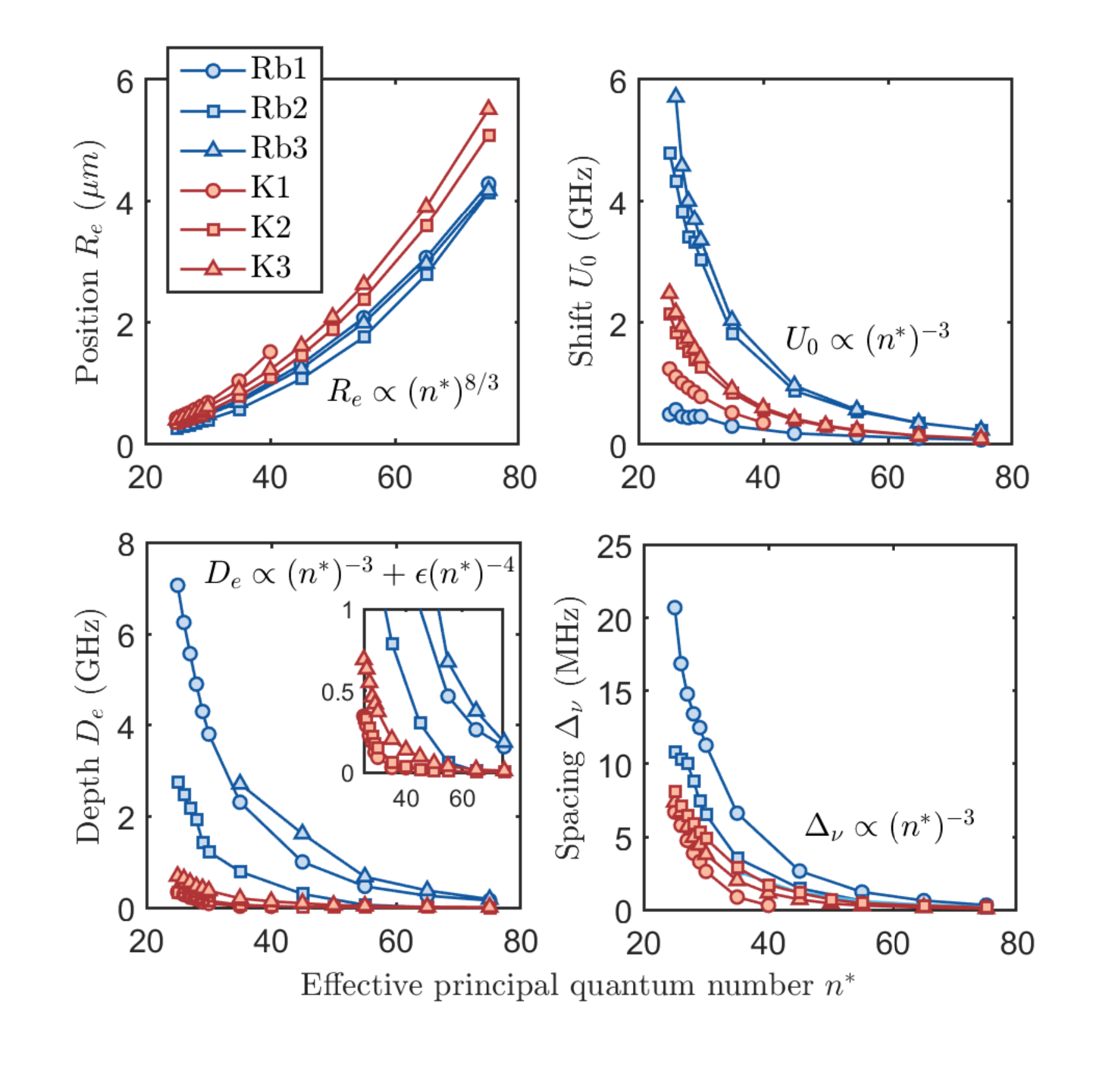}
\put(3,88) {\small(a)}
\put(48,88) {\small(b)}
\put(3,44) {\small(c)}
\put(48,44) {\small(d)}
\end{overpic}
\begin{overpic}[width = 0.48\textwidth, trim = {10 0 12 -10}, clip]{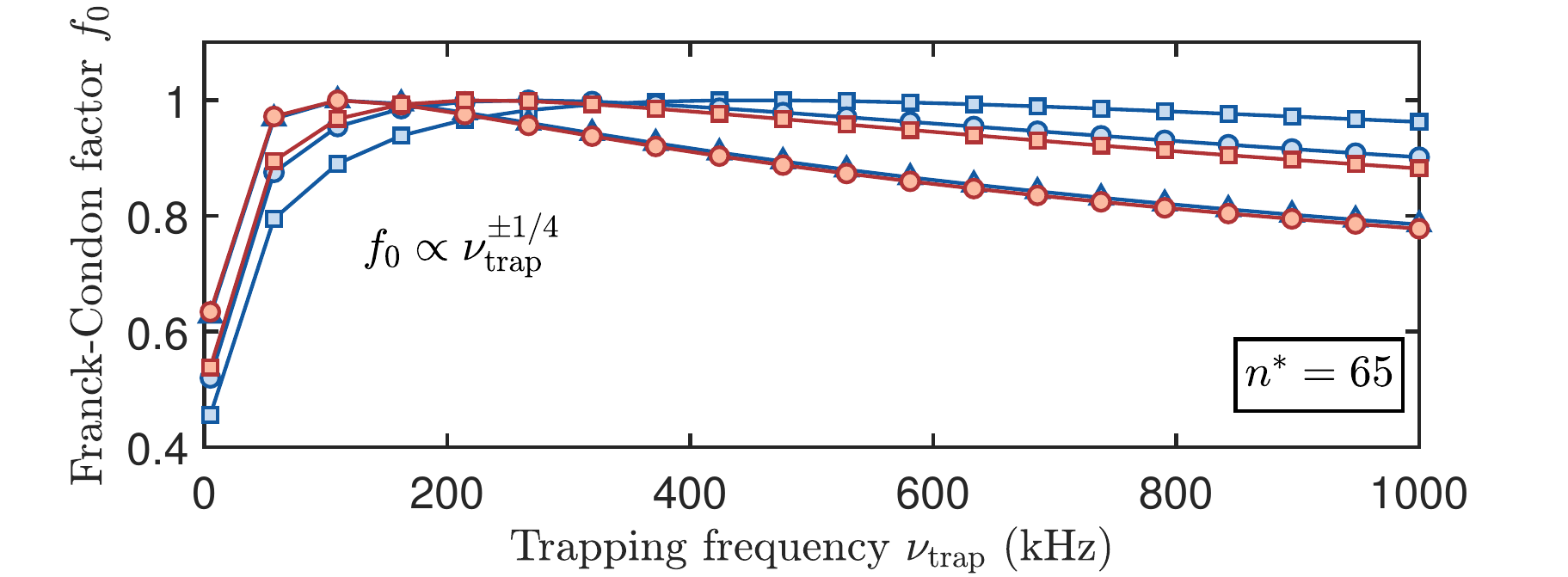}
\put(2,40) {\small(e)}
\end{overpic}
\caption{(a-d) Position of the potential well $R_e$ (a), shift $U_0$ (b), potential depth $D_e$ (c) and spacing $\Delta_\nu$ (d) versus effective principal quantum number $n^*$. The scaling relations of $R_e$ and $D_e$ fit a previous study ~\cite{MacrodimerScaling}. (e) Franck-Condon factor $f_0$ versus trapping frequency $\nu_{\text{trap}}$ for fixed effective principal quantum number $n^* = 65$.}
\label{fig:scalings}
\end{figure}
\subsection*{Appendix A: Dressing Hamiltonian and effective interaction}
In order to determine the effective interaction of the dressed regime, we write the Hamiltonian of the system in the basis $\mathcal{B} = \{|gg\rangle, |\Psi_\text{mol}^0\rangle, |\Psi_\text{mol}^\nu\rangle\}$, after adiabatic elimination of the intermediate state, which reads
\begin{equation}
\hat{H} = \frac{\hbar}{2}
\begin{pmatrix}                                
0 & \tilde \Omega_0 & \tilde \Omega_1 & \cdots \\                                               
\tilde \Omega_0  & \delta_0 & 0 & \cdots \\                                               
\tilde \Omega_1  & 0 & \delta_1 & \cdots \\
\vdots & \vdots & \vdots & \ddots \\                                               
\end{pmatrix}
\end{equation}
with $\delta_\nu = \delta + \Delta_\nu$ and $\Delta_\nu$ being the energy spacing between the lowest and the $\nu$-th vibrational bound state, and $\delta$ the two-photon detuning to the lowest vibrational state $\nu = 0$. As our laser frequency is close to the lowest vibrational resonance, all other vibrational states become far-off detuned $|\delta| \gg \tilde \Omega_\nu$. Additionally the Franck-Condon integral $f_\nu = \int \Phi^*_\nu(R) \Phi_g(R) \text{d}R$ maximizes for the lowest vibrational state. For all potentials studied here, we obtain Franck-Condon factors $f_\nu$ which decrease with increasing vibrational quantum number $\nu$ (see FIG. \ref{fig:f0}), implying weaker coupling for higher vibrational states. Within the dressing regime $|\delta| \gg \tilde \Omega_0$, we determine the dressed interaction $V = \sum_{\nu} V_\nu$, where $V_\nu = \hbar \tilde \Omega_{\nu}^2/4\delta_\nu$ is the contribution of the molecular state with vibrational quantum number $\nu$ to the full interaction. The correction to the approximated dressed interaction $V_0$ of the lowest vibrational state by including higher lying vibrational states up to state $\nu'$ can be calculated via $\sum_{\nu' = 0}^{\nu} V_{\nu'}/V_0$, which is valid for both the single-color and two-color scheme. For a two-photon detuning $\delta = 2\pi \times 3$ MHz to the lowest vibrational state, we obtain an additional factor of two through the contribution of higher vibrational bound states. The additional contributions mainly come from other lower vibrational states, while states with very high vibrational quantum number have negligible contribution. We also find that the contribution of higher vibrational states slightly depend on other parameters such as the trapping frequency.
We furthermore want to mention that for small intermediate state detuning the coupling to the motional states has to be taken into account.

\subsection*{Appendix B: Further scaling properties of the potential wells}
We compute various properties of the potential well minimum, namely the position of the potential well $R_e$, the shift $U_0$, the potential well depth $D_e$ and the spacing $\Delta_\nu$ of the vibrational levels, and identify their scaling relations with respect to $n^*$ (see FIG. \ref{fig:scalings}). With growing $n^*$ the energy scales of the potential curve diminish, meaning that asymptotic pair states become energetically closer. This results in more shallow potential wells $D_e$ and consequently narrower spacing $\Delta_\nu$. We also see the position $R_e$ of the potential well increasing with higher effective principal quantum numbers. We confirm previously obtained scaling laws for the position and the depth ~\cite{MacrodimerScaling}.\\
We further investigate the dependence of the Franck-Condon factor $f_0$ on the trapping frequency $\nu_\text{trap}$ for high principal quantum number $n = 75$ (see FIG. \ref{fig:scalings}. (e)). We obtain a scaling law of $f_0 \propto \nu_{\text{trap}}^{\pm 1/4}$, which is identical to the scaling of the width of the ground state wave function. The maximum implies that both the motional ground state and the lowest vibrational state wave function are identical. Increasing or decreasing the trapping frequency narrows or broadens the ground state wave function and consequently diminishes the overlap between the wave functions.

\subsection*{Appendix C: Dressing quality of coupling schemes and typical scattering rates}
Within the conventional one-photon dressing scheme defined through the Rabi coupling $\Omega$ and the detuning $\Delta$ to the bare Rydberg state, the decoherence rate is defined as $\Gamma_\text{Ryd} = P_\text{Ryd} \gamma$ with the Rydberg admixture $P_\text{Ryd} = \Omega^2/(2 \Delta)^2$ and the scattering rate $\gamma$ of the bare Rydberg state. With the previously defined interaction strength we obtain $|V|/\Gamma_\text{Ryd} = \Omega^2/(2|\Delta|\gamma)$. For a weak admixture we require $\Omega/\Delta \ll 1$, which strongly inhibits the dressing quality.\\
In the macrodimer dressing scheme, the decoherence rate is defined through the decoherence rate of both the intermediate state and the molecular state as $\Gamma = \Gamma_\text{Ryd} + \Gamma_\text{mol} = P_\text{Ryd} \gamma + P_\text{mol} \gamma_\text{mol}$. The admixture of the molecular state reads $P_\text{mol} = \tilde \Omega_0^2/4\delta^2$.  The scattering rate of the macrodimer state can be approximated by twice the scattering rate of the bare Rydberg state involved in forming the macrodimer \cite{Alpha}. Hence we assume the molecular scattering rate to be approximately $\gamma_\text{mol} = 2\gamma$.\\
Within the single-color dressing scheme, the Rydberg admixture $P^\text{1C}_\text{Ryd} = \Omega^2/(2\Delta_\text{1C})^2$ and the effective coupling $\tilde \Omega^\text{1C}_0 = \alpha f_0 \Omega^2/\Delta_\text{1C}$ yield the same dressing quality as the one obtained from the two-color dressing scheme with corresponding Rydberg admixture $P^\text{2C}_\text{Ryd} = (\varepsilon \Omega)^2/(2\Delta_\text{2C})^2$ and effective coupling $\tilde \Omega^\text{2C}_0 = \alpha \varepsilon f_0 \Omega^2/\Delta_\text{2C}$, i.e. $|V^\text{1C}|/\Gamma^\text{1C} = |V^\text{2C}|/\Gamma^\text{2C}$. This hints that the dressing quality is independent of the coupling to the intermediate state and has to be optimized through the coupling to the molecular state. At the optimum detuning $\delta_\text{opt} = \alpha f_0 \Omega$, the admixture of the molecular state and the intermediate state are equal, since $P_\text{mol} = \tilde \Omega_0^2/4\delta^2 = \alpha^2 f_0^2 \Omega^4/(4 \delta^2 \Delta_\text{1C}^2) = \Omega^2/(4 \Delta_\text{1C}^2) \equiv P_\text{Ryd}$ (analogous calculation for the two-color coupling scheme).\\
For the calculation of the typical values of the decoherence rate $\Gamma$, we assume lifetimes to be around $1/\gamma \in [20,200]$ $\mu$s for $n \in [30,85]$ and temperature $T \in [0,300]$ K \cite{DissipationI}. For $\delta = \alpha f_0 \Omega = 2\pi \times 3$ MHz and $\Delta_\text{2C} = 2\pi \times 50$ MHz, we obtain a bare Rydberg scattering rate around $\Gamma_\text{Ryd} \in [1, 10]$ s$^{-1}$ and a macrodimer scattering rate around $\Gamma_\text{mol} \in [2, 15]$ s$^{-1}$.

\begin{figure}
\includegraphics[width = 0.48\textwidth, trim = {10 10 10 0}, clip]{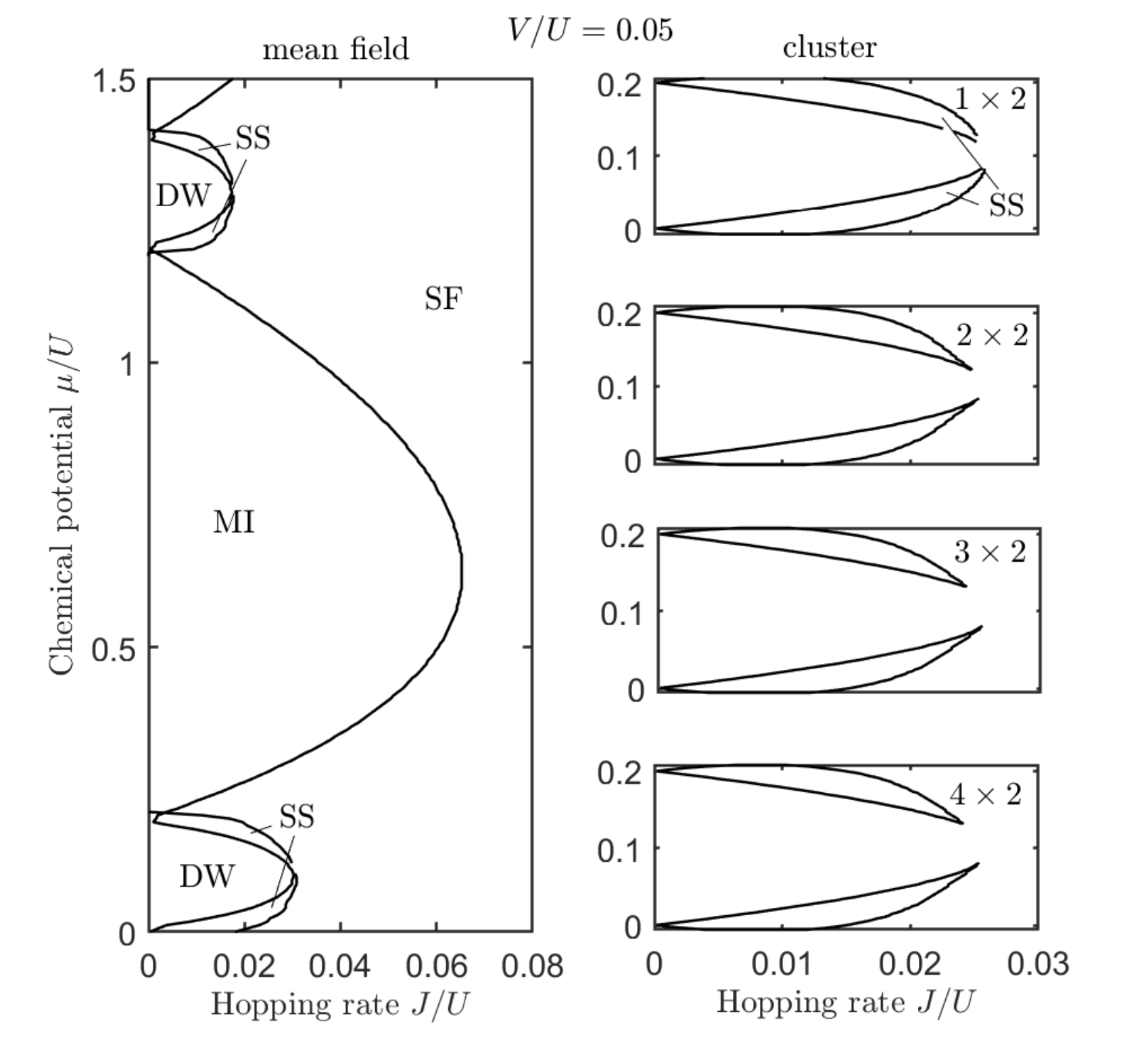}
\caption{Single-site-Gutzwiller mean-field calculation of the extended Bose-Hubbard model (left) and cluster Gutzwiller calculations of the SS regime (right) for comparison within the grand canonical ensemble. Using small clusters incorporates additional quantum fluctuations, which reduce the size of the SS regime. The difference is insignificant between the clusters of size $3\times2$ and $4\times2$ and we assume further increasing of the cluster size will not influence the phase boundaries.}
\label{fig:clusters}
\end{figure}

\subsection*{Appendix D: Cluster sizes and the influence of quantum fluctuations}
By treating the system with the CGA we are able to include non-local quantum fluctuations within the cluster in the computation of the ground state \cite{ClusterLuhmann}. Bigger cluster sizes allow for the inclusion of quantum fluctuations at larger length scales, hereby rendering the method more exact. For the extended Bose-Hubbard model studied in this work, we compute the phase boundaries between the Mott-insulating (MI), SF, DW and SS regimes for various cluster sizes (see FIG. \ref{fig:clusters}). We find a marginal shift of the phase boundaries of the SS regime by going from the single-site to the cluster Gutzwiller approximation, but do not see further changes of the phase boundaries beyond a certain cluster size.\\
In this work, we choose a cluster size of $4 \times 4$, which is sufficiently large for including important non-local quantum fluctuations. This cluster size is used for the equilibrium phase diagram computation and time evolution simulation.\\

\subsection*{Appendix E: Order parameters, mean fields and phase distinction}
We define the condensate order parameter $\phi_i = |\langle \Psi | \hat{b}_i | \Psi \rangle_\mathcal{C}|$ and the occupation number $\rho_i = \langle \Psi| \hat{b}^\dag_i \hat{b}_i | \Psi \rangle_\mathcal{C}$ at a lattice site $i$ in the cluster with the cluster wavefunction $|\Psi\rangle_\mathcal{C}$. For lattice sites on the border $i\in \partial C$, we determine the mean-fields $\varphi_i = \sum_{j \notin \mathcal{C}} \phi_j$ with $j$ being nearest neighbor of site $i$ across the border and $\eta_i = \sum_{k \notin \mathcal{C}} \rho_k$ with $k$ being either NN or NNN of site $i$ across the border, dependent on the type of interaction. These values are determined self-consistently within the iterative procedure.\\
\begin{figure}[h]
\includegraphics[width = 0.45\textwidth, trim = {15 0 0 0}, clip]{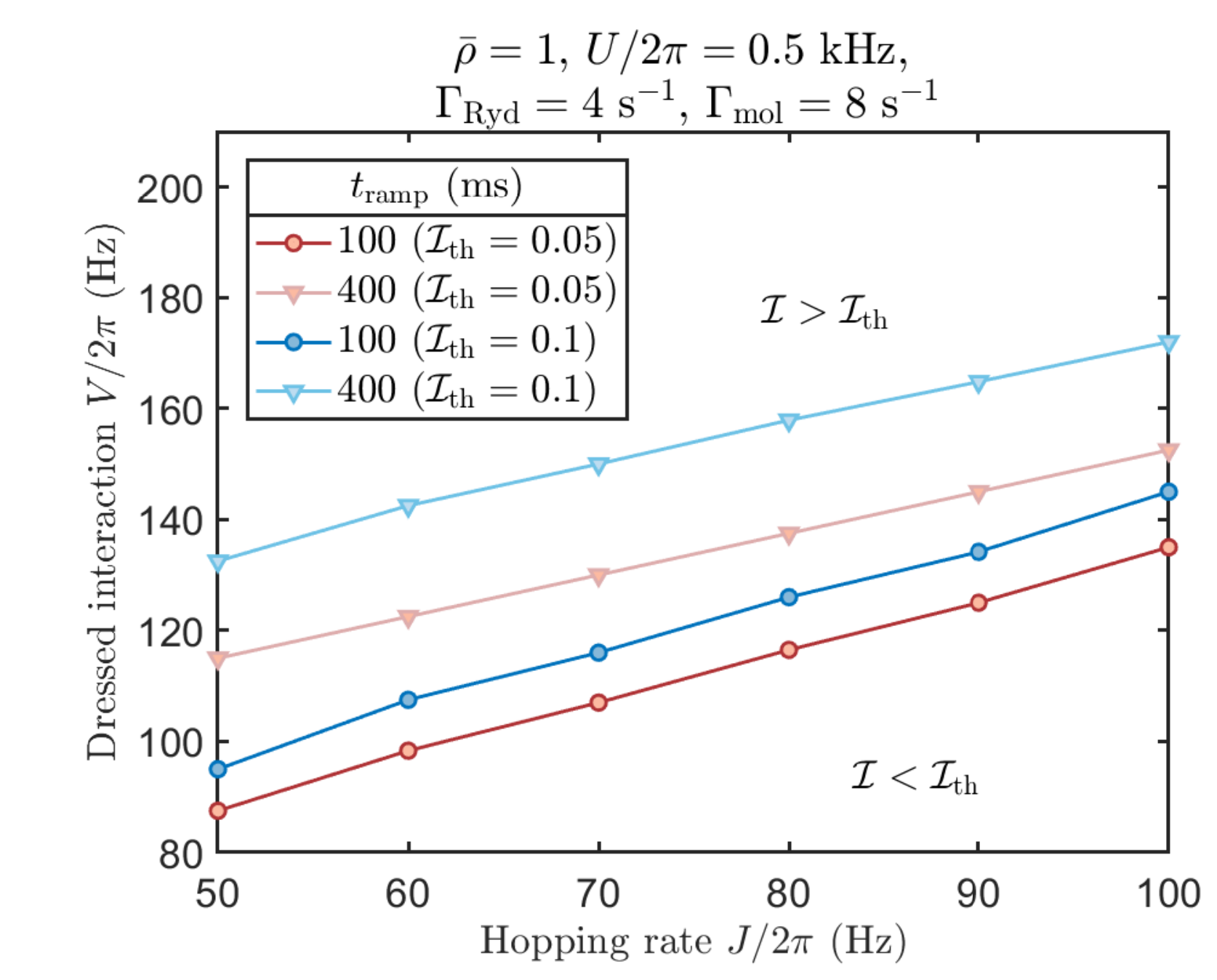}
\caption{Time evolution boundaries for different ramping times $t_\mathrm{ramp}$ and numerical thresholds $\mathcal{I}_\mathrm{th}$. A larger threshold ($\mathcal{I}_\mathrm{th} = 0.1$) shifts the boundaries previously obtained for a smaller threshold ($\mathcal{I}_\mathrm{th} = 0.05$) to higher dressed interactions $V$. The difference between the boundaries does not seem to depend on the hopping rate $J$, but increases with ramping time $t_\mathrm{ramp}$.}
\label{fig:threshold}
\end{figure}
\noindent
Suppose we split the system into $M$ unique clusters of size $N$. We thus define the mean observables $\bar{\phi} = 1/(MN) \sum_\mathcal{C} \sum_{i \in \mathcal{C}} \phi_i$ and $\bar{\rho} = 1/(MN) \sum_\mathcal{C} \sum_{i \in \mathcal{C}} \rho_i$. In the case of an external harmonic confinement, the average filling is determined in the center of the confinement, where the potential is quasi-homogeneous. We also introduce the staggered order parameter $\phi_\text{stag}$, with which we identify supersolid phases \cite{NN6,NN7}. The type of staggered order parameter depends on the expected density-wave order of the phase. In the case of NN and NNN interaction, we expect either checkerboard or stripe modulation. We therefore introduce the checkerboard order parameter $\phi_\text{stag}^\text{CH} = 1/(MN) \sum_\mathcal{C} |\sum_{i \in \mathcal{C}} (-1)^{x(i)+y(i)} \phi_i|$ and the stripe order parameter $\phi_\text{stag}^\text{STR} = 1/(MN) \sum_\mathcal{C} |\sum_{i \in \mathcal{C}} (-1)^{x(i)} \phi_i|$.\\
We classify the phases by these observables
\begin{center}
\begin{tabular}{ c|c c c c } 
 Phase & $\bar{\rho}$ & $\bar{\phi}$ & $\phi_\text{stag}$ \\  
 \hline
 \hline
 Mott insulator & $\mathbb{N}$ & 0 & 0 \\ 
 \hline
 Superfluid & $\mathbb{R}$ & $\mathbb{R}$  & 0\\ 
 \hline
 Density wave & $\mathbb{Q}$ & 0  & 0\\ 
 \hline
 Supersolid & $\mathbb{R}$ & $\mathbb{R}$  & $\mathbb{R}$ \\
\end{tabular}
\end{center}
Following the table, we identify the various phases of the equilibrium phase diagrams.\\
In the time evolution calculation, we characterize the various simulations through the imbalance $\mathcal{I}$. Since numerical fluctuations can accumulate up to $10^{-3}$, we need to define a numerical threshold $\mathcal{I}_\text{th}$, such that $\mathcal{O}(\mathcal{I}_\text{th}) > 10^{-3}$, with which we determine whether translational symmetry has been spontaneously broken or not during the evolution. In FIG. \ref{fig:time} (a), we determined the boundaries for the different ramping times with the threshold set to $\mathcal{I}_\text{th} = 0.05$. Although the chosen value is large enough for characterizing translational symmetry broken time evolution, the boundaries are not fully robust to the choice of the threshold (see FIG. \ref{fig:threshold}). A larger threshold of $\mathcal{I}_\text{th} = 0.1$ requires the dressed interaction to be larger and thus shifts the boundaries to bigger dressed interaction strengths. The shift is approximately $10 (20$) Hz for the ramping time $t_\text{ramp} = 100 (400$) ms and thus does not drastically alter the boundaries.


\end{document}